\documentclass[11pt,a4paper]{article}
\usepackage{amsmath,amssymb,epsfig,a4,latexsym}
\usepackage{graphicx}
\usepackage{cite}
\usepackage{slashed}
\usepackage{color}
\usepackage{axodraw2}
\usepackage{pifont}
\usepackage{feynarts}
\usepackage{stackrel}
\usepackage{jheppub}
\usepackage{hyperref}

\newcommand{\Eqn}[1]{Eq.~(\ref{#1})}
\newcommand{\Eqns}[2]{Eqs.~(\ref{#1}) and (\ref{#2})}

\newcommand{\eps}{\epsilon}

\def\dim{{d}}
\def\ds{{d_{s}}}
\def\dsix{{d_{e}}}
\def\CF{{C_F}}

\def\MS{{\overline{\text{\scshape{ms}}}}}
\def\OS{{\text{\scshape{os}}}}
\def\BPHZ{{\text{\scshape{bphz}}}}
\def\eq{{Q_{q}}}
\def\eqSq{{Q_{q}^{2}}}
\def\muDR{{\mu_{\text{\DR}}}}
\def\muDRpow{{\mu_{\text{\DR}}^{4-\dim}}}
\def\muDRSq{{\mu_{\text{\DR}}^{2}}}

\def\DR{{\scshape ds}}
\def\HV{{\scshape hv}}
\def\FDH{{\scshape fdh}}
\def\FDF{{\scshape fdf}}
\def\SDF{{\scshape sdf}}
\def\DRED{{\scshape dred}}
\def\CDR{{\scshape cdr}}

\def\IReg{{\scshape ireg}}
\def\FDR{{\scshape fdr}}
\def\FDU{{\scshape fdu}}
\def\LORE{{\scshape lore}}
\def\R{{\scshape r}}
\def\CT{{\scshape ct}}
\def\UV{{\scshape uv}}
\def\IR{{\scshape ir}}

\def\mDRED{{\rm\scriptscriptstyle DRED}}
\def\mCDR{{\rm\scriptscriptstyle CDR}}
\def\mHV{{\rm \scriptscriptstyle HV}}
\def\mFDH{{\rm \scriptscriptstyle FDH}}

\newcommand\sss{\scriptscriptstyle}
\newcommand{\bqa}{\begin{eqnarray}}
\newcommand{\eqa}{\end{eqnarray}}

\newcommand{\qbar}{\bar q}
\newcommand\mur{\mu_{\sss\rm R}}

\newcommand{\QQ}{Q}
\newcommand{\GG}{G}
\newcommand{\GH}{\hat G}
\newcommand{\GA}{\Gamma}

\graphicspath{{fdf.files/figs/}}

\newcommand{\Ne}{n_\epsilon}
\newcommand{\Neps}{n_\epsilon}

\newcommand{\QS}[1]{\text{QS}_{[#1]}}

\newcommand{\la}{\langle}
\newcommand{\ra}{\rangle}
\def\qb{\mathbf{q}}
\def\nn{\nonumber}
\def\td#1{\tilde{\delta}\left(#1\right)}

\def\ep{\epsilon}
\def\Eq#1{Eq.~(\ref{#1})}
\def\beq{\begin{equation}}
\def\eeq{\end{equation}}
\def\beqa{\begin{eqnarray}}
\def\eeqa{\end{eqnarray}}

\def\bq{q\hspace{-.42em}/\hspace{-.07em}}

\def\g{g_{s}}
\def\gs{g_{s}}

\def\as{\alpha_{s}}
\def\aas{\frac{\as}{4\pi}}

\def\qqg{{q \bar q g}}

\allowdisplaybreaks

\begin{document}
\thispagestyle{empty}
\begin{flushright}
PSI-PR-17-06;
ZU-TH 10/17\\
TUM-HEP-1081/17;
IFIC/17-17\\
TIF-UNIMI-2017-4
\end{flushright}
\vspace{2em}
\begin{center}
{\Large\bf To $\dim$, or not to $\dim$:\\[10pt]
Recent developments and comparisons of\\[5pt]
regularization schemes}
\\
\vspace{3em}
{\sc
  C.\,Gnendiger$^{a,}$\footnote{e-mail: Christoph.Gnendiger@psi.ch},
  A.\,Signer$^{a,b}$,
  D.\,St\"ockinger$^{c}$,\\
  A.\,Broggio$^{d}$,
  A.\,L.\,Cherchiglia$^{e}$,
  F.\,Driencourt-Mangin$^{f}$,\\
  A.\,R.\,Fazio$^{g}$,
  B.\,Hiller$^{h}$,
  P.\,Mastrolia$^{i,j}$,
  T.\,Peraro$^{k}$,
  R.\,Pittau$^{l}$,\\
  G.\,M.\,Pruna$^{a}$,
  G.\,Rodrigo$^{f}$,
  M.\,Sampaio$^{m}$,
  G.\,Sborlini$^{f,n,o}$,\\
  W.\,J.\,Torres Bobadilla$^{f,i,j}$,
  F.\,Tramontano$^{p,q}$,
  Y.\,Ulrich$^{a,b}$,
  A.\,Visconti$^{a,b}$
}\\[2em]
{\sl
  ${}^a$ Paul Scherrer Institut,
    CH-5232 Villigen PSI\\
  ${}^b$ Physik-Institut, Universit\"at Z\"urich,
    CH-8057 Z\"urich\\
  ${}^c$ Institut f\"ur Kern- und Teilchenphysik, TU Dresden,
    D-01062 Dresden\\
  ${}^d$ Physik Department T31, Technische Universit\"at M\"unchen,
    D-85748 Garching
    \\
  ${}^e$ Centro de Ci{\^e}ncias Naturais e Humanas, UFABC,
    09.210-170 Santo Andr{\'e}
    \\
  ${}^{f}$ Insituto de F\'{\i}sica Corpuscular, UVEG--CSIC,
    Universitat de Val\`encia, E-46980 Paterna\\
  ${}^g$ Departamento de F\'{\i}sica, Universidad Nacional de
    Colombia,  Bogot\'a D.C.
    \\
  ${}^h$ CFisUC, Department of Physics, University of Coimbra,
    3004-516 Coimbra\\
  ${}^i$Dipartimento di Fisica ed Astronomia, Universit\`a di Padova,
    I-35131 Padova\\
  ${}^j$INFN, Sezione di Padova, 
    I-35131 Padova\\
  ${}^k$ Higgs Centre for Theoretical Physics, The University of Edinburgh,
    EH9\,3FD Edinburgh\\
  ${}^l$ Dep.\,de\,F\'{i}sica\,Te\'orica\,y\,del\,Cosmos\,and\,CAFPE,\,
    Universidad\,de\,Granada,\,E-18071 Granada\\
  ${}^m$ Departamento de F{\"i}sica, ICEX, UFMG,
    30.161-970 Belo Horizonte\\
  ${}^{n}$ Dipartimento di Fisica, Universit\`a di Milano,
    I-20133 Milano\\
  ${}^{o}$ INFN, Sezione di Milano,
    I-20133 Milano\\
  ${}^{p}$ Dipartimento di Fisica, Universit\`a di Napoli,
    I-80126 Napoli\\
  ${}^{q}$ INFN, Sezione di Napoli,
    I-80126 Napoli\\
}
\setcounter{footnote}{0}
\end{center}
\vspace{2ex}
\begin{abstract}
{} We give an introduction to several regularization schemes that deal
with ultraviolet and infrared singularities appearing in higher-order
computations in quantum field theories. Comparing the computation of
simple quantities in the various schemes, we point out similarities
and differences between them.
\end{abstract}
\newpage
\setcounter{page}{1}

 \noindent\hrulefill
 \tableofcontents
 \noindent\hrulefill

\setcounter{footnote}{0}
\renewcommand{\thefootnote}{\arabic{footnote}}

\section{Introduction}

Higher-order calculations in quantum field theories usually involve ultraviolet
(UV) and/or infrared (IR) divergences which need to be regularized at intermediate
steps. Only after renormalization and proper combination of real and virtual
corrections, a finite and regularization-scheme independent result can be obtained.
The choice of the regularization scheme matters in several respects of conceptual
and practical relevance:
\begin{itemize}
\item mathematical consistency: It must be excluded that the calculational rules
  lead to internal inconsistencies such as final expressions contradicting each
  other.
\item unitarity and causality: The final finite result must be
  compatible with the basic quantum field theoretical properties of
  unitarity and causality. In practice this compatibility can be shown
  by proving the equivalence of a given scheme with $\MS$-
  or \BPHZ\ renormalization, which are known to have these properties.
\item symmetries: It is desirable that symmetries like Lorentz invariance,
  non-Abelian gauge invariance, or supersymmetry are manifestly preserved by
  the regularization to the largest possible extent. Symmetry breaking by the
  regularization which does not correspond to anomalies must be compensated
  by special, symmetry-restoring counterterms.
\item quantum action principle: The regularized quantum action
  principle is a relation between symmetries of the regularized
  Lagrangian and Ward/Slavnov-Taylor identities of regularized Green
  functions. If it is valid in a given regularization scheme, the
  study of symmetry properties is strongly simplified.
\item computational efficiency: The regularization scheme should allow
  for efficient calculational techniques and ideally reduce the technical
  complexity as much as possible.
\end{itemize}

\noindent
In recent years, the understanding of traditional regularization schemes has
further improved, and novel schemes have been proposed and developed. The
motivation for this progress has been to broaden the conceptual basis as well
as to enable new efficient, automated analytical and numerical calculational
methods. It appears timely to present a uniform and up-to-date description of
all schemes and to collect and compare all established properties, definitions,
and calculational procedures. This is the goal of the present report. The covered
schemes are the following:
\begin{itemize}
\item traditional dimensional schemes:
  conventional dimensional regularization (\CDR),
  the 't~Hooft-Veltman scheme (\HV),
  the four-dimensional helicity scheme (\FDH), and
  dimensional reduction (\DRED),
\item new, distinctive (re-)formulations of dimensional schemes:
  the four-dimensional formulation of the \FDH\ scheme (\FDF),
  the six-dimensional formalism (\SDF),
\item non-dimensional schemes:
  implicit regularization (\IReg),
  four-dimensional regularization/renormalization (\FDR),
  four-dimensional unsubtraction (\FDU).
\end{itemize}

\noindent
In the following we present introductions to all these schemes. Having
applications and practitioners in mind we will perform some simple
calculations to illustrate the differences as well as common features
of the schemes. In particular, we aim to sketch the computation of the
cross section for $e^+e^-\to\gamma^*\to q\bar{q}$ at next-to-leading
order and the fermion self energy. The quantities are chosen such that
potential technical disadvantages of the traditional schemes are
exposed and the properties of novel schemes with respect to UV and IR
divergences and (sub)renormalization can be illustrated. In a number
of footnotes we will directly compare intermediate results and
features of the different schemes and comment on their relation.

Of course, much more detailed information is available in the
literature and we refer to the references listed in the individual
sections for a more in-depth discussion. However, we also have to warn
the reader that, unfortunately, the nomenclature and notation used in
the literature is far from being unique.  This often leads to
misunderstandings. In an attempt to avoid these in the future, we have
adopted a unified description in this article. As a result, the
notation and terms used here will differ in parts from the notation
used in the specialized literature referred to. To help further with
clearing out some of the misunderstandings and elucidating the
relation between the schemes, we will conclude in
Sec.~\ref{sec:summary} by giving a list of concrete statements.

\section{DS: Dimensional schemes CDR, HV, FDH, DRED
\label{sec:fdh}}

\subsection{Integration in $\dim$ dimensions and dimensional schemes}

Dimensional regularization \cite{Bollini:1972ui,'tHooft:1972fi} and
variants are the most common regularization schemes for practical
calculations in gauge theories of elementary particle physics. In the 
following we summarize the basic definitions common to all dimensional
schemes (\DR) discussed in Secs.~\ref{sec:fdh} and \ref{sec:fdf} and then
provide specific definitions for four variants of \DR\ which differ by
the rules for the numerator algebra in analytical expressions.

The basic idea of all \DR\ is to regularize divergent integrals by
formally changing the dimensionality of space-time and of momentum space.
In the present report we always denote the modified space-time dimension
by $\dim$, and we set
\begin{align}
\dim &\equiv 4-2\epsilon\,.
\label{eq:d}
\end{align}
Correspondingly, a four-dimensional loop integration is replaced by a
$d$-dimensional one%
\footnote{
  In this section and in Sec.~\ref{sec:fdf}, the (quasi)dimensionality
  ${dim}$ of an object is indicated by a subscript $[{dim}]$.
  In Secs.~\ref{sec:ireg}, \ref{sec:fdr}, and \ref{sec:fdu}, where
  loop integrations are performed in strictly four dimensions, the
  subscript is suppressed unless stated otherwise.},
\begin{align}
\int \frac{d^4k_{[4]}}{(2\pi)^4}
&\quad\to\quad\muDRpow
\int \frac{d^\dim k_{[\dim]}}{(2\pi)^\dim}\,,
\label{eq:DefIntk}
\end{align}
including the scale of dimensional regularization, $\muDR$.
After this replacement, UV and IR divergent integrals lead to poles of
the form $1/\epsilon^n$.
In Refs.\ \cite{Wilson:1972cf, Collins:1984xc}, it is shown that such an
operation can indeed be defined in a mathematical consistent way and
that this operation has the expected properties such as linearity and
invariance under shifts of the integration momentum.

To define a complete regularization scheme for realistic quantum field
theories, it must be specified how to deal with $\gamma$ matrices,
metric tensors, and other objects appearing in analytical expressions.
Likewise, it should be specified how to deal with vector
fields in the regularized Lagrangian. On a basic level, two decisions
need to be made,
\begin{itemize}
\item regularize only those parts of diagrams which can lead
  to divergences, or regularize everything;
\item regularize algebraic objects like metric tensors, $\gamma$ matrices,
  and momenta in $\dim$ dimensions, or in a different dimensionality.
\end{itemize}

It turns out that there is an elegant way to unify essentially all
common variants of \DR\ in a single framework, where all definitions can
be easily formulated and where the differences and relations between the
schemes become transparent. This framework is based on distinguishing
strictly four-dimensional objects, formally $\dim$-dimensional objects,
and formally $\ds$-dimensional objects%
\footnote{In many original references, objects such as $\hat\gamma^\mu$,
  $\tilde{\gamma}^\mu$, $\breve{\gamma}^\mu$ are introduced with specific
  meanings which differ, depending on the paper, the scheme, and the context.
  Hence, we avoid such short-hand notations here and stick with a
  more explicit one to make the meaning of expressions more apparent.}.
These objects can be mathematically realized~%
\cite{Wilson:1972cf, Collins:1984xc, Stockinger:2005gx} by introducing
a strictly four-dimensional Minkowski space $\text{S}_{[4]}$ and
\textit{infinite}-dimensional vector spaces $\QS{\ds}$, $\QS{\dim}$,
$\QS{\Ne}$, which satisfy the relations
\begin{align}
 \QS{\ds}=\QS{d}\oplus \QS{\Ne}\,,
 \qquad\quad
 \text{S}_{[4]}\subset \QS{d} \, .
 \label{eq:vsDecomp}
\end{align}
The space $\QS{d}$ is the natural domain of \CDR\ and of momentum
integration in all considered schemes. Using
\vspace{-.25cm}
\begin{align}
 \dim_s \equiv d +\Neps=4-2\epsilon+\Neps\,,
\end{align}
it is enlarged to $\QS{\ds}$ via a direct (orthogonal) sum with $\QS{\Ne}$%
\footnote{In \FDH\ and \DRED, $d_s$ is usually taken to be 4, and therefore
$\Neps=2\epsilon$.}.

The structure of the vector spaces in Eq.~\eqref{eq:vsDecomp} gives rise
to the following decomposition of metric tensors and $\gamma$ matrices
\vspace{-.25cm}
\begin{align}
 g_{[\ds]}^{\mu\nu}=g_{[\dim]}^{\mu\nu}+g_{[\Neps]}^{\mu\nu}\,,\qquad\quad
 \gamma_{[\ds]}^{\mu}=\gamma_{[\dim]}^{\mu}+\gamma_{[\Neps]}^{\mu}\,.
 \label{eq:dsDecomp}
\end{align}
Since the quantities in Eq.~\eqref{eq:dsDecomp} do not have a finite-dimensional
representation, in most of the practical calculations only their algebraic
properties are relevant,
\begin{subequations}
\label{eq:algRel}
 \begin{align}
  (g_{[{dim}]})^{\mu}_{\phantom{\mu}\mu}
    &\ =\ {dim}\,,&
  (g_{[\dim]}\,g_{[\Neps]})^{\mu}_{\phantom{\mu}\nu}
    &\ =\ 0\,,&
   \label{eq:algRel1}
   \\[.25cm]
  \big\{\gamma_{[{dim}]}^{\mu},\,\gamma_{[{dim}]}^{\nu\phantom{\mu}}\big\}
    &\ =\ 2\,g_{[{dim}]}^{\mu\nu}\,,&
  \big\{\gamma_{[\dim]}^{\mu},\,\gamma_{[\Neps]}^{\nu\phantom{\mu}}\big\}
    &\ =\ 0\,,&
    \label{eq:algRel2}
 \end{align}
\end{subequations}
with ${dim}\in\{4,\,\ds,\,\dim,\,\Neps\}$. 

\begin{table}
\begin{center}
\begin{tabular}{lcccc}
			&\CDR			&\HV			&\FDH			&\DRED\phantom{\Big|}
\\
\hline
singular VF	&$g_{[\dim]}^{\mu\nu}$	&$g_{[\dim]}^{\mu\nu}$	&$g_{[\ds]}^{\mu\nu}$	&$g_{[\ds]}^{\mu\nu}$
\phantom{\Bigg|}
\\[.25cm]
regular VF	&$g_{[\dim]}^{\mu\nu}$	&$g_{[4]}^{\mu\nu}$	&$g_{[4]}^{\mu\nu}$	&$g_{[\ds]}^{\mu\nu}$
\phantom{\big|}
\end{tabular}
\end{center}
\caption{
Treatment of vector fields in the four different regularization schemes,
i.\,e.\ prescription which metric tensor has to be used in propagator numerators
and polarization sums. The quantity $d_s$ is usually taken to be 4.
This table is taken from Ref.~\cite{Signer:2008va}.
\label{tab:RSs}
}
\end{table}

Furthermore, a complete definition of the various dimensional schemes
requires to distinguish two classes of vector fields (VF)%
\footnote{Note that compared to Ref.~\cite{Signer:2008va} we replaced the
  terms 'internal' and 'external' by 'singular' and 'regular', respectively,
  to avoid possible confusion in later considerations.}:
\begin{itemize}
 \item Vector fields associated with particles in 1PI diagrams or
  with soft and collinear particles in the initial/final state
  are in the following called \emph{singular} VF.
 \item All other vector fields are called \emph{regular} VF. 
\end{itemize}
Since UV and IR divergences are only related to
\textit{singular} VF there is some freedom in the treatment of
the regular ones. In this report, we distinguish the following four \DR:
\begin{itemize}
\item \CDR\ and \HV\ are two flavours of what is commonly called
  'dimensional regularization'. They regularize algebraic objects in
  $\dim$ dimensions, $\Neps$-dimensional objects are not used.
  In \CDR, all VF are regularized, in \HV\ only singular ones.
\item \FDH\ and \DRED\ are two flavours of what is commonly called
  'dimensional reduction'. They regularize algebraic objects in
  $\ds$ dimensions. Sometimes $\ds$ is identified as $\ds\equiv4$ from
  the beginning, but it is possible to keep it as a free parameter, which
  is set to $4$ only at the end of a calculation. In \DRED, all VF
  are regularized, in \FDH\ only singular ones.
\end{itemize}
The definitions of these four schemes can be essentially reduced to
the treatment of vector fields, see Tab.~\ref{tab:RSs}.
This unified formulation of the four schemes makes obvious that
a calculation in \DRED\ covers all elements of a calculation in the
other schemes.

In \FDH\ and \DRED, where singular vector fields are treated in $\ds$
dimensions, the split of Eq.~\eqref{eq:dsDecomp} can be applied to the
regularized Lagrangian and to covariant derivatives. As an illustration,
we provide here the regularized covariant derivatives in QED and QCD,
\begin{subequations}
\label{eq:covDer}
\begin{align}
 \text{QED:}\qquad
 D_{[\ds]}^{\mu}\,\psi_i
 & =\partial_{[d]}^{\mu}\,\psi_i
 +i\Big(e\,A_{[\dim]}^{\mu}+e_e\,A_{[\Neps]}^{\mu}\Big)\,Q\,\psi_i \, ,
 \label{eq:covDerQED}
\\*
 \text{QCD:}\qquad
 D_{[\ds]}^{\mu}\,\psi_i
 & =\partial_{[d]}^{\mu}\,\psi_i
 +i\Big(g_s\,A_{[\dim]}^{\mu,a}+g_e\,A_{[\Neps]}^{\mu,a}\Big)\,T^a_{ij}\,\psi_j \, .
\label{eq:covDerQCD}
\end{align}
\end{subequations}
It is important that the gauge-field part is {\em not} written as a
complete $\ds$-dimensional entity but is split into $\dim$-dimensional
and $\Ne$-dimensional parts, and particularly with independent couplings.
Conventionally, the $\Ne$-dimensional fields are called '$\epsilon$-scalars',
the associated couplings are called 'evanescent couplings'. This split is
strictly necessary at the multi-loop level in non-supersymmetric theories
since the evanescent couplings are not protected by $\dim$-dimensional
Lorentz and gauge invariance and renormalize differently compared to the
corresponding gauge couplings.
As an example, we provide the (minimal) renormalization of the QED gauge
coupling and the corresponding evanescent coupling in \FDH/\DRED,
\begin{subequations}
\label{eq:betaFunctionsFDH}
\begin{align}
\beta_{\phantom{e}}
&=\mu^2\frac{\text{d}}{\text{d}\mu^2}\Big(\frac{e}{4\pi}\Big)^2
=-\Big(\frac{e}{4\pi}\Big)^4
  \Big[\!-\!\frac{4}{3}N_F\Big]
  +\dots\,,
\label{eq:betaQED}
\\*
\beta_e
&=\mu^2\frac{\text{d}}{\text{d}\mu^2}\Big(\frac{e_e}{4\pi}\Big)^2
=-\Big(\frac{e_e}{4\pi}\Big)^4 \big[\!-\!4\!-\!2\,N_F\big]
-\Big(\frac{e}{4\pi}\Big)^2\Big(\frac{e_e}{4\pi}\Big)^2 \big[\!+\!6\,\big]
+\dots\,.
\label{eq:betaeQED}
\end{align}
\end{subequations}
These values can be obtained e.\,g.\ from Ref.~\cite{Harlander:2006rj} by
setting $C_A\to 0$, $N_F\to 2\,N_F$. It is obvious that even for $e_e=e$,
the values of $\beta$ and $\beta_e$ are not the same.

\subsection{Application example 1: Electron self-energy at NLO}
\label{sec:fdhSE}

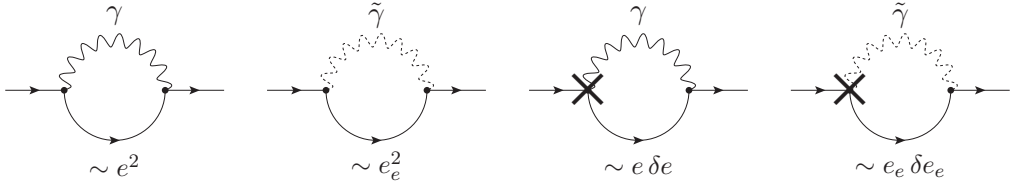
\begin{figure}[t]
   \begin{center}
   \scalebox{.63}{
   \begin{picture}(130,90)(0,0)
   \ArrowArc(65,45)(30,180,360)
   \ArrowLine(0,45)(35,45)
   \ArrowLine(95,45)(130,45)
   \Vertex(35,45){2}
   \Vertex(95,45){2}
   \PhotonArc(65,45)(30,0,180){4}{9}
   \Text(65,90){\scalebox{1.59}{$\gamma$}}
   \Text(65,0){\scalebox{1.43}{$\sim e^2$}}
   \end{picture}
   \qquad
   \begin{picture}(130,90)(0,0)
   \ArrowArc(65,45)(30,180,360)
   \ArrowLine( 0,45)( 35,45)
   \ArrowLine(95,45)(130,45)
   \Vertex(35,45){2}
   \Vertex(95,45){2}
   \DashPhotonArc[](65,45)(30,0,180){4}{9}{1.5}
   \Text(65,90){\scalebox{1.59}{$\tilde{\gamma}$}}
   \Text(65,0){\scalebox{1.43}{$\sim e_e^2$}}
   \end{picture}
   \qquad
   \begin{picture}(130,90)(0,0)
   \ArrowArc(65,45)(30,180,360)
   \ArrowLine(0,45)(35,45)
   \ArrowLine(95,45)(130,45)
   \Vertex(35,45){2}
   \Vertex(95,45){2}
   \PhotonArc(65,45)(30,0,180){4}{9}
   \Text(65,90){\scalebox{1.59}{$\gamma$}}
   \Text(35,36)[b]{\scalebox{2.5}{\ding{53}}}
   \Text(65,0){\scalebox{1.43}{$\sim e\,\delta e$}}
   \end{picture}
   \qquad
   \begin{picture}(130,90)(0,0)
   \ArrowArc(65,45)(30,180,360)
   \ArrowLine( 0,45)( 35,45)
   \ArrowLine(95,45)(130,45)
   \Vertex(35,45){2}
   \Vertex(95,45){2}
   \DashPhotonArc[](65,45)(30,0,180){4}{9}{1.5}
   \Text(65,90){\scalebox{1.59}{$\tilde{\gamma}$}}
   \Text(35,36)[b]{\scalebox{2.5}{\ding{53}}}
   \Text(65,0){\scalebox{1.43}{$\sim e_e\,\delta e_e$}}
   \end{picture}
   }
   \end{center}
   \caption{
   \label{fig:quarkSE}
   Diagrams contributing to the electron self-energy at the one- and two-loop
   level including a quasi $\dim$-dimensional photon (solid wavy line) and a
   quasi    $\Neps$-dimensional $\epsilon$-scalar (dashed wavy line), respectively.
   The insertion of a coupling counterterm is denoted by a cross.
   The $\epsilon$-scalar diagrams only exist in \FDH\ and \DRED.}
\end{figure}

To illustrate the different treatment of the Lorentz algebra in the
various \DR, we consider the electron self-energy at NLO in \DRED,
see Fig.~\ref{fig:quarkSE}. As mentioned in the previous section,
this can be seen as the most comprehensive case of the four considered
\DR. For simplicity, we use massless QED as underlying theory.
On the one hand, the Lorentz algebra can then be evaluated
by applying the split of Eq.~\eqref{eq:dsDecomp},
\begin{align}
-i\,\Sigma_{\text{\DRED}}^{(1)}&=
-i\,\Big\{\Sigma^{(1)}(e^2)+\tilde{\Sigma}^{(1)}(e_e^2)\Big\}
\nonumber\\*
&=\muDRpow\!\int\frac{d^{\dim}k_{[\dim]}}{(2\pi)^{\dim}}
 \bigg\{
  e^2\,
    \gamma^{\mu\phantom{\mu}}_{[\dim]}\,
    \gamma^{\nu\phantom{\mu}}_{[\dim]}\,
    \gamma^{\rho\phantom{\mu}}_{[\dim]}\,
    (g_{[\dim]}^{\phantom{\mu}})_{\mu\rho}
  \!+\!e_e^2\,
    \gamma^{\mu\phantom{\mu}}_{[\Neps]}\,
    \gamma^{\nu\phantom{\mu}}_{[\dim]}\,
    \gamma^{\rho\phantom{\mu}}_{[\Neps]}\,
    (g_{[\Neps]}^{\phantom{\mu}})_{\mu\rho}
    \bigg\}\,
    \frac{(k_{[\dim]}^{\phantom{\mu}})_{\nu}\,}
    {[k_{[\dim]}^{2}]\,[(k_{[\dim]}^{\phantom{2}}
      +p_{[\dim]}^{\phantom{2}})^{2}]}
\notag\\&=
\muDRpow\!\int\frac{d^{\dim}k_{[\dim]}}{(2\pi)^{\dim}}
  \bigg\{
   e^2\,\Big(
    \!-\!\gamma^{\mu\phantom{\mu}}_{[\dim]}\,
    (\gamma_{[\dim]}^{\phantom{\mu}})_{\mu}\,
    \gamma^{\nu\phantom{\mu}}_{[\dim]}\,
    \!+\!2\,\gamma^{\nu\phantom{\mu}}_{[\dim]}
    \Big)
  \!+\!e_e^2\,\Big(
    \!-\!\gamma^{\mu\phantom{\mu}}_{[\Neps]}\,
    (\gamma_{[\Neps]}^{\phantom{\mu}})_{\mu}\,
    \gamma^{\nu\phantom{\mu}}_{[\dim]}\,
    \Big)
  \bigg\}\,
\frac{(k_{[\dim]}^{\phantom{\mu}})_{\nu}}{[\,\dots]\,[\,\dots]}
\notag\\&=
\muDRpow\!\int\frac{d^{\dim}k_{[\dim]}}{(2\pi)^{\dim}}
\bigg\{
  e^2\,\Big(\!-\!\dim\!+\!2\Big)
  +e_e^2\,\Big(\dim\!-\!\ds\Big)
  \bigg\}\,
\frac{\gamma^{\nu}_{[\dim]}\,(k_{[\dim]}^{\phantom{\mu}})_{\nu}}
  {[\,\dots]\,[\,\dots]}\,,
\label{eq:se1}
\end{align}
where Feynman gauge and the equality $\Neps\!=\!(d_s\!-\!\dim)$ have
been used. Setting $\Neps\!=\!0$ then corresponds to the results in \CDR\
and \HV.

On the other hand, for $e_e=e$, the amplitude can also be evaluated
more directly by using a quasi $\ds$-dimensional algebra,
\begin{align}
-i\,\Sigma_{\text{\DRED}}^{(1)}&=
\muDRpow\int\frac{d^{\dim}k_{[\dim]}}{(2\pi)^{\dim}}
 \bigg\{
  e^2\,
    \gamma^{\mu\phantom{\mu}}_{[\ds]}\,
    \gamma^{\nu\phantom{\mu}}_{[\dim]}\,
    \gamma^{\rho\phantom{\mu}}_{[\ds]}\,
    (g_{[\ds]}^{\phantom{\mu}})_{\mu\rho}
    \bigg\}\,
    \frac{(k_{[\dim]}^{\phantom{\mu}})_{\nu}}
    {[k_{[\dim]}^{2}]\,[(k_{[\dim]}^{\phantom{2}}
      +p_{[\dim]}^{\phantom{2}})^{2}]}
\notag\\&=
\muDRpow\int\frac{d^{\dim}k_{[\dim]}}{(2\pi)^{\dim}} \bigg\{
   e^2\,\Big(
    \!-\!\gamma^{\mu\phantom{\mu}}_{[\ds]}\,
    (\gamma_{[\ds]}^{\phantom{\mu}})_{\mu}\,
    \gamma^{\nu\phantom{\mu}}_{[\ds]}\,
    \!+\!2\,\gamma^{\nu\phantom{\mu}}_{[\ds]}
    \Big)
  \bigg\}
  \frac{(k_{[\dim]}^{\phantom{\mu}})_{\nu}}
    {[\,\dots]\,[\,\dots]}
\notag\\&=
\muDRpow\int\frac{d^{\dim}k_{[\dim]}}{(2\pi)^{\dim}} \bigg\{
   e^2\,\Big(\!-\!\ds\!+\!2\Big)
 \bigg\}
 \frac{\gamma^{\nu}_{[\dim]}\,(k_{[\dim]}^{\phantom{\mu}})_{\nu}}
    {[\,\dots]\,[\,\dots]}\,.
 \label{eq:se2}
\end{align}
In the second line, the identity
$\gamma^{\nu}_{[\dim]}\,(k_{[\dim]}^{\phantom{\nu}})_{\nu}
=\gamma^{\nu}_{[\ds]}\,(k_{[\dim]}^{\phantom{\nu}})_{\nu}$
is used which directly follows from the structure of the vector
spaces in Eq.~\eqref{eq:vsDecomp}.

When setting $d_s\!=\!4$, one obtains the result in \FDH/\DRED.
Moreover, setting $e_e\!=\!e$ with $\alpha\!=\!e^2/(4\pi)$, it follows
that the different treatment of the algebra in Eqs.~\eqref{eq:se1} and
\eqref{eq:se2} yields the same result,
\begin{align}
-i\,\Sigma_{\text{\DRED}}^{(1)}
&=i\,\slashed{p}_{[\dim]}\Big(\frac{\alpha}{4\pi}\Big)\,\Big[\,
  \frac{1}{\epsilon}+2-\text{ln}
  \Big(\!-\!\frac{p_{[\dim]}^2}{\muDRSq}\Big)
  + \mathcal{O}(\epsilon) \Big]\,.
\label{eq:SEdr}
\end{align}
As long as no distinction between gauge and evanescent couplings is
required, both approaches are therefore equivalent.

At the two-loop level, however, the different UV renormalization of
$e$ and $e_e$ enters via the counterterm diagrams shown on the right
of Fig.~\ref{fig:quarkSE},
\begin{align}
-i\,\Sigma^{(2,\text{\CT})}_{\text{\DRED}}
=-i\,\Big\{
  \delta^{(1)} e^2\!\times\!\Sigma^{(1)}(e^2)
  +\delta^{(1)} e_e^2\!\times\!\tilde{\Sigma}^{(1)}(e_e^2)\Big\}\,.
\end{align}
Since no distinction between the couplings is possible when using
a quasi $\ds$-dimensional algebra, in this case it is mandatory
to apply the split of Eq.~\eqref{eq:dsDecomp}. Generalizing to an arbitrary
$\ell$-loop calculation, the introduction and separate treatment of
$\epsilon$-scalars has to be considered up to $(\ell\!-\!1)$ loops.
Genuine $\ell$-loop diagrams, on the other hand, can either be evaluated by
using the split of Eq.~\eqref{eq:dsDecomp} or by using a quasi $\ds$-dimensional
Lorentz algebra. Further details regarding the UV renormalization in the various
\DR\ can be found in Refs.~\cite{Capper:1979ns,Jack:1993ws,Jack:1994bn,
Harlander:2006rj,Harlander:2006xq}.

\subsection{Application example 2:
$e^{+}\,e^{-}\to\gamma^{*}\to\,q\bar q$ at NLO}

Any physical observable has to be independent of the regularization
scheme. What is usually done in computing NLO cross sections is to
obtain the virtual corrections in \CDR\ (either directly, or first in
another scheme and then translated to \CDR) and combine them with the
real corrections calculated in \CDR. As shown in Ref.~\cite{Signer:2008va},
it is also possible to compute the real corrections directly in schemes
other than \CDR.

We use the very simple process $e^{+}e^{-}\!\to\!\gamma^{*}\!\to\!q\bar{q}$
with massless quarks to illustrate the interplay between the scheme
dependence in the real and virtual corrections at NLO in QCD. To simplify
further, we average over the directions of the incoming leptons (with
momenta $p$ and $p'$) and actually consider only
$\gamma^{*}\!\to\!q \bar{q}$. This is achieved by replacing the
(regularization-scheme dependent) leptonic tensor by
\begin{align}
\label{leptonavg}
L_{\text{\DR}}^{\mu\nu} = (i e)^2\, 
\text{Tr}[\slashed p'\gamma^{\mu} \slashed p \gamma^\nu] \to 
4\,e^2 \frac{{dim}-2}{2\,({dim}-1)} 
\left(s\, g^{\mu\nu}_{[{dim}]} - q^\mu q^\nu\right)
\to \frac{4\, e^2}{3} s\, g^{\mu\nu}_{[{dim}]}\,,
\end{align}
where $s\equiv q^2=(p+p')^2$. In the first step, the average is taken
in ${dim}$ dimensions. However, the prefactor will be an overall
factor of the full cross section. Hence, for this prefactor we set
${dim}=4$ from the beginning and the only scheme dependence
that is left in $L_{\text{\DR}}^{\mu\nu}$ is in the one in
$g^{\mu\nu}_{[{dim}]}$.
The following discussion might create the impression that schemes
other than \CDR\ are complicated to use. However, this is simply
because we will give the details of the field-theoretic background.
This results in many apparent `complications' that can actually be
avoided at a practical level.

\begin{figure}[t]
\begin{center}
\scalebox{.7}{
\begin{picture}(135,90)(0,0)
\Vertex(45,45){2}
\Vertex(90,45){2}
\ArrowLine(45,45)(15,15)
\ArrowLine(15,75)(45,45)
\ArrowLine(120,15)(90,45)
\ArrowLine(90,45)(120,75)
\Photon(45,45)(90,45){4}{4}
\Text(0,75){\scalebox{1.43}{$e^{-}$}}
\Text(0,15){\scalebox{1.43}{$e^{+}$}}
\Text(135,75){\scalebox{1.43}{$q^{\phantom{-}}$}}
\Text(135,15){\scalebox{1.43}{$\bar{q}^{\phantom{+}}$}}
\Text(65,60){\scalebox{1.43}{$\gamma$}}
\Text(65,0){\scalebox{1.43}{$\sim e^2$}}
\end{picture}
}
\qquad\qquad
\scalebox{.7}{
\begin{picture}(135,90)(0,0)
\Vertex(45,45){2}
\Vertex(90,45){2}
\ArrowLine(45,45)(15,15)
\ArrowLine(15,75)(45,45)
\ArrowLine(120,15)(90,45)
\ArrowLine(90,45)(120,75)
\DashPhoton(45,45)(90,45){4}{4}{1.5}
\Text(0,75){\scalebox{1.43}{$e^{-}$}}
\Text(0,15){\scalebox{1.43}{$e^{+}$}}
\Text(135,75){\scalebox{1.43}{$q^{\phantom{-}}$}}
\Text(135,15){\scalebox{1.43}{$\bar{q}^{\phantom{+}}$}}
\Text(65,60){\scalebox{1.43}{$\tilde{\gamma}$}}
\Text(65,0){\scalebox{1.43}{$\sim e_e^2$}}
\end{picture}
}
\end{center}
\caption{Tree-level diagrams contributing to the process
  $e^{+}e^{-}\!\to\!\gamma^{*}\!\to\!q\bar{q}$.
  The interaction is mediated by a photon $\gamma$ (left) and an
  $\epsilon$-scalar photon $\tilde{\gamma}$ (right), respectively.
  The left diagram is present in all considered schemes, whereas
  the right one only exists in \DRED.}
\label{fdh:FdiagTree}
\end{figure}
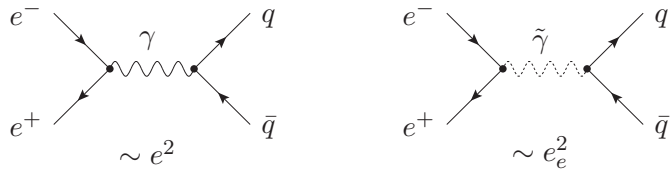

Let us begin with the most straightforward case of \CDR, where the
regular photon is treated in $\dim$ dimensions. Here, only the left
diagram in Fig.~\ref{fdh:FdiagTree} contributes. According to
Tab.~\ref{tab:RSs}, the metric tensor of the photon propagator -- and
hence in \Eqn{leptonavg} -- is $g^{\mu\nu}_{[\dim]}$, the coupling
at the vertices is the gauge coupling $e$. Using \Eqn{leptonavg}, we
get for the (spin summed/averaged) squared matrix element
$M_{\text{\DR}}^{(0)}=\langle\mathcal{A}_{\text{\DR}}^{(0)}\,
|\,\mathcal{A}_{\text{\DR}}^{(0)}\rangle$
\begin{subequations}
\label{fdh:m0}
\begin{align}
M_{\mCDR}^{(0)}
=\frac{\eqSq\,N_c}{3\,s}\,e^4\,(d-2)
\equiv\omega^{(0)} \, e^4 \, (d-2)\, ,
\end{align}
where $\eq\!=\!-1/3,\,2/3$ and $N_c$ are the electric charge and
the colour number of the quark, respectively, and the flux factor
$1/(2\,s)$ is included.

In \HV\ and \FDH, the regular photon is kept unregularized;
the related metric tensor is therefore $g^{\mu\nu}_{[4]}$.
The squared amplitudes are then given by
\begin{align}
M_{\mHV}^{(0)}
=M_\mFDH^{(0)}
=\omega^{(0)} \, e^4 \, (4-2)\,.
\end{align}
In contrast to this, in \DRED, the regular photon is treated in $\dim_s$
dimensions and thus contains a gauge-field part and an $\epsilon$-scalar
part. It is therefore possible to decompose the Born amplitude into the
two diagrams of Fig.~\ref{fdh:FdiagTree}.
The crucial point is that the diagrams involve different couplings; the
left diagram is proportional to the square of the electric gauge coupling
$e$ as in the other schemes, whereas the right diagram is proportional to
$e_e^2$. The result of the squared matrix element in \DRED\ therefore reads
\begin{align}
M_\mDRED^{(0)}
= M_\mDRED^{(0,\gamma)} +  M_\mDRED^{(0,\tilde{\gamma})}
= M_\mCDR^{(0)} +  M_\mDRED^{(0,\tilde{\gamma})}
= \omega^{(0)}\, \Big[ e^4\, (d-2) + e^4_{e}\,(\Neps)\Big]\,.
\end{align}
\end{subequations}
The appearance of a second contributions in \DRED\ is one of those
apparent complications mentioned above. In practice, one usually
sets $e_e=e$ from the beginning and computes the two processes in
a combined way like in Eq.~\eqref{eq:se2}. This is possible since the
different UV renormalization of $e$ and $e_e$ is irrelevant in this case.

Using the results in Eqs.~\eqref{fdh:m0} and integrating over the
phase space, we obtain the (scheme-independent) Born cross section
\begin{align}
\sigma^{(0)} &
=\frac{\Phi_2(\epsilon)}{8\pi}\, M_{\text{\DR}}^{(0)} \Big|_{d\to 4} 
=\frac{\eqSq\,N_c}{3\,s}\Big(\frac{e^4}{4\pi}\Big) \,,
\label{sig_yqq}
\end{align}
where we separate the $\dim$-dependent two-body phase space
\begin{align}
\Phi_2(\epsilon)
=\Big(\frac{4\pi}{s}\Big)^\epsilon
  \frac{\Gamma(1-\epsilon)}{\Gamma(2-2\epsilon)}
=1+{\cal O}(\epsilon)\,.
\label{eq:Phi2}
\end{align}

\subsubsection*{Virtual contributions}
\label{sec:fdhVirtual}

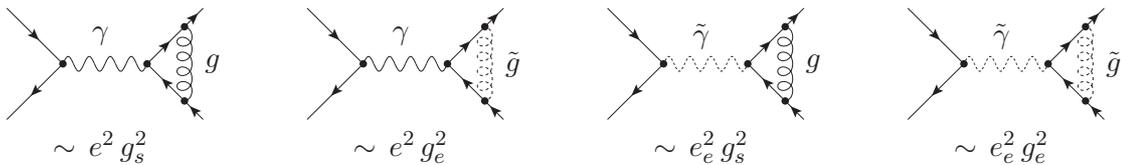
\begin{figure}[t]
\begin{center}
\scalebox{.7}{
\begin{picture}(135,90)(0,0)
\Vertex(45,45){2}
\Vertex(90,45){2}
\ArrowLine(45,45)(15,15)
\ArrowLine(15,75)(45,45)
\Photon(45,45)(90,45){4}{4}
\Text(65,60){\scalebox{1.43}{$\gamma$}}
\Text(65,0){\scalebox{1.43}{$\sim\,e^2\,g_s^2$}}
\Vertex(110,65){2}
\Vertex(110,25){2}
\ArrowLine(110,25)(90,45)
\ArrowLine(120,15)(110,25)
\ArrowLine(90,45)(110,65)
\ArrowLine(110,65)(120,75)
\Gluon(110,65)(110,25){4}{4}
\Text(125,45){\scalebox{1.43}{$g$}}
\end{picture}
\qquad
\begin{picture}(135,90)(0,0)
\Vertex(45,45){2}
\Vertex(90,45){2}
\ArrowLine(45,45)(15,15)
\ArrowLine(15,75)(45,45)
\Photon(45,45)(90,45){4}{4}
\Text(65,60){\scalebox{1.43}{$\gamma$}}
\Text(65,0){\scalebox{1.43}{$\sim\,e^2\,g_e^2$}}
\Vertex(110,65){2}
\Vertex(110,25){2}
\ArrowLine(110,25)(90,45)
\ArrowLine(120,15)(110,25)
\ArrowLine(90,45)(110,65)
\ArrowLine(110,65)(120,75)
\DashGluon(110,65)(110,25){4}{4}{1.5}
\Text(125,45){\scalebox{1.43}{$\tilde{g}$}}
\end{picture}
\qquad
\begin{picture}(135,90)(0,0)
\Vertex(45,45){2}
\Vertex(90,45){2}
\ArrowLine(45,45)(15,15)
\ArrowLine(15,75)(45,45)
\DashPhoton(45,45)(90,45){4}{4}{1.5}
\Text(65,60){\scalebox{1.43}{$\tilde{\gamma}$}}
\Text(65,0){\scalebox{1.43}{$\sim\,e_e^2\,g_s^2$}}
\Vertex(110,65){2}
\Vertex(110,25){2}
\ArrowLine(110,25)(90,45)
\ArrowLine(120,15)(110,25)
\ArrowLine(90,45)(110,65)
\ArrowLine(110,65)(120,75)
\Gluon(110,65)(110,25){4}{4}
\Text(125,45){\scalebox{1.43}{$g$}}
\end{picture}
\qquad
\begin{picture}(135,90)(0,0)
\Vertex(45,45){2}
\Vertex(90,45){2}
\ArrowLine(45,45)(15,15)
\ArrowLine(15,75)(45,45)
\DashPhoton(45,45)(90,45){4}{4}{1.5}
\Text(65,60){\scalebox{1.43}{$\tilde{\gamma}$}}
\Text(65,0){\scalebox{1.43}{$\sim\,e_e^2\,g_e^2$}}
\ArrowLine(110,25)(90,45)
\ArrowLine(120,15)(110,25)
\ArrowLine(90,45)(110,65)
\ArrowLine(110,65)(120,75)
\DashGluon(110,65)(110,25){4}{4}{1.5}
\Vertex(110,65){2}
\Vertex(110,25){2}
\Text(125,45){\scalebox{1.43}{$\tilde{g}$}}
\end{picture}
}
\end{center}
\caption{Virtual diagrams for $e^{+}e^{-}\!\to\!\gamma^{*}\!\to\!q\bar{q}$
  including a gluon $g$ or an $\epsilon$-scalar $\tilde{g}$.
  In \CDR\ and \HV, only the first diagram contributes, whereas in \FDH\ also
  the second diagram is present. In \DRED, all diagrams contribute.}
\label{fdh:FdiagVirtual}
\end{figure}

\begin{figure}[t]
\begin{center}
\scalebox{.7}{
\begin{picture}(135,90)(0,0)
\Vertex(45,45){2}
\Vertex(90,45){2}
\ArrowLine(45,45)(15,15)
\ArrowLine(15,75)(45,45)
\DashPhoton(45,45)(90,45){4}{4}{1.5}
\Text(65,60){\scalebox{1.43}{$\tilde{\gamma}$}}
\ArrowLine(120,15)(90,45)
\ArrowLine(90,45)(120,75)
\Text(90,37.5)[b]{\scalebox{2}{\ding{53}}}
\end{picture}
}
\end{center}
\vspace{-.75cm}
\caption{Counterterm diagram for $e^{+}e^{-}\!\to\!\gamma^{*}\!\to\!q\bar{q}$
  which only contributes in \DRED.}
\label{fdh:FdiagCT}
\end{figure}
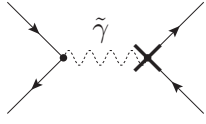

In a next step we consider the virtual corrections to the (spin
summed/averaged) squared matrix element, $M_{\text{\DR}}^{(1)}=2\operatorname{Re}
\langle\mathcal{A}_{\text{\DR}}^{(0)}\,|\,\mathcal{A}_{\text{\DR}}^{(1)}\rangle$.
To obtain the results of the corresponding one-loop amplitudes, we
have to evaluate the diagrams shown in Fig.~\ref{fdh:FdiagVirtual}.
There are two different vector fields in the one-loop diagrams, a
virtual photon that is 'regular' and a virtual gluon that is
'singular'. According to this, the treatment of the photon is as for
the Born amplitude. For \DRED, this results in two contributions, one
proportional to the gauge coupling~$e$, the other proportional to the
evanescent coupling~$e_e$. Due to the Ward identity, only the latter
coupling gets renormalized. In the $\MS$ scheme, we obtain
\begin{align}
(\eq\,e_{e})^2\ \to\ (\eq\,e_{e})^2\, \bigg\{1 
+ \Big(\frac{\alpha_s}{4\pi}\Big)\,\CF\,\Big[\!-\!\frac{3}{\epsilon}\Big]
+ \Big(\frac{\alpha_e}{4\pi}\Big)\,\CF\,\frac{4-\Ne}{2\,\epsilon}\bigg\}\,.
\end{align}
We remark that in schemes other than \CDR, the $\MS$ counterterms in general 
can have $\mathcal{O}(\Neps)$ terms, as discussed e.\,g.\ in
Ref.~\cite{Gnendiger:2014nxa}. In \DRED, one therefore has to consider
the (finite) counterterm
\begin{equation}
{\rm CT}_\mDRED =  M_\mDRED^{(0,\tilde{\gamma})}\,\CF\,
\bigg\{\Big(\frac{\alpha_s}{4\pi}\Big)\,\Big[\!-\!\frac{6}{\epsilon}\Big]
+\Big(\frac{\alpha_e}{4\pi}\Big)\,\frac{4-\Ne}{\epsilon}\bigg\}\,,
\label{fdh:uvct}
\end{equation}
see also Fig.~\ref{fdh:FdiagCT}.
In the same way, when using \FDH\ or \DRED, the gluon can be split according
to Eq.~\eqref{eq:dsDecomp}. Thus, in these schemes we get terms proportional to
$\alpha_s\!=\!g_s^2/(4\pi)$ and terms proportional to $\alpha_e\!=\!g_e^2/(4\pi)$.
The unrenormalized virtual one-loop corrections are given by
\begin{subequations}
\label{fdh:m1bare}
\begin{align}
M_{\mCDR\phantom{J}}^{(1)}&=\
      \omega^{(1)}\, M_\mCDR^{(0)}\, \Big(\frac{\alpha_s}{\pi}\Big)
      \Big[\!-\frac{1}{\epsilon^2}-\frac{3}{2\,\epsilon}-4\Big]
       + \mathcal{O}(\epsilon)\,,\phantom{\bigg\|}
      \\*
M_{\mHV\phantom{JJ}}^{(1)}&=\
      \omega^{(1)}\, M_\mHV^{(0)}\, \Big(\frac{\alpha_s}{\pi}\Big)
      \Big[\!-\frac{1}{\epsilon^2}-\frac{3}{2\,\epsilon}-4\Big]
      + \mathcal{O}(\epsilon)\,,\phantom{\bigg\|}
      \\
M_{\mFDH\phantom{J}}^{(1)}&=\
      \omega^{(1)}\, M_\mFDH^{(0)}\, \bigg\{\Big(\frac{\alpha_s}{\pi}\Big)
      \Big[\!-\frac{1}{\epsilon^2}-\frac{3}{2\,\epsilon}-4\Big]
      +\Big(\frac{\alpha_e}{\pi}\Big)
      \Big[\frac{\Ne}{4\,\epsilon}\Big]
      \bigg\}
      +\mathcal{O}(\epsilon)\,,
      \label{fdh:m1bareFDH}
      \\
M_\mDRED^{(1)}&=\
      \omega^{(1)}\,M_\mDRED^{(0,\gamma)}\, \bigg\{\Big(\frac{\alpha_s}{\pi}\Big)
      \Big[\!-\frac{1}{\epsilon^2}-\frac{3}{2\,\epsilon}-4\Big]
      +\Big(\frac{\alpha_e}{\pi}\Big)
      \Big[\frac{\Ne}{4\,\epsilon}\Big]
      \bigg\}
      \nonumber\\*
     &\ \, + \omega^{(1)}\,M_\mDRED^{(0,\tilde{\gamma})}\,
     \bigg\{\Big(\frac{\alpha_s}{\pi}\Big)
     \Big[\!-\frac{1}{\epsilon^2}\Big] +
      \Big(\frac{\alpha_e}{\pi}\Big)
      \Big[\!-\frac{1}{\epsilon}\Big] 
      \bigg\} + \mathcal{O}(\epsilon)\,,
 \end{align}
\end{subequations}
with
\begin{subequations}
\begin{align}
\label{cgammadef}
\omega^{(1)}
&\equiv \CF \, c_\Gamma(\epsilon)\, {\rm Re}\,
  (-s)^{-\epsilon} 
= \CF\,c_\Gamma(\epsilon)\,s^{-\epsilon}\,
  \Big[1-\epsilon^2\frac{\pi^2}{2}+\mathcal{O}(\epsilon^4)\Big] ,
\\*
c_\Gamma(\epsilon)
&=(4\pi)^{\epsilon}
  \frac{\Gamma(1+\epsilon)\,\Gamma^2(1-\epsilon)}{\Gamma(1-2\epsilon)}
  =1+\mathcal{O}(\epsilon)\,.
\label{cgammadef2}
\end{align}
\end{subequations}
In Eqs.~\eqref{fdh:m1bare}, we have dropped $\Ne$~terms that vanish after
setting $\Ne = 2\epsilon$ and taking the subsequent limit $\epsilon\to 0$.

In particular, the \DRED\ result looks awfully complicated. However, from
a practical point of view the situation is much simpler. As discussed in
the previous section, the virtual contributions can be computed without
distinguishing the various couplings and without splitting the photon or
the gluon.
We can simply evaluate the algebra of the single vertex diagram according to
the scheme and perform the integration. The only part where the split is
crucial so far is to obtain the UV counterterm, \Eqn{fdh:uvct}.
Thus, the computation in schemes other than \CDR\ is not significantly
more extensive.

Computing the (IR divergent) virtual cross section by integrating
the properly(!) renormalized matrix element squared over the two-parton
phase space, \Eqn{eq:Phi2}, we get
\begin{subequations}
\label{fdh:virtual}
\begin{align}
\sigma^{(v)}_\mCDR &=
\sigma^{(0)}\,\Big(\frac{\alpha_s}{\pi}\Big)\,\CF\,
  \Phi_2(\epsilon)\,c_\Gamma(\epsilon)\,
  s^{-\epsilon} \Big[
    \!-\frac{1}{\epsilon^2}
    -\frac{1}{2\, \epsilon}
    -\frac{5-\pi^2}{2} + \mathcal{O}(\epsilon) \Big]\,,
    \phantom{\bigg|}
\label{sig_eev_cdr}
\\
\sigma^{(v)}_{\mHV\phantom{J}} &=
\sigma^{(0)}\,\Big(\frac{\alpha_s}{\pi}\Big)\,\CF\,
  \Phi_2(\epsilon)\,c_\Gamma(\epsilon)\,
  s^{-\epsilon} \Big[
  \!-\frac{1}{\epsilon^2}
    -\frac{3}{2\, \epsilon}
    -\frac{8-\pi^2}{2} + \mathcal{O}(\epsilon) \Big]\,,
    \phantom{\bigg|}
\label{sig_eev_hv}
\\
\sigma^{(v)}_\mFDH =\ \sigma^{(v)}_\mDRED & =
\sigma^{(0)}\,\Big(\frac{\alpha_s}{\pi}\Big)\,\CF\,
  \Phi_2(\epsilon)\,c_\Gamma(\epsilon)\,
  s^{-\epsilon} \Big[
    \!-\frac{1}{\epsilon^2}
    -\frac{3}{2\, \epsilon}
    -\frac{7-\pi^2}{2} + \mathcal{O}(\epsilon)\Big]\,,
    \phantom{\bigg|}
\label{sig_eev_four}
\end{align}
\end{subequations}
where we have set $\Ne=2\epsilon$ and $g_e=g_s$.

\subsubsection*{Real contributions}

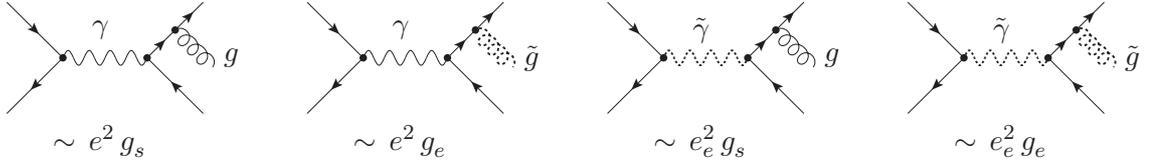
\begin{figure}[t]
\begin{center}
\scalebox{.7}{
\begin{picture}(135,90)(0,0)
\Vertex(45,45){2}
\Vertex(90,45){2}
\ArrowLine(45,45)(15,15)
\ArrowLine(15,75)(45,45)
\Photon(45,45)(90,45){4}{4}
\Text(65,60){\scalebox{1.43}{$\gamma$}}
\Text(65,0){\scalebox{1.43}{$\sim\,e^2\,g_s^{\phantom{2}}$}}
\Vertex(105,60){2}
\ArrowLine(120,15)(90,45)
\ArrowLine(90,45)(105,60)
\ArrowLine(105,60)(120,75)
\Gluon(105,60)(125,40){4}{3}
\Text(135,45){\scalebox{1.43}{$g$}}
\end{picture}
\qquad
\begin{picture}(135,90)(0,0)
\Vertex(45,45){2}
\Vertex(90,45){2}
\ArrowLine(45,45)(15,15)
\ArrowLine(15,75)(45,45)
\Photon(45,45)(90,45){4}{4}
\Text(65,60){\scalebox{1.43}{$\gamma$}}
\Text(65,0){\scalebox{1.43}{$\sim\,e^2\,g_e^{\phantom{2}}$}}
\Vertex(105,60){2}
\ArrowLine(120,15)(90,45)
\ArrowLine(90,45)(105,60)
\ArrowLine(105,60)(120,75)
\DashGluon[width=1](105,60)(125,40){4}{3}{1.5}
\Text(135,45){\scalebox{1.43}{$\tilde{g}$}}
\end{picture}
\qquad
\begin{picture}(135,90)(0,0)
\Vertex(45,45){2}
\Vertex(90,45){2}
\ArrowLine(45,45)(15,15)
\ArrowLine(15,75)(45,45)
\DashPhoton[width=1](45,45)(90,45){4}{4}{1.5}
\Text(65,60){\scalebox{1.43}{$\tilde{\gamma}$}}
\Text(65,0){\scalebox{1.43}{$\sim\,e_e^2\,g_s^{\phantom{2}}$}}
\Vertex(105,60){2}
\ArrowLine(120,15)(90,45)
\ArrowLine(90,45)(105,60)
\ArrowLine(105,60)(120,75)
\Gluon(105,60)(125,40){4}{3}
\Text(135,45){\scalebox{1.43}{$g$}}
\end{picture}
\qquad
\begin{picture}(135,90)(0,0)
\Vertex(45,45){2}
\Vertex(90,45){2}
\ArrowLine(45,45)(15,15)
\ArrowLine(15,75)(45,45)
\DashPhoton[width=1](45,45)(90,45){4}{4}{1.5}
\Text(65,60){\scalebox{1.43}{$\tilde{\gamma}$}}
\Text(65,0){\scalebox{1.43}{$\sim\,e_e^2\,g_e^{\phantom{2}}$}}
\Vertex(105,60){2}
\ArrowLine(120,15)(90,45)
\ArrowLine(90,45)(105,60)
\ArrowLine(105,60)(120,75)
\DashGluon[width=1](105,60)(125,40){4}{3}{1.5}
\Text(135,45){\scalebox{1.43}{$\tilde{g}$}}
\end{picture}
}
\end{center}
\caption{Real diagrams for $e^+ e^- \to q\bar{q} g$ and $e^+ e^- \to
  q\bar{q} \tilde{g}$. In \CDR\ and \HV\ there is only the first
  diagram, whereas in \FDH\ also the second diagram is present. In
  \DRED, all diagrams contribute. An analogous diagram where the gluon
  couples to the other quark leg is understood.}
\label{fdh:FdiagReal}
\end{figure}

Finally we have to face the real corrections. In \CDR, the
amplitude consists of two diagrams (one of which is depicted in
Fig.~\ref{fdh:FdiagReal}). The matrix element squared, expressed
in terms of $s_{ij} \equiv 2\, p_i\cdot p_j$ reads
\begin{subequations}
 \label{eq:Mqqg}
\begin{align}
\label{Mqqg}
M^{(0)}_\mCDR(q\bar{q} g) &=
\omega^{(r)}\,e^4\,g_s^2 \,  (d-2) \bigg\{
\bigg[\frac{(s_{12}+s_{13})^2}{s_{13}\,s_{23}} + 
\frac{d-4}{2} \frac{s_{13}+s_{23}}{s_{23}}\bigg]  
+ \big[1\leftrightarrow 2\big] \bigg\}\,,
\end{align}
where $\omega^{(r)}\!=\!\omega^{(0)}\,2\,\CF/s$. In \HV, the same
diagrams contribute. One might be tempted to assume that
$M^{(0)}_\mHV(q\bar{q} g)$ can be obtained from \Eqn{Mqqg} simply by
setting $d\to 4$. However, this is incorrect.  In the regime where
the gluons become collinear, they have to be treated as singular
gluons. Thus, in \HV\ they are $d$-dimensional. The same is true in
principle for the soft region, but at one loop, there is no scheme
dependence in the soft singularities. This corresponds to the statement
that the cusp anomalous dimension is scheme independent at the one-loop
level \cite{Kilgore:2012tb,Broggio:2015dga}. Treating the gluons properly,
we obtain
\begin{align}
\label{MqqgHV}
M^{(0)}_\mHV(q\bar{q} g) &= \frac{2}{d-2} M^{(0)}_\mCDR(q\bar{q} g)\,. 
\end{align}
In the case of \FDH\ we get contributions $\sim g_s$ and $\sim g_e$.
Again, the gluon has to be treated as a singular one. Hence, it is
split into a $d$-dimensional gluon and an $\epsilon$-scalar, resulting
in
\begin{align}
\label{MqqgFDH}
M^{(0)}_\mFDH(q\bar{q} g) + M^{(0)}_\mFDH(q\bar{q} \tilde{g}) &= 
 M^{(0)}_\mHV(q\bar{q} g) + 
\omega^{(r)}\,e^4\,g_e^2 \, \Ne 
\frac{(s_{13} + s_{23})^2}{s_{13}\, s_{23}}\,. 
\end{align}
Finally, as illustrated in Fig.~\ref{fdh:FdiagReal}, in \DRED\ the
matrix element squared is formally decomposed into four parts,
\begin{align}
M^{(0)}_\mDRED(q\bar{q} g) =\ & M^{(0, \gamma)}_\mDRED(q\bar{q} g) 
+ M^{(0,\gamma)}_\mDRED(q\bar{q}\tilde{g}) 
+ M^{(0, \tilde{\gamma})}_\mDRED(q\bar{q} g) 
+ M^{(0,\tilde{\gamma})}_\mDRED(q\bar{q}\tilde{g})
\phantom{\bigg|}
\notag\\
=\ &  M^{(0)}_\mCDR(q\bar{q} g) + 
e_e^4 \,g_s^2 \, \Ne \,
  \frac{4\, s\, s_{12} + (2-\Ne) (s_{13} + s_{23})^2}{2\, s_{13}\,s_{23}}
\notag\\&
+\frac{d-2}{2} M^{(0)}_\mFDH(q\bar{q} \tilde{g})
+ e_e^4\,g_e^2\, \Ne \,
\frac{-4\, s_{13}\, s_{23} + \Ne (s_{13} + s_{23})^2}{2\, s_{13}\, s_{23}}\,.
\label{MqqgDRED}
\end{align}
\end{subequations}
Note, that if we set $e_e\!=\!e$ and $g_e\!=\!g_s$, the matrix element
in \DRED\ corresponds to the usual four-dimensional matrix element,
\begin{align}
M^{(0)}_\mDRED(q\bar{q} g) \Big|_{\substack{e_e=e_{\phantom{s}}\\g_e=g_s}}
=\ & M^{(0)}_\mCDR(q\bar{q} g) \Big|_{d=4}  \notag\\*
=\ &
\omega^{(r)}\,e^4\,\gs^2\,4\,
\Big(
  \!-\!\frac{1}{y_{13}}
  \!-\!\frac{1}{y_{23}}
  \!+\!\frac{y_{13}}{2\,y_{23}}
  \!+\!\frac{y_{23}}{2\,y_{13}}
  \!+\!\frac{1}{y_{13}\,y_{23}}
  \Big)\,,
\label{mrealfour}
\end{align}
with $y_{ij} \equiv s_{ij}/s$.  This is generally true for aribtrary
tree-level amplitudes in \DRED, but not necessarily in any of the other
schemes. For the considered process, it happens to be true also in \FDH.

The real cross section can now be obtained in any scheme by
integrating the corresponding matrix element over the $d$-dimensional
real phase space,
\begin{subequations}
\label{PS_three_parton}
\begin{align}
\sigma^{(r)}_{\text{\DR}}
&= \frac{s}{2\,(4\pi)^3}\,\Phi_3(\epsilon)\,
  \int_0^1 dy_{13}
  \int_{0}^{1-y_{13}} dy_{23}\
  y_{13}^{-\epsilon}\
  y_{23}^{-\epsilon}\
  (1-y_{13}-y_{23})^{-\epsilon}\,
  M^{(0)}_{\text{\DR}}(q\bar{q} g)
\\*
&\equiv
\frac{s}{2\,(4\pi)^3}\,\Phi_3(\epsilon)
  \iint\displaylimits_{y_{13}\,y_{23}}\,
  y_{13}^{-\epsilon}\
  y_{23}^{-\epsilon}\
  (1-y_{13}-y_{23})^{-\epsilon}\,
  M^{(0)}_{\text{\DR}}(q\bar{q} g)\,.
\end{align}
\end{subequations}
Similar to the two-particle phase space, we extract a $\dim$-dependent factor
\begin{align}
\Phi_3(\epsilon)
=\Big(\frac{4\pi}{s}\Big)^{2\epsilon}\frac{1}{\Gamma(2-2\epsilon)}
= 1 + {\cal O}(\epsilon)\, .
\end{align}
For future reference, we explicitly list the integrals needed to
evaluate Eqs.~\eqref{PS_three_parton},
\begin{subequations}
\label{psintegrals}
\begin{align}
\iint\displaylimits_{y_{13}\,y_{23}}\,
y_{13}^{-\epsilon}\,  y_{23}^{-\epsilon}\,
(1-y_{13}-y_{23})^{-\epsilon}\, \frac{1}{y_{13}}
&= -\frac{1}{\epsilon} - 3
+\mathcal{O}(\epsilon)\, ,
\\*
\iint\displaylimits_{y_{13}\,y_{23}}\,
y_{13}^{-\epsilon}\,  y_{23}^{-\epsilon}\,
(1-y_{13}-y_{23})^{-\epsilon}\, \frac{y_{23}}{y_{13}}
&= -\frac{1}{2\,\epsilon} - \frac{7}{4}
+\mathcal{O}(\epsilon)\, ,
\\*
\iint\displaylimits_{y_{13}\,y_{23}}\,
y_{13}^{-\epsilon}\,  y_{23}^{-\epsilon}\,
(1-y_{13}-y_{23})^{-\epsilon}\, \frac{1}{y_{13} y_{23}}
&= \frac{1}{\epsilon^2} - \frac{\pi^2}{2}
+\mathcal{O}(\epsilon)\, .
\end{align}
\end{subequations}
Using these results for the calculation of the real corrections in the
various schemes and setting $e_e = e$, $g_e=g_s$, $\Ne=2\epsilon$, we obtain
\begin{subequations}
\label{fdh:real}
\begin{align}
\sigma^{(r)}_\mCDR &=
\sigma^{(0)}\,\Big(\frac{\alpha_s}{\pi}\Big)\,\CF\,
  \Phi_3(\epsilon)\,\Big[
  \frac{1}{\epsilon^2}
  +\frac{1}{2\,\epsilon}
  +\frac{13}{4}-\frac{\pi^2}{2} + \mathcal{O}(\epsilon)
  \Big]\,,\phantom{\bigg|}
\label{sig_eer_cdr}
\\*
\sigma^{(r)}_{\mHV\phantom{J}} &=
\sigma^{(0)}\,\Big(\frac{\alpha_s}{\pi}\Big)\,\CF\,
  \Phi_3(\epsilon)\,\Big[
  \frac{1}{\epsilon^2}
  +\frac{3}{2\, \epsilon}
  +\frac{19}{4}-\frac{\pi^2}{2} + \mathcal{O}(\epsilon)
  \Big]\,,\phantom{\bigg|}
\label{sig_eer_hv}
\\*
\sigma^{(r)}_\mFDH =
\sigma^{(r)}_\mDRED &= 
\sigma^{(0)}\,\Big(\frac{\alpha_s}{\pi}\Big)\,\CF\,
  \Phi_3(\epsilon)\,\Big[
  \frac{1}{\epsilon^2}
  +\frac{3}{2\, \epsilon}
  +\frac{17}{4}-\frac{\pi^2}{2} + \mathcal{O}(\epsilon)
  \Big]\,.
  \phantom{\bigg|}
\label{sig_eer_fdh}
\end{align}
\end{subequations}
And, at long last, we find the well-known regularization-scheme
independent physical cross section
\begin{align}
\sigma^{(1)}
= \sigma^{(0)} + \sigma^{(v)}_{\text{\DR}} +
\sigma^{(r)}_{\text{\DR}}\Big|_{d\to 4}
=  \frac{\eqSq\,N_c}{3\, s}\Big(\frac{e^4}{4\pi}\Big)
\Big[\,1+\Big(\frac{\alpha_s}{4\pi}\Big) \, 3\, \CF \Big] \, .
\label{fdh:xs}
\end{align}
The expressions for the virtual and the real cross sections,
\Eqns{fdh:virtual}{fdh:real}, have been obtained setting $e_e=e$ and
$g_e=g_s$. We reiterate that the \FDH/\DRED\ computation can be done
in a much simpler way by directly identifying these couplings from the
beginning. The only place where it is crucial to distinguish them is for
the proper UV (sub)renormalization, i.\,e.\ to obtain the counterterm
in \Eqn{fdh:uvct}. If we had kept the couplings apart to the very end,
the final result would have been unaffected. In other words, terms
involving the `unphysical' couplings $e_e$ and $g_e$ drop out
when adding the virtual, the real, and the counterterms contributions.
For our example this can easily be verified by using the expressions in
Eqs.~\eqref{fdh:m0},  \eqref{fdh:uvct}, \eqref{fdh:m1bare}, and
\eqref{eq:Mqqg}.

\subsection{Established properties and future developments of DS}

As mentioned in the introduction, regularization schemes should not only
simplify practical calculations but also satisfy certain basic requirements.
For decades, dimensional regularization in the two flavours \CDR\ and \HV\
has been the most commonly used regularization, not only because it allows for
the use of powerful calculational techniques but also because many all-order
statements have been rigorously proved in these schemes.

Using an infinite-dimensional vector space as domain, a definition of
the formally $d$-dimensional objects and operations is given in
Refs.~\cite{Wilson:1972cf, Collins:1984xc}. Among the implications are
mathematical consistency and the absence of possible ambiguities. The
equivalence to \BPHZ\ renormalization and the regularized and renormalized
quantum action principle is shown in
Refs.~\cite{Speer:1974cz, Breitenlohner:1977hr}.
As a caveat, however, in chiral theories these statements rely on the use of
a non-anticommuting $\gamma_5$ as defined e.\,g.\ in Refs.~\cite{'tHooft:1972fi,
Breitenlohner:1977hr}. In non-chiral theories like QCD, the quantum action
principle makes it obvious that non-Abelian gauge invariance is manifestly
preserved such that the regularized QCD Green functions automatically satisfy
the Slavnov-Taylor identities at all orders.

The situation regarding \DRED\ and \FDH\ has been considerably more complicated
in the past. However, now these schemes have reached a similar status as \CDR\
and \HV. After first one- and two-loop applications of \DRED~\cite{Capper:1979ns},
the equivalence of \FDH/\DRED\ and \CDR\ is shown in Refs.~\cite{Jack:1993ws,
Jack:1994bn}, indirectly proving that these schemes are compatible with
unitarity and causality. In Ref.~\cite{Stockinger:2005gx}, it is shown how the
spaces in Eq.~\eqref{eq:vsDecomp} can be defined in a rigorous way,
avoiding mathematical ambiguities and excluding the possible inconsistency
found before in Ref.~\cite{Siegel:1980qs}. In this way also an earlier puzzle
regarding unitarity of \DRED\ discussed in Ref.~\cite{vanDamme:1984ig} is
resolved. The key ingredient for the solution is the introduction and separate
treatment of $\epsilon$-scalar fields. 
One important consequence of the additional scalars is the need to distinguish
gauge couplings from evanescent couplings during the renormalization procedure,
as indicated in Eqs.~\eqref{eq:covDer}. The relation between unitarity and the
correct renormalization of evanescent couplings in \FDH/\DRED\ has been further
stressed and exemplified with explicit calculations in
Refs.~\cite{Harlander:2006rj, Harlander:2006xq}.

Apart from the UV properties of the dimensional schemes also IR
divergences and their scheme dependence have been investigated up to
the multi-loop regime.  The separate treatment of $\epsilon$-scalars
has been used in Ref.~\cite{Signer:2005} to clarify a seeming
non-factorization of QCD amplitudes observed earlier in
Refs.~\cite{Beenakker:1988bq, Beenakker:1996dw, Smith:2004ck}.  In
Refs.~\cite{Kunszt:1994, Signer:2008va}, it is shown how \DRED\ and
\FDH\ can be applied in the computation of NLO cross sections in
massless QCD.  The scheme independence of a cross section at NLO has
also been studied in Ref.~\cite{Catani:1996pk}. Regarding virtual
contributions, these considerations have been extended to NNLO in
Refs.~\cite{Kilgore:2011ta, Kilgore:2012tb} and
\cite{Gnendiger:2014nxa, Broggio:2015ata, Broggio:2015dga}.  Moreover,
the latter references provide NNLO transition rules for translating
UV-renormalized virtual amplitudes from one dimensional scheme to
another.  The IR factorization properties of QCD including massive
partons have been investigated at NLO in Ref.~\cite{Catani:2000ef} and
recently up to NNLO in Ref.~\cite{Gnendiger:2016cpg}. For the real
corrections, a formulation of the sector-improved residue subtraction
scheme in the \HV~scheme is presented in Ref.~\cite{Czakon:2014oma}. 

Regarding supersymmetry, \DRED\ and \FDH\ have significant
advantages as in many cases supersymmetry is manifestly preserved
although an all-order proof does not exist. For reviews
regarding applications of these schemes to supersymmetry,
we refer to Refs.~\cite{Jack:1997sr, Jones:2012gfa}.

\section{FDF, SDF: Four- and Six-dimensional formalism}
\label{sec:fdf}

In the following we discuss some new (re-)formulations of \DR. In
Secs.~\ref{sec:fdfDef}--\ref{sec:FDFproperties}, we describe \FDF,
a strictly four-dimensional formulation of the \FDH\ scheme.
The remaining two subsections are dedicated to topics that are not
directly \FDF\ but that are closely related to it, namely automated
NLO calculations using GoSam and the six-dimensional formalism.

\subsection{FDF: four-dimensional formulation of FDH}
\label{sec:fdfDef}

The four-dimensional formulation of the \FDH\ scheme (\FDF) is a novel
implementation of \FDH. Its aim is to achieve the $\dim$-dimensional
regularization of one-loop scattering amplitudes in a purely
four-dimensional framework \cite{Fazio:2014xea}.
The starting point for the formulation of the scheme is the structure
of the quasi $\dim_s$-dimensional \FDH\ space, Eq.~\eqref{eq:vsDecomp},
which we write as
\begin{align}
 \QS{\ds}
 =\QS{\dim}\oplus\QS{\Ne}
 =\text{S}_{[4]}\oplus\QS{-2\epsilon}\oplus\QS{\Ne}
 \equiv\text{S}_{[4]}\oplus\QS{\Ne-2\epsilon}\, .
 \label{eq:vsDecompFDF}
\end{align}
Accordingly, the underlying space of the \FDH\ scheme is written as an
orthogonal sum of a strictly four-dimensional space $\text{S}_{[4]}$ and
a quasi $(\Neps\!-\!2\epsilon)$-dimensional space $\QS{\Ne-2\epsilon}$.
Similar to Eq.~\eqref{eq:dsDecomp}, metric tensors and $\gamma$ matrices
can then be decomposed as
\begin{align}
g_{[d_s]}^{\mu \nu}\
=\ g_{[4]}^{\mu \nu} + g_{[\Ne-2\eps]}^{\mu \nu}\,,
\phantom{\Big|}
\qquad\quad
\gamma_{[d_s]}^{\mu}\
=\ \gamma_{[4]}^{\mu} + \gamma_{[\Ne-2\eps]}^{\mu}\,,
\phantom{\Big|}
\end{align}
with
\begin{subequations}
\label{Eq:OrthoGs}
\begin{align}
(g_{[4]}^{\phantom{\mu}})^{\mu}_{\phantom{\mu}\mu}
&= 4\,\qquad
(g_{[4]}^{\phantom{\mu}}\,g_{[\Ne-2\eps]}^{\phantom{\mu}})^{\mu}_{\phantom{\mu}\nu}
=0 \, ,
\label{Eq:OrthoGs1}
\\
(g_{[\Ne-2\eps]}^{\phantom{\mu}})^{\mu}_{\phantom{\mu}\mu}
&=(\Ne-2\eps)\,
  \stackrel[]{d_s\to4}{\longrightarrow}\, 0\,.\qquad
\label{Eq:OrthoGs2}
\end{align}
\end{subequations}
The algebraic properties of the matrices $\gamma^\mu_{[\Ne-2\eps]}$
can be obtained from Eqs.~\eqref{Eq:OrthoGs} and read
\begin{subequations}
\label{Eq:Gamma}
\begin{align}
\big\{\gamma^{\mu}_{[\Ne-2\eps]},\,
  \gamma^{\nu\phantom{\mu}}_{[\Ne-2\eps]} \big\}
&= 2 \,   g^{\mu \nu}_{[\Ne-2\eps]} \, .
\phantom{\Big|}
\label{Eq:Gamma02}
\\
\big\{ \gamma^{\mu}_{[4]}, \, \gamma^{\nu}_{[\Ne-2\eps]}  \big\}
&=0 \,,\qquad\quad
\big[ \gamma^{5}_{[4]},\, \gamma^{\mu}_{[\Ne-2\eps]} \big]
= 0 \, ,
\phantom{\Big|}
\label{Eq:Gamma01}
\end{align}
\end{subequations}
Loop momenta, on the other hand, are treated in $d$ dimensions
like in any dimensional scheme,
\begin{align}
k_{[\dim]}^{\mu} = k_{[4]}^{\mu} + k_{[-2\eps]}^{\mu} \,,
\label{eq:dmomentum}
\end{align}
with
\begin{align}
k_{[\dim]}^2
  \ =\ \big(k_{[4]}+k_{[-2\epsilon]}\big)^2
  \ =\ k_{[4]}^2+k_{[-2\epsilon]}^2
  \ \equiv\ k_{[4]}^2-\mu^2\,.
\label{eq:kSqFDF}
\end{align}
Here and in the following, the square of the $(\!-\!2\epsilon)$-dimensional
component of a loop momentum is identified with $\!-\!\mu^2$.
The decomposition of the space-time dimension in Eq.~\eqref{eq:kSqFDF}
then suggests that any integral of the form
\begin{align}
I_{i_{1}\cdots i_{k}}^{d}[\mathcal{N}(k_{[\dim]})] &
=\int \frac{d^{\dim}k_{[\dim]}}{(2\pi)^{\dim}}\,
  \frac{\mathcal{N}_{i_{1}\cdots i_{k}}(k_{[\dim]})}{D_{i_{1}}\cdots D_{i_{k}}}
\label{eq:intd}
\end{align}
can be split according to
\begin{align}
I_{i_{1}\cdots i_{k}}^{d}[\mathcal{N}(k_{[4]},\mu^{2})]
& =
  \int \frac{d^{4}k_{[4]}}{(2\pi)^{4}}
  \int \frac{d^{-2\epsilon}k_{[-2\eps]}}{(2\pi)^{-2\epsilon}}\,
  \frac{\mathcal{N}_{i_{1}\cdots i_{k}}
    (k_{[4]},\mu^{2})}{D_{i_{1}}\cdots D_{i_{k}}}\,,
\label{eq:intdsplit}
\end{align}
where $i_{1}\dots i_{k}$ are indices labeling the loop propagators.
With the decomposition of the integral measure in Eq.~\eqref{eq:intdsplit},
any one-loop integral in $\dim$ dimensions has a four-dimensional integrand,
depending on an additional length $\mu^2$. The (radial) integration over
$\mu^2$ can be carried out algebraically by redefining the number of
dimensions~\cite{Bern:1995db},
\begin{align}
I_{i_{1}\cdots i_{k}}^{d}[(\mu^{2})^{r}]
& =(2\pi)^{r}I_{i_{1}\cdots i_{k}}^{d+2r}\left[1\right]
\prod_{j=0}^{r-1}(d-4-2j)\,,
\label{eq:intdprop}
\end{align}
so that powers of $\mu^2$ in the numerator of the integrand generate integrals
in shifted dimensions which are responsible for the \textit{rational} terms
of one-loop amplitudes.

We remark that an $(\Neps\!-\!2\epsilon)$-dimensional metric tensor can
not have a four-dimensional representation. This is due to the fact that
according to Eq.~\eqref{Eq:OrthoGs2}, its square vanishes.
Additionally, in four dimensions the only non-null matrices compatible with
conditions~\eqref{Eq:Gamma} are proportional to $\gamma^5_{[4]}$,
\begin{align}
\gamma_{[\Ne-2\eps]}^{\phantom{\mu}}
\sim \gamma^5_{[4]}\,. 
\label{Eq:GA0}
\end{align}
However, the matrices $ \gamma_{[\Ne-2\eps]}$ fulfill the Clifford
algebra~\eqref{Eq:Gamma02}, and thus
\begin{align} 
\gamma^{\mu}_{[\Ne-2\eps]} \, (\gamma_{[\Ne-2\eps]}^{\phantom{\mu}})_{\mu}
=(\Ne-2\eps)\,\stackrel[]{d_s\to4}{\longrightarrow}\, 0\, ,
\quad \mbox{ while }  \quad   (\gamma^5_{[4]})^2 = \mathbb{I}_{[4]} \,.
\label{Eq:GA1}
\end{align}
Eqs.~\eqref{Eq:GA0} and \eqref{Eq:GA1} are therefore not compatible with
each other.
Finally, the component $k_{[-2\eps]}^{\mu}$ of the loop momentum vanishes
when contracted with a strictly four-dimensional metric tensor, i.\,e.\
$k^\mu_{[-2\eps]} \,(g_{[4]}^{\phantom{\mu}})_{\mu \nu}= 0$.
In four dimensions, the only four vector fulfilling this relation
is the null one.

The above arguments exclude any four-dimensional representation of the
\mbox{$(\Neps\!-\!2\epsilon)$}- and \mbox{$(\!-\!2\epsilon)$}-%
dimensional subspaces. It is possible, however, to find a representation by
introducing additional rules, in the following called $(\!-\!2\epsilon)$
{\it selection rules}, $(\!-\!2\epsilon)$-SRs.
Indeed, the Clifford algebra~(\ref{Eq:Gamma02}) is equivalent to
\begin{align}
  \cdots
  (\gamma_{[\Ne-2\eps]}^{\phantom{\mu}})^{\mu}
  \cdots
  (\gamma_{[\Ne-2\eps]}^{\phantom{\mu}})_{\mu}
  \cdots \,\stackrel[]{d_s\to4}{\longrightarrow}\,0\,, \qquad    
 \slashed k_{[-2\eps]} \slashed k_{[-2\eps]} =  -\mu^2\, .
\label{Eq:Gamma03}
\end{align}
Therefore, any regularization scheme which is equivalent of \FDH\ has to
fulfill conditions~\eqref{Eq:OrthoGs}--\eqref{eq:kSqFDF},
and~\eqref{Eq:Gamma03}. The orthogonality conditions \eqref{Eq:OrthoGs}
and \eqref{eq:kSqFDF} are fulfilled by splitting a $d_s$-dimensional vector
field into a strictly four-dimensional one and a scalar field, while the
other conditions are fulfilled by performing the substitutions
\begin{align}
 g^{\alpha \beta}_{[\Ne-2\eps]} \to   G^{AB}\,, \qquad
 \gamma^\alpha_{[\Ne-2\eps]} \to \gamma^5_{[4]} \, \GA^A, \qquad
 k^{\alpha}_{[-2\eps]} \to i \, \mu \, \QQ^A \, .
 \phantom{\Big|}
\label{Eq:SubF}
\end{align} 
The $(\Neps\!-\!2\epsilon)$-dimensional and $(\!-\!2\epsilon)$-dimensional
indices are thus traded for ($\!-\!2\epsilon$)-SRs such that
\begin{align}
\GG^{AB}\GG^{BC} &= \GG^{AC}\,,&
\GG^{AA} &= 0\,,& 
\GG^{AB} &= \GG^{BA}\,,&
\notag\\*
\GG^{AB} \GA^A &= \GA^B\,,&
\GA^{A} \GA^A &=0\,,&
\{\Gamma^A,\Gamma^B\}&=2\,G^{AB}\,,&
\notag\\*
\GG^{AB} \QQ^A &= \QQ^B\,,&
\QQ^A \QQ^{A} &=1\,,&
\QQ^A \GA^{A} &=1\,.&
\label{Eq:2epsA}
\end{align}
The exclusion of terms containing odd powers of $\mu$ completely defines
the \FDF\ scheme. It allows one to build integrands which, upon integration,
yield the same results as in the \FDH\ scheme. As mentioned before, the \FDF\
scheme is closely connected to the introduction of an additional scalar field.
The role of this field and its relation to the $\epsilon$-scalar present in
the \FDH\ scheme will be discussed in Sec.~\ref{sec:fdfEquiv}.

The rules in Eq.~\eqref{Eq:2epsA} constitute an abstract algebra which is
similar to an algebra related to internal symmetries. For instance, in a Feynman
diagrammatic approach, the ($\!-\!2\epsilon$)-SRs can be handled as the colour
algebra and  performed for each diagram once and for all. In each diagram,
the indices  of the ($\!-\!2\epsilon$)-SRs are fully contracted and the outcome
of their manipulation is  either  $0$ or $\pm1$. It is worth to remark that
the  replacement of   $\gamma^{\alpha}_{[\Ne-2\eps]}$  with $\gamma^5_{[4]}$
takes care of the $d_s$-dimensional Clifford algebra automatically. Thus,
we do not need to introduce any additional scalar field for each fermion
flavour. 

Depending on the gauge we use, further simplifications can arise.
In Feynman gauge, for example, there are no contributions coming from
scalar loops, which is due to the $(\!-\!2\epsilon)$-SRs,
\begin{align}
\GG^{A_1A_2}\GG^{A_2A_3}\dots\GG^{A_{k}A_1} = \GG^{A_1A_1}=0\,.
\end{align}
Similarly, for diagrams with internal scalars and fermions
we get the same cancellation,
\begin{align}
\GA^{A_1}\GG^{A_1A_2}\ldots\GG^{A_{k-1}A_k}\GA^{A_{k}}
= \GA^{A_1}\GA^{A_1}=0\,.
\end{align}
With the use of axial gauge, we obtain the opposite behaviour since
contributions from internal scalars have to be taken in account, 
\begin{align}
\GG^{A_1A_2}\GH^{A_2A_3}\ldots\GG^{A_{k-1}A_k}\GH^{A_{k}A_1}
= \GG^{A_{1}A_2}\GH^{A_2A_1} = - Q^{A_1}Q^{A_1} =-1\,,
\end{align}
where $\GH^{AB} \equiv  \GG^{AB} - \QQ^A \QQ^B$.
Diagrams that contain interactions between generalized gluons
and scalars are dropped according to the $(\!-\!2\epsilon)$-SRs,
\begin{align}
Q^{A_1}\GH^{A_1A_2}\ldots Q^{A_{m}}\ldots\GH^{A_kA_1}
= \GH^{A_{1}A_2}Q^{A_2} =0\,.
\end{align}

\subsection{Wave functions in FDF}
\label{sec:onshell}

Generalized-unitarity methods in dimensional regularization require an
explicit representation of the polarization vectors and the spinors of
$d_s$-dimensional particles.  The latter ones are essential ingredients 
for the construction of the tree-level amplitudes that are sewn along
the generalized cuts.  In this respect, the \FDF\ scheme  is suitable for
the four-dimensional formulation of $d$-dimensional generalized unitarity.
The main advantage of \FDF\ is that the four-dimensional expression
of the propagators in the loop admits an explicit representation in terms
of generalized spinors and polarization expressions which is collected below. 

In the following discussion, the $d$-dimensional momentum $k_{[\dim]}$
will be put on-shell and decomposed according to Eq.~\eqref{eq:dmomentum}.
Its four-dimensional component, $k_{[4]}$, will be expressed as 
\begin{align}
k_{[4]} = k^\flat_{[4]} + \hat q_{[4]} \, ,
\qquad\text{with}\qquad
\hat q_{[4]} \equiv \frac{m^2+\mu^2 }{2\, k_{[4]} \cdot q_{[4]}} q_{[4]}  \,  ,
\label{Eq:Dec}
\end{align}
in terms of the two massless momenta $k^\flat_{[4]}$ and $q_{[4]}$.

\subsubsection*{Spinors}
The spinors of a $d_s$-dimensional fermion have to fulfill a completeness
relation which reconstructs the numerator of the cut  propagator,
\begin{subequations}
\label{Eq:CompFD}
\begin{align}
\sum_{\lambda=1}^{2^{(d_s-2)/2}}
  u_{\lambda,[\dim_s]}(k_{[\dim]})\,\bar{u}_{\lambda,[\dim_s]}(k_{[\dim]})
  & =\slashed k_{[\dim]}+m \, ,
\\* 
\sum_{\lambda=1}^{2^{(d_s-2)/2}}
  v_{\lambda,[\dim_s]}(k_{[\dim]})\,\bar{v}_{\lambda,[\dim_s]}(k_{[\dim]})
  & = \slashed k_{[\dim]}-m \, .
\end{align}
\end{subequations}
The substitutions~(\ref{Eq:SubF}) allow one to express the r.\,h.\,s.\
of Eqs.~(\ref{Eq:CompFD}) as,
\begin{subequations}
\begin{align}
\slashed{k}_{[\dim]}+m
= \slashed{k}_{[4]} +\slashed k_{[-2\epsilon]}+m = 
\slashed{k}_{[4]} + i\, \mu\, \gamma^5_{[4]} + m & 
= \sum_{\lambda=\pm}u_{\lambda}\left(k_{[4]}\right)\bar{u}_{\lambda}
  \left(k_{[4]} \right) \, , \\
\slashed{k}_{[\dim]} -m = \slashed k_{[4]} +\slashed k_{[-2\epsilon]}-m = 
 \slashed{k}_{[4]} + i\, \mu\, \gamma^5_{[4]}  - m  & 
= \sum_{\lambda=\pm}v_{\lambda}\left(k_{[4]}  \right)\bar{v}_{\lambda}\left(k_{[4]} \right)\,,
\label{Eq:CompF4}
\end{align}
\end{subequations}
in terms of generalized four-dimensional massive spinors defined as
\begin{subequations}
\label{Eq:SpinorF}
\begin{align}
u_{+}\left(k_{[4]} \right)
& =\left| k^{\flat}_{[4]}\right\rangle
+\frac{\left(m - i\,\mu\right)}{\left[ k^{\flat}_{[4]}\,q_{[4]} \right]}\left|q_{[4]} \right]\, ,  &
u_{-}\left(k_{[4]} \right)
& =\left| k^{\flat}_{[4]}\right]
+\frac{\left(m  +  i\,\mu\right)}{\left\langle k^{\flat}_{[4]}\, q_{[4]} \right\rangle }\left|q_{[4]} \right\rangle , \notag\\
v_{-}\left(k_{[4]} \right)
& =\left| k^{\flat}_{[4]}\right\rangle
-\frac{\left(m  +  i\,\mu\right)}{\left[ k^{\flat}_{[4]}\, q_{[4]} \right]}\left|q_{[4]} \right]\, ,  &
\label{Eq:SpinorG}
v_{+}\left(l \right)
& =\left| k^{\flat}_{[4]}\right]
-\frac{\left(m -  i\,\mu\right)}{\left\langle k^{\flat}_{[4]}\, q_{[4]} \right\rangle }\left|q_{[4]} \right\rangle , \\[3ex]
\bar{u}_{+}\left(k_{[4]} \right)
& =\left[k^{\flat}_{[4]}\right|
+\frac{\left(m + i\,\mu\right)}{\left\langle q_{[4]}\, k^{\flat}_{[4]}\right\rangle }\left\langle q_{[4]}\right|\, , &
\bar{u}_{-}\left(k_{[4]} \right)
& =\left\langle k^{\flat}_{[4]}\right|
+\frac{\left(m -  i\,\mu\right)}{\left[q_{[4]}\,  k^{\flat}_{[4]}\right]}\left[q_{[4]}\right| \, , \notag\\
\bar{v}_{-}\left(k_{[4]} \right)
& =\left[k^{\flat}_{[4]}\right|
-\frac{\left(m  -  i\,\mu\right)}{\left\langle q_{[4]}  \, k^{\flat}_{[4]}\right\rangle }\left\langle q_{[4]} \right| \, ,  &
\bar{v}_{+}\left(k_{[4]} \right)
& =\left\langle k^{\flat}_{[4]}\right|
-\frac{\left(m + i\,\mu\right)}{\left[q_{[4]}\,  k^{\flat}_{[4]}\right]}\left[q_{[4]} \right| \, .
\end{align}
\end{subequations}
The spinors in Eqs.~(\ref{Eq:SpinorG}) are solutions of the tachyonic Dirac
equations~\cite{PTP.5.14,Leiter,Trzetrzelewski:2011vr,Jentschura:2012vp}
\begin{align}
\left(\slashed k_{[4]} + i\,\mu\,\gamma^{5}_{[4]} + m \right)\, u_\lambda\left(k_{[4]} \right)  &= 0 \, , &
\left(\slashed k_{[4]} + i\,\mu\,\gamma^{5}_{[4]}  - m \right)\, v_\lambda \left(k_{[4]} \right) &=0 \, .
\end{align}
It is worth to notice that  the spinors in Eqs.~(\ref{Eq:SpinorF}) fulfill
the Gordon identities
\begin{align}
\frac{\bar u_\lambda(k_{[4]}^{\phantom{\nu}}) \;
\gamma^\nu_{[4]}  \;  u_\lambda(k_{[4]}^{\phantom{\nu}}) }{2}
= 
\frac{\bar v_\lambda(k_{[4]}^{\phantom{\nu}}) \;
\gamma^\nu_{[4]}  \;  v_\lambda(k_{[4]}^{\phantom{\nu}}) }{2}
= k_{[4]}^\nu \, .
\label{Eq:SpinMom}
\end{align}

\subsubsection*{Polarization vectors}
The $d_s$-dimensional polarization vectors of a spin-1 particle 
fulfill the relation
\begin{align}
\sum_{i=1}^{d_s -2} \, \varepsilon_{i, [d_s]}^\mu( k_{[\dim]},\eta)
\varepsilon_{i, [d_s]}^{\ast \nu}( k_{[\dim]} ,  \eta)
= -  g^{\mu \nu}_{[d_s]}
  +\frac{ k_{[\dim]}^\mu \,
    \eta^\nu +  k_{[\dim]}^\nu \,
    \eta^\mu}{ k_{[\dim]} \cdot  \eta} \,,
\label{Eq:CompGD}
\end{align}
where $ \eta$ is an arbitrary $d$-dimensional massless momentum such that
$ k \cdot  \eta \neq 0$.  Gauge invariance in $d$ dimensions guarantees that
the cut is independent of  $ \eta$. In particular the choice
\begin{align}
 \eta^\mu = k^\mu_{[4]} -  k^\mu_{[-2\eps]} \, ,
\end{align} 
allows one to disentangle the four-dimensional contribution from the
$(\!-\!2\epsilon)$-dimensional one:  
\begin{align}
\sum_{i=1}^{d_s -2} \,
  \varepsilon_{i\, (d_s)}^\mu\left ( k ,  \eta \right )
  \varepsilon_{i\, (d_s)}^{\ast \nu}\left (k ,  \eta \right )
&=\left (   - g^{\mu \nu}_{[4]}
    +\frac{ k^\mu_{[4]} k^\nu_{[4]}}{\mu^2} \right) 
  -\left ( g^{\mu \nu}_{[\Ne-2\eps]}
    +\frac{  k^\mu_{[-2\eps]}  k^\nu_{[-2\eps]}}{\mu^2} \right ) \, .
\label{Eq:CompGD2}
\end{align}
The first term is  related to the cut propagator of a massive gluon and
can be expressed as
\begin{align}
- g^{\mu \nu}_{[4]}  +\frac{ k^\mu_{[4]} k^\nu_{[4]}}{\mu^2}
&=  \sum_{\lambda=\pm,0}\varepsilon_{\lambda}^{\mu}(k_{[4]}) \,
  \varepsilon_{\lambda}^{*\nu}(k_{[4]})
\label{flat}
 \end{align}
in terms of the four-dimensional polarizations of a vector boson of mass
$\mu$~\cite{Mahlon:1998jd},  
\begin{align}
&\varepsilon_{+}^{\mu}\left(k_{[4]} \right)
  = -\frac{\left[k^{\flat}_{[4]}
    \left|\gamma^{\mu}\right|  \hat q_{[4]} \right\rangle }{\sqrt{2}\mu}\,,
&&\varepsilon_{-}^{\mu}\left(k_{[4]} \right)
  = - \frac{\left\langle k^{\flat}_{[4]}
    \left|\gamma^{\mu}\right| \hat q_{[4]} \right]}{\sqrt{2}\mu}\,,
&&\varepsilon_{0}^{\mu}\left(k_{[4]}   \right)
  =   \frac{k^{\flat\mu}_{[4]}-\hat q_{[4]}^{\mu}}{\mu} \, .
\label{emass1}
\end{align}
The latter  fulfill the  well-known relations
 \begin{align}
\varepsilon^2_{\pm}(k_{[4]})&=\phantom{-} 0\, ,
& \varepsilon_{\pm}(k_{[4]})\cdot\varepsilon_{\mp}(k_{[4]})&=-1\, ,
& \varepsilon_{0}^2(k_{[4]})&=-1\, ,
\notag\\
\varepsilon_{\pm}(k_{[4]})\cdot\varepsilon_{0}(k_{[4]})&=\phantom{-} 0\, ,
&\varepsilon_{\lambda}(k_{[4]}) \cdot k_{[4]} &=\phantom{-} 0 \, .
\label{Eq:propEps}
\end{align}
The second term of the r.\,h.\,s.\ of Eq.~(\ref{Eq:CompGD2}) is related
to the numerator of cut propagator of the scalar and can be expressed
in terms  of the $(\!-\!2 \epsilon)$-SRs as:
\begin{align}
g^{\mu \nu}_{[\Ne-2\eps]} +\frac{  k^\mu_{[-2\eps]}  k^\nu_{[-2\eps]}}{\mu^2}
\quad \to  \quad  \GH^{AB} \equiv  \GG^{AB} - \QQ^A \QQ^B  \, .
\label{Eq:Pref}
\end{align}
Therefore, we can define the cut propagators as
\vspace{-0.4cm}
\begin{align}
\parbox{20mm}{
\unitlength=0.20bp%
\begin{feynartspicture}(300,300)(1,1)
\FADiagram{}
\FAProp(4.,10.)(16.,10.)(0.,){/ScalarDash}{0}
\FALabel(5.5,8.93)[t]{\tiny $a,A$}
\FALabel(14.5,8.93)[t]{\tiny $b, B$}
\FAVert(4.,10.){0}
\FAVert(16.,10.){0}
\FAProp(10.,6.)(10.,14.)(0.,){/GhostDash}{0}
\end{feynartspicture}
}
=  \hat G^{AB}\,\delta^{ab} \, . 
\label{Eq:Rules2a}
\end{align}
The generalized four-dimensional spinors and polarization vectors defined
above can be used for constructing tree-level amplitudes with full
$\mu$-dependence.

\subsection{Established properties and future developments of FDF}
\label{sec:FDFproperties}

At one-loop, \FDF\ has been successfully applied to compute the scattering
amplitudes for multi-gluon scattering $g g\to n$ gluons with $n=2,3,4$, and
for $g g\to H + n\ \text{gluons}$ with $n=2,3$
\cite{Bobadilla:2015wma,Bobadilla:2016scr}.
The use of dimensionally regularized tree-amplitudes within \FDF\ has been
employed to study the colour-kinematics duality \cite{Mastrolia:2015maa} for
one-loop dimensionally regularized amplitudes \cite{Primo:2016omk}.

The extension of \FDF\ beyond the one-loop level is currently under
investigation. In particular at two loops, \FDF\ should be able to capture
the dependence of the integrand on the extra dimensional terms of the loop
momenta, namely on two mass-like variables, say $\mu_1^2$ and  $\mu_2^2$,
as well as on the scalar product $\mu_1\cdot\mu_2$.

\subsubsection*{Equivalence of FDF and FDH at NLO: virtual contributions to
$e^{+}e^{-}\to\gamma^{*}\to q\bar{q}$}

\begin{figure}[t]
\begin{center}
\scalebox{.9}{
\begin{picture}(135,75)(0,0)
\Vertex(90,45){2}
\Photon(45,45)(90,45){4}{4}
\Vertex(110,65){2}
\Vertex(110,25){2}
\ArrowLine(110,25)(90,45)
\ArrowLine(120,15)(110,25)
\ArrowLine(90,45)(110,65)
\ArrowLine(110,65)(120,75)
\Photon(110,65)(110,25){4}{4}
\Text(65,58){\scalebox{1.11}{$\gamma$}}
\Text(125,45){\scalebox{1.11}{$\gamma$}}
\end{picture}
\qquad\quad
\begin{picture}(135,75)(0,0)
\Vertex(90,45){2}
\Photon(45,45)(90,45){4}{4}
\Vertex(110,65){2}
\Vertex(110,25){2}
\ArrowLine(110,25)(90,45)
\ArrowLine(120,15)(110,25)
\ArrowLine(90,45)(110,65)
\ArrowLine(110,65)(120,75)
\DashLine[width=1](110,65)(110,25){4}
\Text(65,58){\scalebox{1.11}{$\gamma$}}
\Text(125,45){\scalebox{1.11}{$\gamma'$}}
\end{picture}
}
\end{center}
\vspace*{-.9cm}
   \caption{
   \label{fig:virtualsFDF}
   Virtual diagrams contributing to $\gamma^{*}\to q\bar{q}$ at NLO including
   a strictly four-dimensional photon $\gamma$ (wavy line) and an \FDF\ scalar
   $\gamma'$ (dashed line), respectively. Using Feynman gauge, the right diagram
   vanishes according to the $(\!-\!2\epsilon)$-SRs.}
\end{figure}
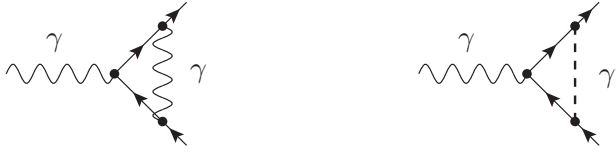

To show that the strictly four-dimensional Feynman rules of \FDF\
together with the $(\!-\!2\epsilon)$-SRs indeed reproduce the corresponding
results in the \FDH\ scheme for $\alpha_e=\alpha_s$,
we consider virtual one-loop contributions to the process
$e^{+}e^{-}\!\to\!\gamma^{*}\!\to\!q\bar{q}$.

According to the discussion in Sec.~\ref{sec:fdfDef}, in \FDF\ each
vector field is split into a strictly four-dimensional field and a
corresponding scalar field. The vertex correction subgraph
$ \gamma^{*}\!\to\!q {\bar q}$ therefore receives two contributions
in \FDF, see Fig.~\ref{fig:virtualsFDF}. The diagram including an
internal \FDF-scalar vanishes according to the $(-2\epsilon)$-SRs
since in Feynman gauge it is proportional to
$\Gamma^A\Gamma^B G^{AB}=\Gamma^A\Gamma^A=0$.
Using only strictly four-dimensional quantities, the amplitude is
then given by
\begin{align}
({\cal A}^{(1)}_{\text{\FDF}})_{\mu}
=-e\,\eq\,g_s^2\,\CF\,\int \frac{d^{\dim} k_{[\dim]}}{(2 \pi)^{\dim}}
  \frac{
    {\bar u}(p_q)\,
    \gamma^\nu \big(
      \slashed{k}_{[4]}^{\phantom{\mu}}
      \!+\!\slashed{p}_{q,[4]}
      \!+i\,\mu\,\gamma_5\big)
    \gamma_\mu \big(
      {\slashed{k}_{[4]}^{\phantom{\mu}}
      \!-\!{\slashed{p}}_{\bar q,[4]}}
      \!+i\,\mu\,\gamma_5\big)
    \gamma_\nu\, u(p_{\bar q})}{
      \big[k_{[4]}^2-\mu^2\big]
      \big[(k_{[4]}+p_{q,[4]})^2-\mu^2\big]
      \big[(k_{[4]}-p_{\bar q,[4]})^2-\mu^2\big]}\,,
\label{eq:FDFvirtuala}
\end{align}
where $p_q$ and $p_{\bar q}$ denote the four-momenta of the massless
quarks. Evaluating the strictly four-dimensional algebra and performing
a tensor integral decomposition in $\dim$ dimensions, the amplitude can
be written as
\begin{align}
({\cal A}^{(1)}_{\text{\FDF}})_{\mu}
=-i\,({\cal A}^{(0)}_{\text{\FDF}})_{\mu}\,g_s^2\,\CF\,
  \bigg\{
    \frac{\dim}{\dim-4}\,I_{2}^{\dim}[1]
    -2\,I_{3}^{\dim}[\mu^2]
  \bigg\}\,,
\label{eq:FDFvirtualb}
\end{align}
with
\begin{subequations}
\label{eq:FDFintegrals}
\begin{align}
 I_{2}^{\dim}[1]
 &=\int \frac{d^{\dim} k_{[\dim]}}{(2 \pi)^{\dim}}
  \frac{1}{
      (k_{[\dim]}+p_{q,[\dim]})^2
      (k_{[\dim]}-p_{\bar q,[\dim]})^2}\,,
 \label{eq:I2one}
 \\
 I_{3}^{\dim}[\mu^2]
 &=\int \frac{d^{\dim} k_{[\dim]}}{(2 \pi)^{\dim}}
  \frac{\mu^2}{
      (k_{[\dim]}+p_{q,[\dim]})^2
      (k_{[\dim]}-p_{\bar q,[\dim]})^2
      (k_{[\dim]})^2}\,.
 \label{eq:I2mu}
\end{align}
\end{subequations}
Note, that in the denominators we used Eq.~\eqref{eq:kSqFDF}. In this way,
the integral in Eq.~\eqref{eq:I2one} is an ordinary $\dim$-dimensional one.
The integral in Eq.~\eqref{eq:I2mu}, on the other hand, can be evaluated by
using Eq.~\eqref{eq:intdprop},
\begin{align}
 I_{3}^{\dim}[\mu^2]
 =(2\pi)(-2\epsilon)I_{3}^{\dim+2}[1]
 =\frac{i}{(4\pi)^2}\frac{1}{2}+\mathcal{O}(\epsilon)\,.
 \label{eq:I2muRes}
\end{align}
For the virtual corrections to the (spin summed/averaged)
squared matrix element $M_{\text{\FDF}}^{(1)}=2\operatorname{Re}
\langle\mathcal{A}_{\text{\FDF}}^{(0)}\,|\,\mathcal{A}_{\text{\FDF}}^{(1)}\rangle$,
we then obtain%
\footnote{Since $M^{(0)}_{\text{\FDF}}\equiv M^{(0)}_{\text{\FDH}}$,
  this result coincides with the one obtained in \FDH\ for $\Neps=2\epsilon$
  and $g_e=g_s$ at least up to $\mathcal{O}(\epsilon^0)$,
  compare with Eq.~\eqref{fdh:m1bareFDH}.}
\begin{align}
 M^{(1)}_{\text{\FDF}}
 =\omega^{(1)}\,M^{(0)}_{\text{\FDF}}\,\Big(\frac{\alpha_s}{\pi}\Big)
  \Big[-\frac{1}{\epsilon^2}-\frac{3}{2\,\epsilon}-\frac{7}{2}
    +\mathcal{O}(\epsilon)\Big]\,.
\end{align}

\subsubsection*{Renormalization of the FDF-scalar--fermion coupling}
\label{sec:fdfEquiv}

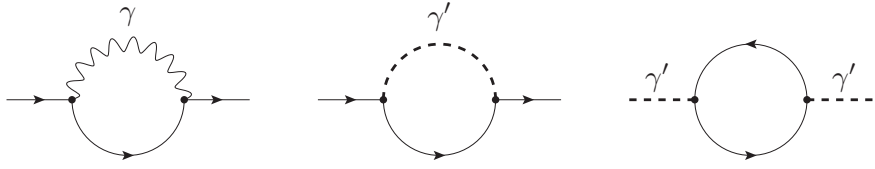
\begin{figure}[t]
   \begin{center}
   \scalebox{.7}{
   \begin{picture}(130,75)(0,0)
   \ArrowArc(65,45)(30,180,360)
   \ArrowLine(0,45)(35,45)
   \ArrowLine(95,45)(130,45)
   \Vertex(35,45){2}
   \Vertex(95,45){2}
   \PhotonArc(65,45)(30,0,180){4}{9}
   \Text(65,90){\scalebox{1.43}{$\gamma$}}
   \end{picture}
   \qquad\quad
   \begin{picture}(130,75)(0,0)
   \ArrowArc(65,45)(30,180,360)
   \ArrowLine( 0,45)( 35,45)
   \ArrowLine(95,45)(130,45)
   \Vertex(35,45){2}
   \Vertex(95,45){2}
   \DashCArc[width=1.5](65,45)(30,0,180){4}
   \Text(65,90){\scalebox{1.43}{$\gamma'$}}
   \end{picture}
   \qquad\quad
   \begin{picture}(130,75)(0,0)
   \ArrowArc(65,45)(30,180,360)
   \DashLine[width=1.5]( 0,45)( 35,45){4}
   \DashLine[width=1.5](95,45)(130,45){4}
   \Vertex(35,45){2}
   \Vertex(95,45){2}
   \ArrowArc(65,45)(30,0,180)
   \Text(15,57){\scalebox{1.43}{$\gamma'$}}
   \Text(115,57){\scalebox{1.43}{$\gamma'$}}
   \end{picture}
   }
   \end{center}
   \vspace*{-.4cm}
   \caption{
   \label{fig:quarkSEfdf}
   One-loop diagrams contributing to the self-energy of the quark
   (left and middle) and of the \FDF-scalar (right). The diagram with the
   internal \FDF\ scalar vanishes according to the $(-2\epsilon)$-SRs.}
\end{figure}

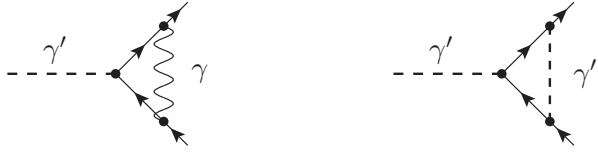
\begin{figure}[t]
\begin{center}
\scalebox{.9}{
\begin{picture}(135,75)(0,0)
\Vertex(90,45){2}
\DashLine[width=1](45,45)(90,45){4}
\Vertex(110,65){2}
\Vertex(110,25){2}
\ArrowLine(110,25)(90,45)
\ArrowLine(120,15)(110,25)
\ArrowLine(90,45)(110,65)
\ArrowLine(110,65)(120,75)
\Photon(110,65)(110,25){4}{4}
\Text(65,55){\scalebox{1.11}{$\gamma'$}}
\Text(125,45){\scalebox{1.11}{$\gamma$}}
\end{picture}
\qquad
\begin{picture}(135,75)(0,0)
\Vertex(90,45){2}
\DashLine[width=1](45,45)(90,45){4}
\Vertex(110,65){2}
\Vertex(110,25){2}
\ArrowLine(110,25)(90,45)
\ArrowLine(120,15)(110,25)
\ArrowLine(90,45)(110,65)
\ArrowLine(110,65)(120,75)
\DashLine[width=1](110,65)(110,25){4}
\Text(65,55){\scalebox{1.11}{$\gamma'$}}
\Text(125,45){\scalebox{1.11}{$\gamma'$}}
\end{picture}
}
\end{center}
\vspace*{-.7cm}
   \caption{
   \label{fdh:FdiagVirtualFDF}
   Diagrams contributing to the interaction of the \FDF\ scalar with
   fermions at the one-loop level. The right diagram vanishes according
   to the $(-2\epsilon)$-SRs.}
\end{figure}

In the following we determine the $\beta$ function related to the coupling
of the \FDF-scalar to fermions in QED with $N_F$ fermion flavours, and
compare it to the known renormalization of the gauge and the evanescent
coupling in the \FDH\ scheme given in Eqs.~\eqref{eq:betaFunctionsFDH}.

To start, we consider the fermion self energy, where two diagrams
contribute at the one-loop level, see Fig.~\ref{fig:quarkSEfdf}.
Using the Feynman rules of Ref.~\cite{Fazio:2014xea} together with the
$(\!-\!2\epsilon)$-SRs, we obtain for the case of massless fermions
\begin{align}
-i\,\Sigma^{(1)}_{\text{\FDF}}&=
\muDRpow\int\frac{d^{\dim}k_{[\dim]}}{(2\pi)^{\dim}}
\bigg\{(\!-i)^4\,e^2\,
  \gamma_{[4]}^{\mu}\,
  \frac{\slashed{k}_{[4]}+i\,\mu\,\gamma_5}{k_{[4]}^2-\mu^2}\,
  \gamma_{[4]}^{\rho}\,
  \frac{(g_{[4]})_{\mu\rho}}{(k_{[4]}+p_{[4]})^2-\mu^2}\,
\bigg\}
\notag\\&=
\muDRpow\int\frac{d^{\dim}k_{[\dim]}}{(2\pi)^{\dim}}
 \bigg\{
  e^2\,
    \gamma^{\mu\phantom{\mu}}_{[4]}\,
    \gamma^{\nu\phantom{\mu}}_{[4]}\,
    \gamma^{\rho\phantom{\mu}}_{[4]}\,
    (g_{[4]}^{\phantom{\mu}})_{\mu\rho}
    \bigg\}
    \frac{(k_{[4]}^{\phantom{\mu}})_{\nu}^{\phantom{\mu}}}
      {[k_{[4]}^2-\mu^2]\,[(k_{[4]}^{\phantom{2}}
      +p_{[4]}^{\phantom{2}})^2-\mu^2]}
\notag\\&=
\muDRpow\int\frac{d^{\dim}k_{[\dim]}}{(2\pi)^{\dim}} \bigg\{
   e^2\,\Big(
    \!-\!\gamma^{\mu\phantom{\mu}}_{[4]}\,
    (\gamma_{[4]}^{\phantom{\mu}})_{\mu}\,
    \gamma^{\nu\phantom{\mu}}_{[4]}\,
    \!+\!2\,\gamma^{\nu\phantom{\mu}}_{[4]}
    \Big)
  \bigg\}
  \frac{(k_{[4]}^{\phantom{\mu}})_{\nu}^{\phantom{\mu}}}
    {[k_{[\dim]}^2]\,[(k_{[\dim]}+p_{[\dim]})^2]}
\notag\\&=
\muDRpow\int\frac{d^{\dim}k_{[\dim]}}{(2\pi)^{\dim}} \bigg\{
   e^2\,\Big(\!-\!4\!+\!2\Big)
 \bigg\}
  \frac{\gamma^{\nu\phantom{\mu}}_{[4]}\,
    (k_{[4]}^{\phantom{\mu}})_{\nu}^{\phantom{\mu}}}
    {[k_{[\dim]}^2]\,[(k_{[\dim]}+p_{[\dim]})^2]}\,.
\label{eq:seFDF1}
\end{align}
In particular, we applied relation \eqref{eq:kSqFDF} and made use of
the fact that terms containing odd powers of $\mu$ are set to zero.
The diagram including an internal \FDF-scalar vanishes according to
the $(\!-\!2\epsilon)$-SRs since it is proportional to
$\Gamma^A\Gamma^B G^{AB}=\Gamma^A\Gamma^A=0$. Evaluating the
$\dim$-dimensional integral in Eq.\,\eqref{eq:seFDF1}, we then obtain%
\footnote{This result can be compared to the one obtained in \FDH,
  see Eq.~\eqref{eq:SEdr}. After subtraction of the UV divergence, the
  limit $\dim\!\to\!4$ can be taken and both results coincide. However,
  due to the vanishing scalar contribution it is clear that the additional
  scalar field in \FDF\ is different from the $\epsilon$-scalar of \FDH.}
\begin{align}
-i\,\Sigma^{(1)}_{\text{\FDF}}
=i\,\slashed{p}_{[4]}\Big(\frac{\alpha}{4\pi}\Big)\,\Big[\,
  \frac{1}{\epsilon}+2-\text{ln}
  \Big(\!-\!\frac{p_{[\dim]}^2}{\muDRSq}\Big)
  +\mathcal{O}(\epsilon)
  \Big]\,.
  \label{eq:SEfdfRes}
\end{align}
Using minimal subtraction, the renormalization of the fermion field
is therefore given by
\begin{subequations}
\label{eq:MSbarRenFDF}
\begin{align}
 Z_2
 =1
  +\Big(\frac{\alpha}{4\pi}\Big)\Big[-\frac{1}{\epsilon}\Big]
  +\mathcal{O}(\alpha^2)\,.
\end{align}
A calculation similar to Eqs.~\eqref{eq:seFDF1} yields for the
renormalization of the \FDF-scalar field
\begin{align}
 Z_{3}'
 =1
  +\Big(\frac{\alpha}{4\pi}\Big)\Big[-\frac{2}{\epsilon}\,N_F\Big]
  +\mathcal{O}(\alpha^2)\,.
\end{align}
Finally, we consider the vertex correction. Again, in \FDF\ two diagrams
contribute at the one-loop level, see Fig.~\ref{fdh:FdiagVirtualFDF}.
According to the $(\!-\!2\epsilon)$-SRs, the diagram with an internal
\FDF-scalar is proportional to $\Gamma^B \Gamma^A \Gamma^B
=-\Gamma^B \Gamma^B \Gamma^A+2\,\Gamma^B G^{AB}= 2\,\Gamma^{A}$.
Evaluating the strictly four-dimensional Lorentz algebra and performing the
$\dim$-dimensional loop integration, the renormalization of the vertex is
given by
\begin{align}
 Z_{1}'
 =1
  +\Big(\frac{\alpha}{4\pi}\Big)\Big[-\frac{4}{\epsilon}\Big]
  +\mathcal{O}(\alpha^2)\,.
\end{align}
\end{subequations}
In a similar way, the renormalization constants can be obtained for the
case of massive fermions. In the on-shell scheme (\OS) they read%
\footnote{The result of $Z_2$ in the on-shell scheme has already been
  obtained in Ref.~\cite{Gnendiger:2016cpg} for the case of \FDH.
  It coincides with Eq.~\eqref{eq:Z2osFDF} for $\Neps=2\epsilon$ and the
  case of equal couplings.},
\begin{subequations}
\label{eq:OSRenFDF}
\begin{align}
Z_{2}\Big|_{\OS}
&=1+\Big(\frac{\alpha}{4\pi}\Big)
  \Big[
    -\frac{3}{\epsilon}
    +\text{ln}\Big(\frac{m_e^2}{\mu^2}\Big)
    -5\Big]
  +\mathcal{O}(\alpha^2)\,,
\label{eq:Z2osFDF}
\\
Z_{3}'\Big|_{\OS}
&=1+\Big(\frac{\alpha}{4\pi}\Big)\,N_F\,
  \Big[
    -\frac{2}{\epsilon}
    +2\,\text{ln}\Big(\frac{m_e^2}{\mu^2}\Big)
    -\frac{2}{3}\Big]
  +\mathcal{O}(\alpha^2)\,,
\\
Z_{1}'\Big|_{\OS}
&=1+\Big(\frac{\alpha}{4\pi}\Big)
  \Big[
    -\frac{4}{\epsilon}
    +4\,\text{ln}\Big(\frac{m_e^2}{\mu^2}\Big)
    -8\Big]
  +\mathcal{O}(\alpha^2)\,.
\end{align}
\end{subequations}
Combining the results in Eqs.~\eqref{eq:MSbarRenFDF} or~\eqref{eq:OSRenFDF},
the $\beta$ function of the \FDF-scalar coupling to fermions is finally
given by
\begin{align}
 \beta'
 =-\Big(\frac{\alpha}{4\pi}\Big)^2\Big[\,2-2\,N_F\Big]
 +\mathcal{O}(\alpha^3)\,,
 \label{eq:betaeFDF}
\end{align}
and therefore identical to the renormalization of the evanescent coupling in
\FDH\ for $e_e=e$, compare with Eq.~\eqref{eq:betaeQED}. According to the
discussion in Sec.~\ref{sec:fdhSE}, the different renormalization of the
couplings in the \FDH\ scheme (and therefore in \FDF) does not play any role
at the one-loop level.
At higher perturbative orders, however, it can lead to a breaking of
unitarity \cite{Kilgore:2011ta}. The way, how the different renormalization
of the scalar coupling can be consistently implemented beyond one loop in
the \FDF\ framework is currently under investigation.

\subsection{Automated numerical computation
\label{sec:gosam}}

To build a fully consistent procedure that is valid for every
Lagrangian is an issue for the complete automation of higher order
computations via numerical recipes.  In the GoSam~\cite{Cullen:2014yla}
actual architecture we adopted a scheme that naturally produces
results in \FDH%
\footnote{The scheme is actually called dimensional reduction in GoSam and
  in Ref.~\cite{Kunszt:1994}, but corresponds to what we call \FDH\ in this
  article.}.
In this scheme, GoSam can generate the full one-loop amplitude for every
process originating from every Lagrangian with the only condition that
the power of the loop momentum in the numerator of a diagram cannot
exceed the number of loop denominators plus one. On the other hand,
we still do not have a completely general procedure for the renormalization.
Technically, the algebraic implementation of our procedure is extremely
simple and can be summarized in the following three points:
\begin{enumerate}
\item Assume that all Lorentz indices are four-dimensional, even if in a
  following step the loop momentum $k$ will be treated as $d$-dimensional.
\item In all fermion chains, also in fermion loops, bring all chiral
  projectors to the left and all loop momenta to the right.
\item Apply the rule
  $\slashed{k}\,\slashed{k} = 
   k_{[d]}\cdot k_{[d]} = k_{[4]} \cdot k_{[4]} - \mu^2.$
\end{enumerate}
This is a simplified version of what is effectively coded, that has
the same algebraic content and produces the same result.  The $\mu^2$
parameter represents the length of the loop momentum into the
$\epsilon$-dependent dimensions.

In GoSam, the generation of amplitudes starts from diagram generation
with QGRAF \cite{Nogueira:1991ex} that searches for topologies and fills
them with fields in all possible ways. This construction paired with the
few rules given above guarantees that no spurious anomalies are
generated and, most important, it provides the correct result for all
the computations that are anomaly free. In full generality, for every
diagram we are then left with two ingredients: a number of non
vanishing integrals with $\mu^2$, and a polynomial of the
four-dimensional part of the loop momentum sitting on every number of
denominators. Loop integrals with $\mu^2$ in the numerator have been
computed analytically since long, so that their implementation is
trivial. Furthermore, reduction programs like
Golem95~\cite{Binoth:2008uq, Cullen:2011kv},
Ninja~\cite{Mastrolia:2012bu, vanDeurzen:2013saa, Peraro:2014cba} or
Samurai~\cite{Mastrolia:2010nb} reduce them easily. The polynomial in
the four-dimensional component of the loop momentum is the optimal
representation of the loop integral for the numerical reduction with
programs like CutTools~\cite{Ossola:2007ax}, Golem95, Ninja or
Samurai.

When we are computing higher-order differential cross sections using
some subtraction scheme~\cite{Frixione:1995ms, Catani:1996vz} to
regularize IR divergences, the choice of the dimensional scheme
adopted is restricted to the virtual
integration, and one can exploit unitarity to derive the transition
rules among renormalized amplitudes computed in different (unitary)
schemes, see Ref.~\cite{Kunszt:1994, Catani:1996pk} for more
details. For this reason it is trivial to derive transition rules from
\FDH\ to \CDR\ for example deducing them from the different
finite part of the integrated dipoles computed in the two schemes. We
refer to the dipoles-subtraction technique, but the reasoning is
completely general and provides the same conversion factors
irrespective of the subtraction scheme.  To be definite, to convert a
one-loop amplitude in the Standard Model, one can start from the
massless gauge-boson emissions from QCD radiation to determine the
shift as $n_{lq}\,C_F/2 + n_g\,C_A/6$ times the underlying tree-level
interference, where $n_{lq}(n_g)$ is the number of the external light
quarks (gluons) being part of the hard scattering amplitude. This
agrees with the shift found in Ref.~\cite{Kunszt:1994}. Similarly,
for QED radiation the shift is again the underlying tree-level
interference times the sum of factors
$\delta_{\text{RS}}=-q_i \sigma_i q_k \sigma_k/2$ for each pair of
emitter (i) with electric charge $q_i$ and spectator (k) with electric
charge $q_k$ and $\sigma$ being $1\,(-1)$ for an incoming fermion and
outgoing anti-fermions (vice versa).

Now we come to the renormalization. In GoSam, this is still not fully
automated.  For the QCD part of the Lagrangian that is renormalized
with the $\MS$ prescription, subtracting only the poles, with
\FDH\ or \DRED\ one is left with a different definition for the
renormalized coupling constants w.\,r.\,t.\ \CDR.  A finite
renormalization is needed to restore the customary definition (\CDR).
There is of course no such problem with the on-shell renormalization
that is often used for electroweak corrections. In GoSam we computed and
implemented all the renormalization constants of the Standard Model
Lagrangian and derived the conversion factors from \FDH\ to
\CDR. They can be found in Ref.~\cite{Chiesa:2015mya}.

To conclude, the \FDH\ scheme appears optimal for numerical computations
and the conversion rules to other schemes can be easily worked out once
and for all exploiting unitarity. Finally, we stress that on the path
towards fully automated computations for every Lagrangian, the
automated computation of the renormalization constants is mandatory.

\subsection{SDF: Six-dimensional formalism}
\label{sec:sixDim}

In this section we discuss the possibility of implementing dimensional
regularization schemes via an embedding of the loop degrees of freedom
in a $\dsix$-dimensional space, where $\dsix$ ($e$ stands for embedding)
is an integer greater than 4 which depends on the loop order. This is
possible in dimensional schemes such as \FDH\ and \HV, where the degrees
of freedom of the external particles live in the genuine four-dimensional
space $S_{[4]}$.  In particular, we focus on the case $\dsix\!=\!6$,
which is sufficient up to two loops~\cite{Bern:2002zk}.

Having a finite integer-dimensional embedding of the loop degrees of
freedom is especially useful in the context of integrand reduction via
generalized unitarity~\cite{Britto:2004nc,Giele:2008ve,Ossola:2006us,
  Kosower:2011ty, Mastrolia:2011pr, Badger:2012dp,Zhang:2012ce,
  Mastrolia:2012an, Badger:2013gxa}, which provides an efficient way
of generating loop integrands from products of tree-level amplitudes
summed over the internal helicity states.  In particular, the
possibility of using a $\dsix$-dimensional spinor-helicity formalism
provides a finite-dimensional (six-dimensional in our case)
representation of both external and internal states.  The
six-dimensional spinor-helicity formalism has been extensively
developed in Ref.~\cite{Cheung:2009dc}, and used in the context of
multi-loop generalized unitarity for producing analytic results for
five- and six-point two-loop all-plus amplitudes in
(non-supersymmetric) Yang-Mills
theory~\cite{Badger:2013gxa,Badger:2015lda,Badger:2016ozq}.

A useful property of this approach is that it gives both internal and
external states an explicit finite-dimensional representation.  This
means that one can perform both analytic and numerical calculations by
working directly with the components of momenta and spinors. Numerical
calculations can in turn be used to infer properties of the
result before a full analytic calculation, or in order to employ
functional reconstruction techniques (see e.\,g.\
Ref.~\cite{Peraro:2016wsq}) which allow to reconstruct full analytic
results from numerical calculations over finite fields.

As mentioned, in this section we focus on a dimensional regularization
scheme where the external states live in the physical four-dimensional
space $\text{S}_{[4]}$, while we keep the dimension $d_s$ of the space
$\QS{d_s}$ undetermined.  The special cases of \FDH\ and \HV\ can be
obtained by setting $d_s\!=4\!$ and $d_s\!=\!d$ respectively at the
end of the calculation.

\subsubsection*{Internal degrees of freedom}
\label{sec:intern-degr-freed}

We consider a generic contribution to an $\ell$-loop amplitude
\begin{align}
  \int\displaylimits_{-\infty}^{\infty} \left(\prod_{i=1}^\ell d^d k_i\right)
  \frac{\mathcal{N}(k_i)}{\prod_j D_j(k_i)}\,,
\end{align}
where $\mathcal{N}$ and $D$ are polynomials in the components of the
loop momenta $k_i$ (a rational dependence on the external kinematic
variables is always understood).  In particular, the denominators
$D_i$ correspond to loop propagators and have the generic quadratic
form
\begin{align}
  D_i = \ell_i^2 - m_i^2\,,\qquad
  l_i^\mu = \sum_{j=1}^\ell \alpha_{ij} k_j^\mu
    + \sum_{j=1}^n \beta_{ij} p_j^\mu\,, \qquad
 \alpha_{ij},\beta_{ij}\in \{0,\pm 1\}\,,
\end{align}
with $p_j$ being the external momenta.
It is often useful to split the loop momenta $k_i^\mu$ into a
four-dimensional part $k_{i,[4]}^\mu$ and a $(d\!-\!4)$-dimensional part
$k_{i,[d-4]}^\mu$ as
\begin{align}
  k_i^\mu = k_{i,[4]}^\mu + k_{i,[d-4]}^\mu\,.
\end{align}
In a regularization scheme where the external states are
four-dimensional, a loop integrand can only depend on the
$(d\!-\!4)$ extra-dimensional components of each loop through
scalar products $\mu_{ij}$ defined as
\begin{align} \label{eq:muij}
  \mu_{ij} = - \left( k_{i,[d-4]}\cdot k_{j,[d-4]} \right)\,.
\end{align}
The scalar products $\mu_{ij}$ can in turn be reproduced by embedding
the loop momenta in an integer-dimensional space with dimension
$\dsix\!\geq\!4\!+\!\ell$.  In particular, as stated, the choice
$\dsix\!=\!6$ is sufficient up to two-loops.  Although we will focus
on the case $\dsix\!=\!6$ and scattering amplitudes at one- or
two-loops, unless stated otherwise our statements are valid for any
multi-loop amplitude, provided that the integer $\dsix$ is
sufficiently large.

In order to correctly reconstruct the dependence of the integrand on
the dimension $d_s$ of the space $\QS{d_s}$ where internal gluon
polarizations live, we add $(\dim_s\!-\!\dsix)$
flavours of scalar particles to the theory, which represent gluon
polarizations orthogonal to both the external and the loop momenta.
The Feynman rules for these scalars can be easily derived from the
ones of gluons (see e.\,g.\ Ref.~\cite{Badger:2013gxa}).

\subsubsection*{Internal states: six-dimensional spinor-helicity formalism}
\label{sec:internal-states:-six}

External states of helicity amplitudes can be efficiently
described using the well-known four-dimensional spinor-helicity
formalism~\cite{Mangano:1987xk,Berends:1987me}.  After a
higher-dimensional embedding of internal states, one can similarly
describe these by means of a higher-dimensional spinor-helicity
formalism.  In particular, the spinor-helicity formalism in six
dimensions has been developed in
Refs.~\cite{Cheung:2009dc,Bern:2010qa,Davies:2011vt}.  While a
comprehensive treatment of the subject is beyond the purpose of
this report (we refer the reader to Ref.~\cite{Cheung:2009dc} for
more details), it is worth pointing out a few properties of
six-dimensional spinors which are useful for providing an
integer-dimensional embedding of the loop internal states, in
particular for applications in the context of integrand reduction
via generalized unitarity, as we shall see in the next section.

Six-dimensional Weyl spinors $|p^a\rangle$ and $|p_{\dot a}]$
(with $a,\dot a \in \{0,1\}\equiv\{+,-\}$) are defined as
independent solutions of the six-dimensional Dirac equation
\begin{align}
  p^\mu\, \sigma^{(6)}_\mu |p^a\rangle
  = p^\mu\, \tilde \sigma^{(6)}_\mu |p_{\dot a}] = 0\,,
\end{align}
where $\sigma^{(6)}_\mu$ and their dual $\tilde \sigma^{(6)}_\mu$ are
six-dimensional generalizations of the Pauli matrices (see
Ref.~\cite{Cheung:2009dc} for an explicit representation).
Six-dimensional momenta can be built from spinors,
\begin{align} \label{eq:6dpmu}
  p^\mu = -\frac{1}{4}\,\langle p^a|\sigma^\mu|p^b\rangle\,
    \epsilon_{a b}\,,\qquad
  p^\mu = -\frac{1}{4}\, [ p_{\dot a} | \tilde \sigma^\mu | p_{\dot b} ] \,
    \epsilon^{\dot a \dot b}\,.
\end{align}
Similarly, given a six-dimensional momentum $p^\mu$, a representation
for the spinors $|p^a\rangle$ and $|p_{\dot a}]$ satisfying the
previous equations, while not unique, is not hard to find.  Note that,
when building loop integrands, the internal spinors always combine as
on the r.\,h.\,s.\ of Eq.~\eqref{eq:6dpmu}, hence the physical results
are always unambiguous and independent of the chosen representation.
Moreover, a subset of the six-dimensional spinor components can be
identified with the components of four-dimensional Weyl spinors
$|p\rangle$ and $|p]$, which ensures a smooth four-dimensional limit.

Internal gluon states are described by six-dimensional polarization
vectors, which can be built out of these spinors
\begin{align}
  \epsilon^\mu_{a \dot a}(p,\eta) = \frac{1}{\sqrt{2}\, (p\cdot \eta)}\,
  \langle p_a | \sigma^\mu | \eta_{b} \rangle\, \langle \eta_c | p _{\dot a} ]\,
  \epsilon^{b c}
\end{align}
with
\begin{align}
 (a\, \dot a) \in \big\{(00), (11), (01), (10)\big\}
 \equiv \big\{(++), (--), (+-), (-+)\big\}\,.
\end{align}
While $(++)$ and $(--)$ correspond to positive and negative helicity in the
four-dimensional limit, respectively, the polarizations $(+-)$,
$(-+)$ only exist in six dimensions.  One can
show~\cite{Cheung:2009dc} that these polarization vectors satisfy all
the expected properties, including the completeness relation
\begin{align}
 \epsilon^\mu_{a \dot a}(p,\eta)\, \epsilon^\nu{}^{a \dot a}(p,\eta)
 = g^{\mu \nu} - \frac{1}{(p\cdot \eta)}
  \left( p^\mu \eta^\nu + p^\nu \eta^\mu \right)\,.
\end{align}
When building an integrand via generalized unitarity, internal
polarization states always combine as on the l.\,h.\,s.\ of the
previous equation.

\subsubsection*{Applications to integrand reduction via generalized unitarity}
\label{sec:appl-integr-reduct}

Integrand reduction methods rewrite loop integrands as a sum of
irreducible contributions,
\begin{align}
  \frac{\mathcal{N}(k_i)}{\prod_j D_j(k_i)}
  = \sum_{T} \frac{\Delta_T(k_i)}{\prod_{j\in T} D_j(k_i)}\,,
\end{align}
where the sum on the r.\,h.\,s.\ runs over the non-vanishing
sub-topologies of the parent topology identified by a set of
denominators $\{D_j\}$.  The on-shell numerators or residues
$\Delta_T$ can be written as a linear combination of polynomials
$\mathbf{q}_T\!=\!\{q_{T,1}, q_{T,2},\ldots\}$ which can be combined to
form an integrand basis up to terms proportional to the denominators
of the corresponding sub-topology $T$,
\begin{align}
\label{eq:deltaT}
  \Delta_{T}(k_i)
  = \sum_{\alpha} c_{T,\alpha}\, \left(\mathbf{q}_T(k_i)\right)^\alpha,\qquad
  \mathbf{q}_T^\alpha\equiv \prod_j q_{T,j}^{\alpha_j}\,,
\end{align}
where $\alpha\!=\!(\alpha_1,\alpha_2,\ldots)$ runs over an appropriate
set of multi-indices.  Techniques for choosing an appropriate integrand
basis have been proposed e.\,g.\ in Refs.~\cite{Badger:2012dp,Zhang:2012ce,
Mastrolia:2012an,Badger:2016ozq}.

The coefficients $c_{T,\alpha}$ only depend on the external kinematics
(they also have a polynomial dependence on $d_s$) and they can be
determined by evaluating the integrand on values of the loop momenta
such that the propagators of the corresponding loop sub-topology are
put on-shell $\{D_j=0\}_{j\in T}$.  These constraints are also known
as multiple cuts.  On these values of the loop momenta, the integrand
factorizes as a product of tree-level amplitudes summed over the
internal helicities corresponding to the cut on-shell loop momenta.
Hence, an efficient way of computing the integrands on the cut
conditions is by sewing together tree-level amplitudes.  This is known
as generalized unitarity.  As explained, by means of a
higher-dimensional spinor-helicity formalism, one can build products
of trees which contain the full dependence of the integrand on the
loop degrees of freedom.

More explicitly, the solutions of the cut conditions in $\dsix$
dimensions can be expressed as a linear combination of terms of a
$\dsix$-dimensional vector basis $\{e_{ij}\}_{j=1}^\dsix$,
\begin{align}
  k_i^\mu = \sum_{j=1}^\dsix y_{ij}\, e_{ij}^\mu\,,
\end{align}
where, in turn, the coefficient of this linear combination can be
expressed as $y_{ij}=y_{ij}(\{\tau_k\})$, where $\{\tau_k\}$ is a set
of free variables which are not constrained by the cut conditions.
From these $\dsix$-dimensional on-shell momenta, we thus build
the corresponding $\dsix$-dimensional spinors, which in turn are
used to evaluate the tree-level helicity amplitudes which define the
integrand on the considered multiple cut.

As we mentioned, the correct dependence of the integrand on $d_s$ is
obtained by adding to the theory $(\dim_s\!-\!\dsix)$ flavours of
scalars representing additional polarizations of the internal gluons.
At two-loops, an integrand can have at most a quadratic dependence on
scalar flavours
\begin{align}
  \Delta_T = \Delta_T^{(\dsix,0)}
  + (d_s-\dsix)\, \Delta_T^{(\dsix,1)}
  + (d_s-\dsix)^2\, \Delta_T^{(\dsix,2)}\,.
\end{align}
More in general, each scalar loop can add at most one power of
$(\dim_s\!-\!\dsix)$.  We stress that the result for $\Delta_T$
does not depend on the dimension $\dsix$ of the chosen embedding, unlike
each of the terms on the r.\,h.\,s.\ of the previous equation.

This setup has been used for the calculation of planar five- and
six-point two-loop amplitudes in Yang-Mills theory presented in
Refs.~\cite{Badger:2013gxa,Badger:2015lda,Badger:2016ozq}, as
well as for the first application of multivariate reconstruction
techniques to generalized unitarity presented in
Ref.~\cite{Peraro:2016wsq}.  The latter includes the calculation of the
on-shell integrands of the maximal cuts of the two-loop planar
pentabox and the non-planar double pentagon topology, for a complete
set of independent helicity configurations.  This shows that this
strategy is suitable for performing complex multi-leg calculations at
two loops, which is currently a very active field of research.

\section{IREG: Implicit regularization
\label{sec:ireg}}

\subsection{Introduction to IREG and electron self energy at NLO}

Implicit regularization (\IReg) is a regularization framework proposed by
the end of the nineties \cite{Battistel:1998sz,BaetaScarpelli:2001ix,
BaetaScarpelli:2000zs} as an alternative to well-known dimensional schemes.
A main characteristic of the method is that it stays in the physical
dimension of the underlying quantum field theory, avoiding, in principle,
some of the drawbacks of \DR\ such as the mismatch between fermionic and
bosonic degrees of freedom which leads to the breaking of supersymmetry.
\IReg\ is proposed to work in momentum space and relies on the following
observation: the UV divergent piece of any Feynman integral should not
depend on physical parameters such as external momenta or particles masses%
\footnote{This point of view is shared by other methods as well, for instance
  by \FDR\ which is described in Sec.~\ref{sec:fdr}. In the latter scheme,
  these intrinsic divergent pieces are called 'vacua'.}.
This simple fact leads to profound consequences as we are going to see.

For ease of the reader, we will develop the basic concepts of \IReg\ by
considering a familiar example of massless QED, the one-loop corrections to
the fermion propagator. We write the initial (unregularized) expression as		
\begin{align}
-i\,\Sigma^{(1)}(p)
=-e^2\int\frac{d^4k}{(2 \pi)^4}
  \gamma^{\mu}\frac{1}{\slashed{k}}\gamma_{\mu}\frac{1}{(k-p)^{2}}\,,
\end{align}
where $p$ is an external momentum.
The first step is to perform simplifications using Dirac algebra in
strictly four dimensions. In this example, the result is particularly simple
\begin{align}
-i\,\Sigma^{(1)}(p)
=2 e^2\,\gamma_{\mu}\int\frac{d^4k}{(2 \pi)^4}
  \frac{k^{\mu}}{k^2(k-p)^{2}}\,.
\end{align}
The next step is just to introduce a fictitious mass in the propagators
which will allow us to control spurious IR divergences introduced
in the course of the evaluation. Thus, the integral can be rewritten as
\begin{align}
-i\,\Sigma^{(1)}(p)
&=\lim_{\mu^{2}\rightarrow 0}
  2 e^2\,\gamma_{\mu}\int\frac{d^4k}{(2 \pi)^4}
  \frac{k^{\mu}}{(k^2-\mu^{2})[(k-p)^{2}-\mu^{2}]}
\equiv\lim_{\mu^{2}\rightarrow 0}\big[\!-\!i\,
  \Sigma^{(1)}_{\text{\IReg}}(p,\mu)\big]\,.
\end{align}
At this point one uses the main observation of \IReg, that the intrinsic
divergent integral should not depend on physical parameters, the external
momentum in this case. To achieve that, one just uses the following
identity as many times as necessary to isolate the physical parameters
in the finite part,
\begin{align}
\frac{1}{(k-p)^2-\mu^2}
=\frac{1}{(k^2-\mu^2)}
+\frac{(-1)(p^2-2\,p \cdot k)}{(k^2-\mu^2)\left[(k-p)^2-\mu^2\right]}\,.
\label{id}
\end{align}
In our example, one ends up with the following divergent expression
\begin{align}
-i\,\Sigma_{\text{\IReg}}^{(1)}(p,\mu)\Big|_{\text{div}}
=2 e^2\,\gamma_{\mu}\Bigg[
  \int\frac{d^4k}{(2 \pi)^4}\frac{k^{\mu}}{(k^{2}-\mu^{2})^{2}}
  +2 p_{\nu}\int\frac{d^4k}{(2 \pi)^4}\frac{k^{\mu}k^{\nu}}{(k^{2}-\mu^{2})^{3}}
  \Bigg]\,,
\end{align}
in which all dependence on the external momenta is only in the numerator.
The latter can be therefore pulled outside the integration.
Focusing on the divergences, one notices the existence of linear and
logarithmic terms. The first piece is automatically null (as in \CDR) and
we are left with only the logarithmic term, whose integral is a particular
example of the general expression   
\begin{align}
 I_{\text{log}}^{\nu_{1}\cdots\nu_{2N}}(\mu^{2})
 \equiv \int\frac{d^4k}{(2 \pi)^4}
  \frac{k^{\nu_{1}}\cdots k^{\nu_{2N}}}{(k^{2}-\mu^{2})^{N+2}}\,.
  \phantom{\Bigg|}
\end{align}
This is a characteristic of \IReg, that the UV divergence can be always
expressed in terms of a precise set of Basic Divergent Integrals (BDI),
composed of scalar and tensorial ones. However, it can be shown that
\textit{all} tensorial integrals can be further expressed in terms of
the scalar ones plus surface terms. In our particular example one has
\begin{align}
\Upsilon_{0}^{\mu\nu}
=\int\frac{d^4k}{(2 \pi)^4}
  \frac{\partial}{\partial k_{\mu}}\frac{k^{\nu}}{(k^{2}-\mu^{2})^{2}}
=g^{\mu\nu}I_{\text{log}}(\mu^{2})
  -4\, I_{\text{log}}^{\mu\nu}(\mu^{2})\equiv g^{\mu\nu}\upsilon_{0,2}\,,
\label{eq:upsilon}
\end{align}
where $\Upsilon_{0}^{\mu\nu}$ is a surface term, arbitrary in principle.
More comments regarding the surface terms and their relation to momentum
routing invariance will be given at the end of this section.

After all UV divergences are taken care of, one needs
to evaluate the finite part, for which we obtain
\begin{align}
-i\,\Sigma_{\text{\IReg}}^{(1)}(p,\mu)\Big|_{\text{fin}}
&=2 e^2\,\gamma_{\mu}\Bigg[
  \!-\!p^2\int\frac{d^4k}{(2 \pi)^4}
  \frac{k^{\mu}}{(k^{2}-\mu^{2})^{3}}
  +\int\frac{d^4k}{(2 \pi)^4}
    \frac{k^{\mu}(p^2-2\,p \cdot k)^{2}}{
    [k^{2}-\mu^{2}]^{3}[(k-p)^2-\mu^2]}\Bigg]
\nn\\
&=e^2\,b\,\slashed{p}\,
  \Bigg[2-\ln\Big(\!-\!\frac{p^{2}}{\mu^{2}}\Big)\Bigg]+O(\mu^2)\,,
\qquad\quad\text{with}\quad
b=\frac{i}{(4\pi)^{2}}\,.
\label{eq:bDefIReg}
\end{align}
It should be noticed that the limit $\mu^{2}\!\to\!0$ has still to be
taken in the final result. However, it can be easily seen that both
$I_{\text{log}}(\mu^{2})$ and the logarithm term then develop an IR
singularity which is spurious since our starting integral was IR safe.
To avoid this issue, one still needs to introduce a scale
$\lambda^{2}\!\neq\!0$, which plays the role of a renormalization
scale in renormalization group equations,
\begin{align}
I_{\text{log}}(\mu^2)
=I_{\text{log}}(\lambda^2)-b\,\ln\Big(\mu^2/\lambda^2\Big)\,.
\end{align}
Combining the divergent and finite part and writing the dimension of the
external momentum explicitly, one finally gets%
\footnote{This result can be compared with the corresponding one obtained
  in \FDH, see Eq.\,\eqref{eq:SEdr}. Setting the surface term $\upsilon_{0,2}$
  to zero which is necessary to preserve gauge invariance, see also
  Sec.~\ref{sec:ireg_gauge_invariance}, the \textit{finite} terms of
  the electron self energy in \IReg\ and \DR\ are the same for
  $\dim=4$ and $\lambda=\muDR$. The relation for the UV divergence is given
  by $b^{-1}I_{\text{log}}(\lambda^2)\leftrightarrow\frac{1}{\epsilon}$.}
\begin{align}
-i\,\Sigma^{(1)}_{\text{\IReg}}(p,\lambda)
=i\,\slashed{p}_{[4]}\,\Big(\frac{\alpha}{4\pi}\Big)\Big[
  b^{-1}I_{\text{log}}(\lambda^2)
  +2
  -\ln\Big(\!-\!\frac{p^{2}_{[4]}}{\lambda^{2}}\Big)
  -b^{-1}\upsilon_{0,2}
  +\mathcal{O}(\lambda)
  \Big]\,.
\end{align}
In summary, the treatment of UV divergent amplitudes in \IReg\ can be
described as follows:
\begin{enumerate}
\item Introduce a fictitious mass $\mu^{2}$ in propagators to avoid
  spurious IR divergences in the course of the evaluation.
\item Use Eq.~(\ref{id}) as many times as necessary to free the divergent part
  from physical parameters like external momenta and masses.
  In the case of massive theories, a similar identity can be applied,
  see Ref.~\cite{Cherchiglia:2010yd} for details.
\item Express the divergent part in terms of scalar and tensorial basic
  divergent integrals.
\item Reduce tensorial BDIs to the scalar ones plus surface terms.
\item Remove the $\mu^{2}$ dependence by introducing a scale $\lambda^{2}$ which
  plays the role of a renormalization scale on renormalization group equations.
\end{enumerate}  
At this point, we would like to emphasize the role played by the surface terms
which, as defined, are just differences between integrals with the same degree
of divergence. As shown in Ref.~\cite{Ferreira:2011cv}, these objects are at the
root of momentum routing invariance (the freedom one has in the assignment of
internal momenta inside a given Feynman diagram). This can only be
respected when the surface terms are set to zero. It can also be shown that
the same conclusion holds for Abelian gauge invariance, allowing one to
conjecture that surface terms are at the root of symmetry breaking in general.
In Ref.~\cite{Ferreira:2011cv}, it is shown that this conjecture may hold for
supersymmetric theories as well. Similar analyses, in many different theories
and contexts, have been carried out in Refs.~\cite{Sampaio:2002ii,Sampaio:2005pc,
Pontes:2007fg,Carneiro:2003id,Souza:2005vf,Ottoni:2006ij,Scarpelli:2008fw,
Dias:2008iz,Cherchiglia:2012zp,Gazzola:2013hba,Felipe:2014gma,
Cherchiglia:2015vaa,Vieira:2015fra,Viglioni:2016nqc}.

\subsection{Application example:
$e^{+}\,e^{-}\to\gamma^{*}\to q\bar q$ at NLO}

In this section we perform the computation of the total cross section of
the process $e^{+}e^{-}\!\to\!\gamma^{*}\!\to q\bar{q}$, showing an
example on how \IReg\ deals with different kinds of divergences.
We divide the presentation in two parts, as usual.

\subsubsection*{Virtual contributions}

The (unregularized) amplitude for the one-loop vertex correction subgraph
$ \gamma^{*}\!\rightarrow q {\bar q}$ reads
\begin{align}
\label{V1}
{\cal A}^{(1)}_{\mu}
=-e\,\eq\,g_s^2\,\CF\,\int \frac{d^4 k}{(2 \pi)^4}
  \frac{
    {\bar u}(p_q)
    \gamma^\nu ({\slashed{k}}+{\slashed{p}}_{q})
    \gamma_\mu ({\slashed{k}}-{\slashed{p}}_{\bar q})
    \gamma_\nu u(p_{\bar q})}{k^2(k+p_q)^2(k-p_{\bar q})^2}\,,
\end{align}
where $p_q$ and $p_{\bar q}$ denote the four-momenta of the massless quarks.
Using the Dirac equation for massless quarks, the integral can be decomposed as
\begin{align}
\label{V2}
{\cal A}^{(1)}_{\mu}\,
= -4e\,\eq\,g_s^2\,\CF\,\Big\{
  &{\bar u}(p_q)\,\gamma_\mu\,u(p_{\bar q})
  \Big[
    (p_q\cdot p_{\bar q})\,I
    -(p_{q,\alpha} -p_{\bar{q},\alpha})\,I^\alpha
    -I_2/2
  \Big]
\nonumber\\
&+
  {\bar u}(p_q)\,\gamma_\alpha\,u(p_{\bar q})
  \Big[
    (p_{q,\mu} -p_{\bar{q},\mu})\,I^\alpha
    +I^\alpha_{\phantom{\alpha}\mu}
    \Big]
 \Big\}\,,
\end{align}
with
\begin{subequations}
\label{V34}
\begin{align}
\label{V3}
\{I,I^\alpha,I^{\alpha\beta}\}
&=\int \frac{d^4 k}{(2 \pi)^4}
\frac{\{1,k^\alpha,k^{\alpha}k^{\beta}\}}
  {k^2(k+p_q)^2(k-p_{\bar q})^2}\,,
\\*
\label{V4}
I_2
&=\int \frac{d^4 k}{(2 \pi)^4}
  \frac{k^2}{k^2(k+p_q)^2(k-p_{\bar q})^2}
= \int \frac{d^4 k}{(2 \pi)^4}
  \frac{1}{(k+p_q)^2(k-p_{\bar q})^2}\,.
\end{align}
\end{subequations}
One notices the prescription of \IReg\ to cancel denominators as in $I_2$
\textit{before} introducing a regulating mass in the propagators%
\footnote{This is a crucial difference compared to \FDR, where $\mu^2$-terms
  remain in the numerators. In a second step, they are then removed by so-called
  'extra integrals'. Further discussions can be found in
  Sec.~\ref{sec:ireg_properties}}.

The integrals in Eqs.~\eqref{V34} are IR divergent for
$p_q^2\!=\!p_{\bar q}^2\!=\!0$. In addition, the integral in Eq.~\eqref{V3}
carrying two Dirac indices and the integral in Eq.~\eqref{V4} are
logarithmically UV divergent. To deal with the latter, a regulating mass
$\mu$ is introduced in all propagators,
\begin{align}
\label{R1}
\{
  I_{\text{\IReg}}^{\phantom{\alpha}},
  I_{\text{\IReg}}^{\alpha},
  I_{\text{\IReg}}^{\alpha\beta}
  \}
=\int\!\frac{d^4 k}{(2\pi)^4} \frac{
  \{1,k^\alpha,k^{\alpha}k^{\beta}\}}{
  [k^2-\mu^2][(k+p_q)^2-\mu^2][(k-p_{\bar q})^2-\mu^2]}\,,
\end{align}
and, after cancellation of one of the denominators, also in
\begin{equation}
\label{CV4}
I_{2,\text{\IReg}}
=\int \frac{d^4 k}{(2 \pi)^4}
\frac{1}{[(k+p_q)^2-\mu^2][(k-p_{\bar q})^2-\mu^2]}\,.
\end{equation}
The limit $\mu^2\!\rightarrow\!0$ in the divergent contributions is only to
be taken after the cross section of the whole process has been evaluated.
Endowed with the regulating mass, all integrals are IR finite. Using
$\mu_0\!\equiv\!\mu^2/s$ and
$s\!\equiv\!(p_q+p_{\bar q})^2\!=\!2\,p_q\cdot p_{\bar q}$, one obtains%
\footnote{The results of the \IReg\ integrals can be
  compared with the corresponding ones in \DR. Setting
  $\muDRSq\!=\!s$, the integrals in Eqs.~\eqref{CV31a} and
  \eqref{CV31b}, for example, are given by
  $I_{\text{\DR}}\big|_{{p_q^2=p_{\bar q}^2=0}}
  \!=\!c_{\Gamma}(\epsilon)\,\frac{i}{(4\pi)^2}\,
  \frac{1}{s}\,\big[
    \frac{1}{\epsilon^2}
    +\frac{i\pi}{\epsilon}
    -\frac{\pi^2}{2}+\mathcal{O}(\epsilon)
    \big]$
  and $I_{\text{\DR}}^{\alpha}\big|_{{p_q^2=p_{\bar q}^2=0}}
  \!=\!c_{\Gamma}(\epsilon)\,\frac{i}{(4\pi)^2}\, \frac{(p_q-p_{\bar
      q})^{\alpha}}{s}\,\big[
    \frac{1}{\epsilon}
    +i\pi
    +2
    +\mathcal{O}(\epsilon)\big]$.
  Using $c_{\Gamma}(\epsilon\!=\!0)\!=\!1$, the one-to-one
  correspondence between double and single IR poles in \DR\ and
  \IReg\ then reads
  $\frac{1}{\epsilon^2}\!\leftrightarrow\!\frac{1}{2}\ln^2(\mu_0)$ and
  $\frac{1}{\epsilon}\!\leftrightarrow\!\ln(\mu_0)$,
  see also Sec.~\ref{sec:propertiesFDR}.}$^{,}$%
  \footnote{Similar results for the integrals are obtained when using
    'loop regularization' (\LORE), a strictly four-dimensional
    regularization prescription \cite{Wu:2002xa,Wu:2003dd,Bai:2017zuw}.
    In \LORE, UV and IR divergences are regularized via logarithms of a regulator
    mass $M_c$ and a soft scale $\mu_s$, respectively; surface terms as appearing
    in Eq.~\eqref{eq:upsilon} are set to zero by definition. For more details
    regarding the definition of \LORE\ we refer to
    Refs.~\cite{Wu:2002xa,Wu:2003dd}.}
\begin{subequations}
\label{CV31}
\begin{align}
\label{CV31a}
I_{\text{\IReg}}\Big|_{{p_q^2=p_{\bar q}^2=0}}
&=\frac{i}{(4\pi)^2}\frac{1}{s}
  \Big[
    \frac{\ln^2(\mu_0)}{2}
    +i\pi\ln(\mu_0)
    -\frac{\pi^2}{2}
  +\mathcal{O}(\mu_0)\Big]\,,
\\
I_{\text{\IReg}}^{\alpha}\Big|_{{p_q^2=p_{\bar q}^2=0}}
&=\frac{i}{(4\pi)^2}\frac{(p_q-p_{\bar q})^{\alpha}}{s}\,
  \big[\ln(\mu_0)+i\pi+2+ \mathcal{O}(\mu_0)\big]\,,
\label{CV31b}
\\
I_{\text{\IReg}}^{\alpha\beta}\Big|_{{p_q^2=p_{\bar q}^2=0}}
&= \frac{g^{\alpha\beta}}{4}\,\Big\{
  I_{\text{log}}(\mu^2)
  + \frac{i}{(4\pi)^2}\big[\ln(\mu_0)+i\pi+3\big]
  \Big\}
\\*&\quad
-\frac{i}{(4\pi)^2}\frac{1}{2s}\Big\{{p_q}^\alpha
  \big(
    {p_{\bar q}}^\beta
    +{p_q}^\beta\big[\ln(\mu_0)+i\pi+2\big]
    \big)
  +(q,\bar q)\!\rightarrow\!(\bar q, q)\Big\}
  + \mathcal{O}(\mu_0)\,,
\notag\\
I_{2,\text{\IReg}}\Big|_{{p_q^2=p_{\bar q}^2=0}}
&= I_{\text{log}}(\mu^2)
  + \frac{i}{(4\pi)^2} \big[\ln(\mu_0)+i\pi+2 +\mathcal{O}(\mu_0)\big]\,.
\end{align}
\end{subequations}
In the UV divergent integrals, the BDI $I_{\text{log}}(\mu^2)$ has been
isolated, according to the rules of \IReg. Inserting the integrals from
Eqs.~\eqref{CV31} into Eq.~\eqref{V2} and performing the remaining contractions,
one obtains for the one-loop vertex correction
\begin{align}
({\cal A}^{(1)}_{\text{\IReg}})_{\mu}
&=({\cal A}^{(0)}_{\text{\IReg}})_{\mu}\,
    \Big(\frac{\alpha_s}{\pi}\Big)\,\CF\,\Big[
      \!-\!\frac{\ln^2(\mu_0)}{4}
      \!-\!\frac{3+2i\pi}{4}\ln(\mu_0)
      \!-\!\frac{7-\pi^2+3i\pi}{4}
      +\mathcal{O}(\mu_0)
      \Big]\,,
\label{eq:AmuRes}
\end{align}
where the UV divergent contributions $\!\sim\!I_{\text{log}}(\mu^2)$ are
dropped. Taking twice the real part of the one-loop correction, the
virtual contribution to the total cross section is then given by%
\footnote{The virtual cross section in \IReg\ can be compared with the ones
  obtained in \DR, see Eqs.~\eqref{fdh:virtual}. Using the aforementioned
  translation rules for IR divergences in \IReg\ and \DR, and
  \mbox{$\Phi_2(\epsilon\!=\!0)=c_\Gamma(\epsilon\!=\!0)=1$}, it follows that
  Eq.~\eqref{crsv} coincides with the results obtained in \FDH\ and \DRED,
  Eq.~\eqref{sig_eev_four}. In Sec.~\ref{sec:CSfdr}, it will be shown that the
  result of the virtual cross section in \IReg\ also coincides with the one
  obtained in \FDR.}
\begin{align}
\label{crsv}
\sigma^{(v)}_{\text{\IReg}}
=\sigma^{(0)}\,\Big(\frac{\alpha_s}{\pi}\Big)\,\CF\,\Big[
  -\frac{\ln^2(\mu_0)}{2}
  -\frac{3}{2} \ln(\mu_0)
  -\frac{7-\pi^2}{2}
  +\mathcal{O}(\mu_0)
  \Big]\,,
\end{align}
with $\sigma^{(0)}$ given in Eq.~\eqref{sig_yqq}.
The divergences occurring in the limit of a vanishing regulator mass $\mu_0$
will be exactly cancelled by the cross section related to the bremsstrahlung
diagrams, as shown in the next section.

\subsubsection*{Real contributions}

In the following we obtain the bremsstrahlung contribution to the
total cross section, using the same regulator mass $\mu$ for the
gluon and the quarks, as in the previous section. At least at NLO,
apart from minor technical differences, the treatment of IR singularities
in \IReg\ is equivalent to the \FDR\ solution proposed in
Ref.~\cite{Pittau:2013qla} (see also Sec.~\ref{sec:realFDR}).

The total cross section pertaining to the real emission process
$e^{+}(p')\,e^{-}(p)\to \gamma^*(q) \to q(k_1)\,{\bar q}(k_2)\,g(k_3)$
is obtained as
\begin{equation}
\label{bs}
\sigma^{(r)}_{\text{\IReg}}=\frac{1}{2\,s}\!
  \int\frac{d^3k_1}{(2\pi)^3 2\omega_1}\!
  \int\frac{d^3k_2}{(2\pi)^3 2\omega_2}\!
  \int\frac{d^3k_3}{(2\pi)^3 2\omega_3}
  (2\pi)^4 \delta^{(4)}(q-k_1-k_2-k_3)\,
  M^{(0)}_{\text{\IReg}}(q\bar{q}g)\,, 
\end{equation}
in terms of $k_i^0 =\omega_i=\sqrt{{\vec k}_i^2 +\mu^2}$.

Let us first analyze how the regulating mass enters the phase space
integration boundaries. Using the CM frame of the virtual photon,
$\delta^{(4)}(q\!-\!k_1\!-\!k_2\!-\!k_3)=
\delta(q_0\!-\!\omega_1\!-\!\omega_2\!-\!\omega_3)\times$
$\delta^{(3)}({\vec k}_1\!+\!{\vec k}_2\!+\!{\vec k}_3)$,
and after integrating out the three-momentum of the gluon, the
phase space integration $P$ reduces to
\begin{subequations}
\begin{align}
P
&=\int\frac{d^3k_1}{(2\pi)^3 2\omega_1}
  \int\frac{d^3k_2}{(2\pi)^3 2\omega_2}
  \int\frac{d^3k_3}{(2\pi)^3 2\omega_3}
  (2\pi)^4\,\delta^{(4)}(q-k_1-k_2-k_3)\,,
\\
&=\int\frac{d^3k_1}{(2\pi)^3 2\omega_1}
  \int\frac{d^3k_2}{(2\pi)^3 2\omega_2}
  \left(\frac{\pi}{\omega_3}\right)
  \delta(q_0-\omega_1-\omega_2-\omega_3)\,,
\end{align}
\end{subequations}
with $\omega_3\!=\!{\sqrt{({\vec k}_1\!+\!{\vec k}_2)^2\!+\!\mu^2}}$.
The integration over the angle $\theta$ between ${\vec k}_1$ and
${\vec k}_2$ is performed, noting that
$\omega_3\,d\omega_3=|{\vec k}_1||{\vec k}_2|\,d\text{cos}(\theta)$.
In addition, with $ |{\vec k}_i|\,d|{\vec k}_i|=\omega_i\, d\omega_i$
we get
\begin{align}
 P
 =\frac{1}{32\pi^3}
 \int_{\omega_{1m}}^{\omega_{1M}} d\omega_1
 \int_{\omega_{2m}}^{\omega_{2M}} d\omega_2
 \int_{\omega_{3m}}^{\omega_{3M}} d\omega_3\
 \delta^{(0)}(q_0-\omega_1-\omega_2-\omega_3)\,.
\end{align}
The boundary values for the $\omega_3$ integration can be traced back
from the range of allowed $\theta$ angle values. At fixed ${\vec k}_1$
and ${\vec k}_2$ one thus obtains
$\omega_{3m}\!=\!{\sqrt{\mu^2\!+\!(|{\vec k}_1|\!-\!|{\vec k}_2|)^2}}$
corresponding to
$\theta\!=\!\pi$ and $\omega_{3M}
\!=\!{\sqrt{\mu^2\!+\!(|{\vec k}_1|\!+\!|{\vec k}_2|)^2}}$
for $\theta\!=\!0$. In the first case, the quark and antiquark have
opposite momenta and thus a soft gluon momentum ${\vec k}_3$ can be
emitted together with hard fermion momenta. In the second case, the
fermions move parallel and soft gluon emission is accompanied with
soft fermion momenta. Introducing now dimensionless variables
\begin{align}
\label{chi}
\chi_i=\frac{(k_i-q)^2}{q^2}-\frac{\mu^2}{q^2}
\end{align}
with $k_i^2\!=\!\mu^2$ and $q^2\!=\!q_0^2$, one gets
$\chi_i\!=\!1\!-\!2\,\frac{\omega_i}{q_0}$ and
$d\chi_i\!=\!-2\,\frac{d\omega_i}{q_0}$.
In these variables, the phase space integral becomes
\begin{align}
P
=\frac{ q_0^2}{(4\pi)^3}
  \int_{\chi_{1m}}^{\chi_{1M}} d\chi_1
  \int_{\chi_{2m}}^{\chi_{2M}} d\chi_2\,,
\end{align}
keeping in mind the interval allowed for non-vanishing contributions of the
$\delta$-integration. The latter restrict the boundaries of the $\chi_2$
integration to
\begin{align}
\label{chi2lim}
\chi_2^{\pm}=\frac{1-\chi_1}{2}
\pm \sqrt{\frac{(\chi_1-3 \mu_0)\,[(1-\chi_1)^2-4\mu_0]}{4\,(\chi_1+\mu_0)}}\,,
\end{align}
with the notation $\chi_2^{+}\!=\!\chi_{2M}$, $\chi_2^{-}\!=\!\chi_{2m}$.
Finally, the $\chi_1$ integration boundaries are obtained as follows.
From $\chi_1\!=\!1-2\,\frac{\omega_1}{q_0}$,  the upper limit is easily
extracted, given when ${\vec k}_1\!=\!0$, 
\begin{subequations}
\label{chi1lim}
\begin{align}
  \chi_{1M} = 1 - 2 \sqrt{\mu_0}\,. 
\end{align}
The lower boundary is obtained for maximal $\omega_1$, i.\,e.\ for 
${\omega_1}_M\!=\!\mu^2\!+\!|{\vec k}_{1M}|^2
\!=\!\mu^2\!+\!(|{\vec k}_2|\!+\!|{\vec k}_3|)^2$,
achieved when the angle ${\theta}_{23}$ between the fermion and the
gluon is zero. Using further that energy conservation is expressed
in the $\chi$ variables as $1\!=\!\chi_1\!+\!\chi_2\!+\!\chi_3$ and
rewriting Eq.~\eqref{chi} as $\frac{|{\vec k}_i|^2}{q_0^2}
\!=\!\frac{(1-\chi_i)^2}{4}\!-\!\frac{\mu^2}{q_0^2}$,
one can express ${\omega_1}_M$ only in terms of the variables
$\chi_1, \chi_2, \mu_0$. The minimum value of $\chi_1$ then occurs for
$\chi_2\!=\!\frac{(1-3\mu_0)}{2}$, leading to
\begin{align}
  \chi_{1m} = 3\mu_0\,. 
\end{align}
\end{subequations}
Using Eqs.~\eqref{chi1lim} and \eqref{chi2lim} together with
$q_0^2\!=\!q^2\!=\!s$, we obtain for the phase space integral
\begin{align}
P
\ =\ \frac{s}{(4\pi)^3}
  \int_{3\mu_0}^{1-2 \sqrt{\mu_0}} d\chi_1
  \int_{\chi_{2m}}^{\chi_{2M}} d\chi_2
\ \equiv\ \frac{s}{(4\pi)^3}
  \iint\displaylimits_{\chi_1\,\chi_2}\,.
\end{align}

We now turn back to Eq.~(\ref{bs}) and evaluate the matrix element
squared. Following Sec.~\ref{sec:fdhVirtual}, it can be written as
\begin{align}
M^{(0)}_{\text{\IReg}}(q\bar{q}g)
= e^2 g_s^2\,\omega^{(r)}\,L_{\mu\nu}\,G^{\mu\nu}\,,
\end{align}
with
\begin{subequations}
\begin{align}
G^{\mu\nu}
&=-\frac{1}{8}\,\text{Tr}\big[\,
  \slashed k_{1}\,
  \Lambda_{\lambda}^{\phantom{\lambda}\mu}\,
  \slashed k_{2}\,
  \Lambda^{\nu\lambda}\,
  \big]\,,
\\
\label{ireg:real-b}
\Lambda_{\lambda \mu}
&=-\frac{1}{(k_{1}+k_{3})^2}\gamma_{\lambda}
    (\slashed k_{1} + \slashed k_{3})\gamma_{\mu}
  +\frac{1}{(k_{2}+k_{3})^2}\gamma_{\mu}
    (\slashed k_{2} + \slashed k_{3})\gamma_{\lambda} \, ,
\end{align}
\end{subequations}
where we use the leptonic tensor of Eq.~\eqref{leptonavg} and
$\omega^{(r)}\!=\!2\,\eqSq\CF/s^2$. 

The result can be simplified by considering gauge invariance, which implies
that $G^{\mu\nu}$, after phase space integration, must be transverse to the
photon momentum $q$. Thus, the total cross section due to real contribution
can be expressed as 
\begin{align}
\sigma_{\text{\IReg}}^{(r)}
= \sigma^{(0)}\,\Big(\frac{\alpha_{s}}{\pi}\Big)\,\CF
  \iint\displaylimits_{\chi_1\,\chi_2}
  \Big[-\frac{1}{2}\,g_{\mu\nu} G^{\mu\nu}\Big]\,.
\end{align} 
After a tedious, yet straightforward computation, one obtains
\begin{align}
-\frac{1}{2}\,g_{\mu\nu}G^{\mu\nu}
= -\!\left[
    \frac{1}{\mu_{0}\!+\!\chi_{1}}
    \!+\!\frac{1}{\mu_{0}\!+\!\chi_{2}}\right]
  \!+\!\frac{1}{2}\!\left[
    \frac{\chi_{2}}{\mu_{0}\!+\!\chi_{1}}
    \!+\!\frac{\chi_{1}}{\mu_{0}\!+\!\chi_{2}}\right]
  \!+\!\frac{1}{(\mu_{0}\!+\!\chi_{1})(\mu_{0}\!+\!\chi_{2})}
  +\mathcal{O}(\mu_{0})\,,
\end{align}
where we use the definition of $\chi_{i}$ in Eq.~\eqref{chi}
and $k_{i}^{2}\!=\!\mu^{2}$. Finally, the integrals can be evaluated with%
\footnote{These integrals are the same in \IReg\ and \FDR, see
  e.\,g.\ Eqs.~(34) and (35) of Ref.~\cite{Pittau:2013qla}. They can be compared
  with the corresponding ones obtained in \DR, see Eqs.~\eqref{psintegrals}.
  Again, the transition rules for the IR divergences  between \DR\ and \IReg/\FDR\
  read $\frac{1}{\epsilon^2}\leftrightarrow\frac{1}{2}\ln^{2}(\mu_{0})$
  and $\frac{1}{\epsilon}\leftrightarrow\ln(\mu_{0})$.}
\begin{subequations}
\label{eq:RealIntegralsIReg}
\begin{align}
\iint\displaylimits_{\chi_1\,\chi_2}
  \frac{1}{\mu_{0}+\chi_{1}}
\ =\ \iint\displaylimits_{\chi_1\,\chi_2}
  \frac{1}{\mu_{0}+\chi_{2}}
&\ =\ -\ln(\mu_{0})-3 + \mathcal{O}(\mu_{0})\,,
\\
\iint\displaylimits_{\chi_1\,\chi_2}
  \frac{\chi_{2}}{\mu_{0}+\chi_{1}}
\ =\ \iint\displaylimits_{\chi_1\,\chi_2}
  \frac{\chi_{1}}{\mu_{0}+\chi_{2}}
&\ =\ -\frac{\ln(\mu_{0})}{2}-\frac{7}{4} + \mathcal{O}(\mu_{0})\,,
\\
\iint\displaylimits_{\chi_1\,\chi_2}
  \frac{1}{(\mu_{0}+\chi_{1})(\mu_{0}+\chi_{2})}
&\ =\ \frac{\ln^{2}(\mu_{0})}{2}-\frac{\pi^{2}}{2} + \mathcal{O}(\mu_{0})\,.
\end{align}
\end{subequations}
Finally, the total cross section due to the real contribution is given by%
\footnote{This result can be compared with the ones obtained in \DR,
  see Eqs.~\eqref{fdh:real}.
  Using the rules for translating IR divergences between \IReg\ and \DR\
  together with $\Phi_3(\epsilon\!=\!0)\!=\!1$, it follows that
  Eq.~\eqref{ireg:real} coincides with the results in \FDH\ and \DRED,
  Eq.~\eqref{sig_eer_fdh}. In Sec.~\ref{sec:CSfdr}, it will be shown that
  Eq.~\eqref{ireg:real} also coincides with the corresponding result in \FDR.}
\begin{align}
\sigma_{\text{\IReg}}^{(r)}
= \sigma^{(0)}\left(\frac{\alpha_{s}}{\pi}\right)
  \CF\Big[\,
    \frac{\ln^{2}(\mu_{0})}{2}
    +\frac{3}{2}\ln(\mu_0)
    +\frac{17}{4}
    -\frac{\pi^{2}}{2} 
    + \mathcal{O}(\mu_0)
    \Big]\,.
\label{ireg:real}
\end{align} 
The procedure of obtaining the real corrections in \IReg\ can be
summarized as follows: compute the matrix element squared for
\textit{massless} external and internal particles as in
Eq.~\eqref{ireg:real-b}. However, the on-shell limit $k_i^2=0$
should not be applied. Instead, wherever a squared momentum appears it
should be replaced by $k_i^2=\mu^2$. The phase-space integration is to
be carried out for massive external particles. IR divergences appear
as $\ln(\mu_0)$ terms%
\footnote{The only technical difference to the
  evaluation of real corrections in \FDR\ is that in \FDR\ the matrix
  elements are computed in the strict massless limit, i.\,e.\ using
  $k_i^2=0$. Thus, at least at NLO the two schemes differ at most by
  terms $\mathcal{O}(\mu_0)$. }.

Finally, adding the virtual contribution, Eq.~\eqref{crsv}, one
obtains the well-known UV and IR finite result
\begin{align}
\sigma^{(1)}
= \sigma^{(0)} + \sigma^{(v)}_{\text{\IReg}} +
\sigma^{(r)}_{\text{\IReg}}\Big|_{\mu_0\to 0}
=  \frac{\eqSq\,N_c}{3\, s}\Big(\frac{e^4}{4\pi}\Big)
\Big[\,1+\Big(\frac{\alpha_s}{4\pi}\Big) \, 3\, \CF \Big] \, .
\label{ireg:xs}
\end{align}

\subsection{Established properties of IREG}
\label{sec:ireg_properties}

\subsubsection*{Gauge invariance}
\label{sec:ireg_gauge_invariance}

In gauge theories, the initial structure of a given Feynman diagram contains
Dirac matrices, Lorentz contractions, etc. These operations may generate terms
with squared momenta in the numerator which must be cancelled against propagators
\textit{before} applying the rules of \IReg. This point was first emphasized in
differential regularization whose rules have a one-to-one correspondence
with the \IReg\ prescription \cite{Sampaio:2002ii}. As an example, consider
the (unregularized) off-shell vacuum polarization tensor in massless QED at
one-loop
\begin{align}
\Pi^{\mu\nu} = - e^2\,i^4\,\text{Tr} \int\frac{d^4k}{(2 \pi)^4}
\gamma^{\mu}\,\frac{1}{\slashed k}\,\gamma^{\nu}\frac{1}{\slashed k-\slashed p}\,,
\end{align}
which, after evaluating the Dirac algebra, can be expressed as 
\begin{subequations}
\label{eq:pi}
\begin{align}
\Pi^{\mu\nu}
&=-4e^{2}\int\frac{d^4k}{(2 \pi)^4}\frac{
  2k^{\mu}k^{\nu}
  -g^{\mu\nu}k^{2}
  -k^{\mu}p^{\nu}-k^{\nu}p^{\mu}
  +g^{\mu\nu}(k\cdot p)
  }{k^{2}(k-p)^{2}},
\\&
\equiv -4e^{2}\big[
  2I^{\mu\nu}
  -g^{\mu\nu}J
  -I^{\mu}p^{\nu}
  -I^{\nu}p^{\mu}
  +g^{\mu\nu}(I_{\alpha}p^{\alpha})
  \big]\,.
\end{align}
\end{subequations}
The integrals, after applying the rules of \IReg, are given as
\begin{subequations}
\begin{align}
J_{\text{\IReg}}&= \int\frac{d^4k}{(2 \pi)^4}\frac{k^2}{k^{2}(k-p)^{2}}
= \int\frac{d^4k}{(2 \pi)^4}\frac{1}{(k-p)^{2}}
=-p^{2}\upsilon_{0,2}\,,
\label{eq:J}
\\
I_{\text{\IReg}}^{\mu}
&= \int\frac{d^4k}{(2 \pi)^4}\frac{k^{\mu}}{k^{2}(k-p)^{2}}
= \frac{p^{\mu}}{2}\left[
  I_{\text{log}}(\lambda^{2})
  -b\ln\left(-\frac{p^{2}}{\lambda^{2}}\right)
  +2\,b
  -\upsilon_{0,2}
  \right]\,,
\\
I_{\text{\IReg}}^{\mu\nu}
&= \int\frac{d^4k}{(2 \pi)^4}\frac{k^{\mu}k^{\nu}}{k^{2}(k-p)^{2}}
= \frac{1}{3}p^{\mu}p^{\nu}\left[
  I_{\text{log}}(\lambda^{2})
  -b\ln\left(-\frac{p^{2}}{\lambda^{2}}\right)
  +\frac{11}{6}b
  \right]
\\*\nonumber
&\ \,
  \!-\!\frac{1}{12}g^{\mu\nu}p^{2}\left[
  I_{\text{log}}(\lambda^{2})
  \!-b\ln\left(\!-\!\frac{p^{2}}{\lambda^{2}}\right)
  \!+\!\frac{4}{3}b
  \right]
  \!-\!\frac{g^{\mu\nu}}{2}\upsilon_{2,2}
  -\!\frac{1}{6}(g^{\mu\nu}p^{2}+2p^{\mu}p^{\nu})\upsilon_{0,4}
  \!+\!\frac{1}{4}g^{\mu\nu}p^{2}\upsilon_{0,2}\,,
\end{align}
\end{subequations}
where we have suppressed quadratic divergences (in the example they
cancel exactly), and $\nu_{i,j}$ are surface terms defined as
\begin{align}
g^{\{\nu_{1}\cdots\nu_{j}\}}\upsilon_{i,j}
\equiv\Upsilon_{i}^{\nu_{1}\cdots\nu_{j}}
\equiv\int{\frac{d^{4}k}{(2 \pi)^4}}
  \frac{\partial}{\partial k_{\nu_{1}}}\frac{k^{\nu_{2}}\cdots k^{\nu_{j}}}
  {(k^{2}-\mu^{2})^{\frac{2+j-i}{2}}}\,,
\end{align}
where we use $g^{\{\nu_{1}\cdots\nu_{j}\}}
\equiv g^{\nu_{1}\nu{2}}\cdots g^{\nu_{j-1}\nu_{j}}$ +
symmetric combinations.
Inserting all results in $\Pi_{}^{\mu\nu}$, one obtains
\begin{align}
\Pi_{\text{\IReg}}^{\mu\nu}
&=-\frac{4}{3}e^{2}\left(
  g^{\mu\nu}p^{2}
  -p^{\mu}p^{\nu}
  \right)\left[
  I_{\text{log}}(\lambda^{2})
  -b\ln\left(-\frac{p^{2}}{\lambda^{2}}\right)
  +\frac{7}{3}b
  \right]
\nonumber\\&\quad
-4\,e^{2}\left[
  -\frac{1}{3}(g^{\mu\nu}p^{2}+2p^{\mu}p^{\nu})\upsilon_{0,4}
  +p^{\mu}p^{\nu}\upsilon_{0,2}
  -g^{\mu\nu}\upsilon_{2,2}
  \right]\,.
\end{align}
As can be seen, to enforce gauge invariance (expressed in the transversality
of $\Pi_{\text{\IReg}}^{\mu\nu}$), surface terms should be null as previously
discussed \cite{Ferreira:2011cv}.

We remark the appearance of a $k^{2}$ term in Eqs.~\eqref{eq:pi}, defined
as the divergent $J$ integral, and the importance of applying \IReg\ rules
only \textit{after} cancelling such term against propagators. Proceeding otherwise,
by rewriting $k^2\!=\!g^{\mu\nu}k_{\mu}k_{\nu}$ for instance, one would obtain
\begin{align}
\int\frac{d^4k}{(2 \pi)^4}\frac{k^2}{k^{2}(k-p)^{2}}
= g^{\mu\nu}\int\frac{d^4k}{(2 \pi)^4}\frac{k_{\mu}k_{\nu}}{k^{2}(k-p)^{2}}
= \frac{p^{2}}{6}b -2\,\upsilon_{2,2} - p^{2}(\upsilon_{0,4}-\upsilon_{0,2})\,, 
\end{align} 
which is different from the $J$ integral, Eq.~\eqref{eq:J}, not only by arbitrary
terms encoded in the $\upsilon_{i,j}$ but also by a finite term. In this way,
gauge invariance would be broken even if the surface terms are systematically
set to zero. It should be emphasized that the discussion above is restricted to
divergent integrals.

\subsubsection*{UV renormalization}

We would also like to briefly show how renormalization-group functions can be
computed in the framework of \IReg. For simplicity, we adopt the background
field method \cite{Abbott:1983zw} which relates the wave function
renormalization of the background field, $B_{0}\!=\!Z_{B}B$, with
the coupling renormalization, $e_{0}\!=\!Z_{e}\,e$, through the
equation $Z_{e}\!=\!Z_{B}^{-1/2}$. Therefore, by applying this method to QED,
the $\beta$ function can be obtained only with the knowledge of the vacuum
polarization tensor. Performing a minimal subtraction, which in \IReg\ amounts
to subtract only basic divergent integrals as $I_{\text{log}}(\lambda^{2})$, and
remembering that $\lambda$ plays the role of a renormalization group scale,
one obtains%
\footnote{This result coincides with the well-known value of the
QED $\beta$ function of the gauge coupling obtained in \DR, see
Eq.~\eqref{eq:betaQED}.}
\begin{align}
\beta
&=\lambda^2\!\frac{\partial}{\partial\lambda^2}\Big(\frac{e}{4\pi}\Big)^2
\!=\frac{e^4}{(4\pi)^2}
  \frac{4}{3}N_F\,
  i\,\lambda^2\!\frac{\partial}{\partial\lambda^2}
    I_{\text{log}}(\lambda^{2})
  \!+\!\mathcal{O}(e^6)
=\!-\Big(\frac{e}{4\pi}\Big)^4\Big[\!-\!\frac{4}{3}N_F\Big]
  \!+\!\mathcal{O}(e^{6})\,.
\end{align}
Further examples can be found in Refs.~\cite{Carneiro:2003id,Sampaio:2005pc,
Brito:2008zn,Fargnoli:2010mf,Cherchiglia:2015vaa}.  

\section{FDR: Four-dimensional regularization/renormalization}
\label{sec:fdr}

\FDR~\cite{Pittau:2012zd} is a fully four-dimensional framework to compute
radiative corrections in QFT. The calculation of the loop corrections is
conceptually simplified with respect to more traditional approaches in that
there is no need to include UV counterterms in the Lagrangian $\mathcal{L}$.
In fact, the outcome of an \FDR\ calculation at any loop order is directly
a UV-renormalized quantity. Moreover, this particular way of looking at the UV
problem may open new perspectives in the present understanding of fundamental
and effective QFTs~\cite{Pittau:2013ica}. In the following, we review the
\FDR\ treatment of UV and IR divergences, also using the
$e^{+} e^{-}\!\to\!\gamma^{*}\!\to\! q\bar{q}(g)$ process as an explicit example.

\subsection{FDR and UV infinities}

Let $J(q_1,\ldots,q_\ell)$ be an integrand depending on $\ell$ integration
momenta $q_1,\ldots,q_\ell$. The \FDR\ integral over $J$ is {\em defined}
as follows:
\begin{align}
\label{eq:FDRdef}
\int [d^4 q_1] \cdots [d^4 q_\ell]\, J(q_1,\ldots,q_\ell,\mu^2) \equiv
\lim_{\mu \to 0} 
\int d^4 q_1 \cdots d^4 q_\ell \,
J_{\rm F}(q_1,\ldots,q_\ell,\mu^2)\,,
\end{align}
where $J_{\rm F}(q_1,\ldots,q_\ell,\mu^2)$ is the UV-finite part of
$J(q_1,\ldots,q_\ell,\mu^2)$ (specified below), $\mu$ is an infinitesimal
mass needed to extract $J_{\rm F}$ from $J$, and $\int[d^4q_i]$ denotes the
\FDR\ integration.
The integrands $J(q_1,\ldots,q_\ell,\mu^2)$ and
$J_{\rm F}(q_1,\ldots,q_\ell,\mu^2)$ are obtained from
$J(q_1,\ldots,q_\ell)$ with the help of the following rules:
\begin{itemize}
\item[i)] Squares of integration momenta appearing both in the
denominators of $J(q_1,\ldots,q_\ell)$ and in contractions generated
in the numerator by Feynman rules are shifted by $\mu^2$,
\begin{align}
\label{eq:shift}
q^2_i \to q^2_i -\mu^2 \equiv \qbar^2_i\,.
\end{align}
This replacement is called {\em global prescription}.
\item[ii)] A splitting
\begin{align}
\label{eq:split}
J(q_1,\ldots,q_\ell,\mu^2)= [J_{\rm INF}(q_1,\ldots,q_\ell,\mu^2)]+
J_{\rm F}(q_1,\ldots,q_\ell,\mu^2)
\end{align}
is performed in such a way that UV divergences are entirely parametrized
in terms of divergent integrands contained in $[J_{\rm INF}]$, that
solely depend on $\mu^2$. By convention, we write divergent integrands
in square brackets and call them {\em \FDR\ vacua}, or simply {\em vacua}.

\item[iii)]  The global prescription in \Eqn{eq:shift} should be made
compatible with a key property of multi-loop calculus:
\begin{align}
\label{eq:subcon}
\begin{tabular}{l}
{\em In an $\ell$-loop diagram,
one should be able to calculate a subdiagram,} \\
{\em insert the integrated
form into the full diagram and get the same answer.} \\
\end{tabular}
\end{align}
We dub this {\em subintegration consistency}.
\end{itemize}
\nopagebreak
Finally, after $\lim_{\mu \to 0}$ is taken, $\ln \mu \to \ln \mur$ is
understood on the r.\,h.\,s.\ of \Eqn{eq:FDRdef}, where $\mur$ is an
arbitrary renormalization scale.  Note that inserting \Eqn{eq:split}
into \Eqn{eq:FDRdef} gives an alternative definition 
\begin{align}
\label{eq:FDRdef1}
&\int [d^4 q_1] \cdots [d^4 q_\ell]\, J(q_1,\ldots,q_\ell,\mu^2)
\notag\\*&\quad
= \lim_{\mu \to 0} \int_{\text{\R}} d^4 q_1 \cdots d^4 q_\ell\,
\Big\{
J(q_1,\ldots,q_\ell,\mu^2)-[J_{\rm INF}(q_1,\ldots,q_\ell,\mu^2)]
\Big\}\,,
\end{align}
where \R\ denotes an arbitrary UV regulator. \Eqn{eq:FDRdef1} tells us that
the UV subtraction is directly encoded in the definition of \FDR\ loop
integration: no divergent integrand is considered separately from
its subtraction term.

\FDR\ integration preserves shift invariance which is easy to prove 
when using \Eqn{eq:FDRdef1} with \R\ =\ \DR,
\begin{align}
\label{eq:shiftinv}
\int [d^4q_1] \ldots [d^4q_\ell]\,  J(q_1,\ldots, q_\ell,\mu^2)
=\int [d^4q_1] \ldots [d^4q_\ell]\,  J(q_1+p_1,\ldots, q_\ell+p_\ell,\mu^2)\,, 
\end{align}
and the possibility of cancelling numerators and denominators
\begin{align}
\label{eq:canc}
\int [d^4q_1] \ldots [d^4q_\ell]\,
  \frac{\qbar^2_i-m^2_i}{ (\qbar^2_i-m^2_i)^m \ldots}
=
\int [d^4q_1] \ldots [d^4q_\ell]\,
  \frac{1}{ (\qbar^2_i-m^2_i)^{m-1}\ldots}\,,
\end{align}
which are properties needed to retain the symmetries of $\mathcal{L}$~%
\cite{Donati:2013iya}.
From \Eqns{eq:shiftinv}{eq:canc} follows that algebraic manipulations
in \FDR\ integrands are allowed as if they where convergent ones. This
authorizes one to reduce complicated multi-loop integrals to a limited
set of Master Integrals (MI) by using four-dimensional tensor
decomposition~\cite{Donati:2013voa} or integration-by-parts
identities~\cite{Pittau:2014tva}. In other words, the definition in
\Eqn{eq:FDRdef} (or \Eqn{eq:FDRdef1}) can be applied just at the end of
 the calculation, when the actual value of the MIs is needed. 

An important subtlety implied by \Eqn{eq:canc} is that the needed cancellation
works only if integrands involving explicit powers of $\mu^2$ in the numerator
are also subtracted {\em as if $\mu^2 = q^2_i$}, where $q^2_i$
is the momentum squared which generates $\mu^2$. For instance, one computes 
\begin{align}
\label{eq:eqextra}
\int [d^4q] \frac{\mu^2}{(q^2-M^2)^3}= \lim_{\mu \to 0} \mu^2 \int
d^4q \bigg\{ \frac{1}{(q^2-M^2)^3}-\bigg[\frac{1}{\qbar^6}\bigg]\bigg\}
= \frac{i\pi^2}{2}\,,
\end{align}
in accordance with \Eqn{eq:FDRdef1}. In this case both integrals on the
r.\,h.\,s.\ are UV convergent and the only contribution which survives
the ${\mu \to 0}$ limit is generated by the subtraction term. As a
consequence, although only one kind of $\mu^2$ exists, one has to keep
track of its origin when it appears in the numerator of
$J(q_1,\ldots,q_\ell,\mu^2)$. For this we use the notation $\mu^2|_i$,
which understands the same subtraction required for the case $\mu^2=
q^2_i$.  \FDR\ integrals with powers of $\mu^2|_i$ in the numerator
are called 'extra integrals'%
\footnote{This is different compared to \IReg\ where no extra integrals
  are introduced. While extra integrals are not strictly needed in \FDR,
  they are introduced for convenience to allow the decomposition of \FDR\
  tensor integrals into MIs and to avoid introducing couterterms in
  ${\cal L}$. }.
Their computation is elementary, as illustrated by \Eqn{eq:eqextra}.
Additional one- and two-loop examples can be found in
Refs.~\cite{Pittau:2012zd,Donati:2013voa}. \FDR\ extra integrals play
an important role in maintaining the theory unitary without the need
of introducing counterterms in ${\cal L}$, as will be discussed in
Sec.~\ref{sec:propertiesFDR}.

\noindent
As a simple example of an \FDR\ integration, we consider the scalar
one-loop integrand
\begin{align}
J(q)= \frac{1}{(q^2-M^2)^2}\,,
\end{align}
which diverges logarithmically for $q\!\to\!\infty$.
The steps to define its \FDR\ integral are as follows:
\begin{itemize}
\item Shift squares of the integration momentum,
  \begin{align}
    \label{eq:ing}
    J(q) \to J(q,\mu^2) \equiv \frac{1}{(\qbar^2-M^2)^2}\,,
    \quad{\rm with}\quad\qbar^2 \equiv q^2-\mu^2\,.
    \end{align}
\item Subtract the divergent part of the integrand
  $[J_{\rm INF}(q,\mu^2)] = \left[\frac{1}{\qbar^4} \right]$
  in the $\mu\!\to\!0$ limit, setting $\mu\!\to\!\mur$ in the logarithms
  \begin{align}
  \label{eq:example}
    \int [d^4q] \frac{1}{(\qbar^2-M^2)^2} \equiv
    \lim_{\mu \to 0} \int_{\text{\R}} d^4q
    \bigg\{
      \frac{1}{(\qbar^2-M^2)^2}
      -\bigg[\frac{1}{\qbar^4}\bigg]
      \bigg\} \Bigg|_{\mu \to \mur}\,.
  \end{align}
\item The dependence on \R\ is eliminated by using the partial
  fraction identity
  \begin{align}
  \label{eq:parf}
    \frac{1}{\qbar^2-M^2}
    = \frac{1}{\qbar^2} \Big(1+ \frac{M^2}{\qbar^2-M^2}\Big)
  \end{align}
in the first integrand on the r.\,h.\,s.\ of Eq.~\eqref{eq:example}.
This exactly cancels the divergent term $\big[\frac{1}{\qbar^4}\big]$
{\em before} integration, leaving the UV finite result%
\footnote{The alternative definition in \Eqn{eq:example} with, for example,
  \R\ =\ \DR\ gives the same result,
  $\int [d^4q] \frac{1}{(\qbar^2-M^2)^2}
    \!=\!\mu_{\text{\DR}}^{4-\dim}\int d^dq \frac{1}{(q^2-M^2)^2}
      -\lim_{\mu \to 0}\mu_{\text{\DR}}^{4-\dim}
	\int d^dq \frac{1}{\qbar^4}\Big|_{\mu \to \mu_{\text{\DR}}}
    \!=\!-i \pi^2 \ln \frac{M^2}{\mu_{\text{\DR}}^2}$\,.
  In the first integral, $\mu$ can be directly set to zero  
  since it is IR convergent.}
  \vspace*{-.1cm}
  \begin{subequations}
  \label{eq:fin}
  \begin{align}
    \int [d^4q] \frac{1}{(\qbar^2-M^2)^2}
    &\equiv\lim_{\mu \to 0}\int d^4q\bigg\{
      \frac{M^2}{\qbar^4(\qbar^2-M^2)}
      +\frac{M^2}{\qbar^2(\qbar^2-M^2)^2}\bigg\}
    \Bigg|_{\mu \to \mur}
    \\&
    = -i \pi^2 \ln \frac{M^2}{\mur^2}\,.
  \end{align}
  \end{subequations}
\end{itemize}
In practice, one can directly start from the integrand in \Eqn{eq:ing}
and expand it by means of \Eqn{eq:parf}. This procedure allows one to
naturally separate $[J_{\rm INF}(q_1,\ldots,q_\ell,\mu^2)]$ from any
integrand $J(q_1,\ldots,q_\ell,\mu^2)$ and write down definitions
analogous to Eqs.~\eqref{eq:fin} at any loop order. Explicit examples
for the extraction of \FDR\ vacua from two-loop integrands are
presented in Ref.~\cite{Donati:2013voa}.

Given the fact that the definition of \FDR\ loop integration is
compatible with a graphical proof of the Slavnov-Taylor identities
through \Eqns{eq:shiftinv}{eq:canc} and can be made congruent with
the subintegration consistency of \Eqn{eq:subcon} without the
need of introducing UV counterterms in ${\cal L}$
(see Sec.~\ref{sec:propertiesFDR} and  Ref.~\cite{Page:2015zca}
for more details on this point), \FDR\ quantities are directly
interpretable as UV-renormalized ones. As an example, the correspondence
between off-shell two-loop QCD correlators computed in \FDR\ and \DR\
has been worked out in Ref.~\cite{Page:2015zca}.

\subsection{FDR and IR infinities}

\begin{figure}[t]
\begin{center}
\begin{picture}(80,50)(0,0)
\SetOffset(60,25)
\SetWidth{0.5}
\SetWidth{2}
\Line(-40,0)(-15,15)
\Line(-15,-15)(-40,0)
\Line(-15,15)(-15,-15)
\SetWidth{0.5}
\Line(-15,15.5)(10,15.5)
\SetWidth{0.5}
\Line(-15,-15.5)(10,-15.5)
\LongArrow(5,23)(-5,23)
\Text(13,23)[l]{\small $ p^2_1= 0 $}
\LongArrow(-80,0)(-53,0)
\Text(-85,12)[l]{\small $ (p_2-p_1)^2$}
\LongArrow(-5,-23)(5,-23)
\Text(13,-23)[l]{\small $ p^2_2= 0 $}
\LongArrow(-7,-6)(-7,6)
\Text(-1,0)[l]{$q$}
\end{picture}
\end{center}
\vspace*{-10pt}
\caption{\label{fig1}
Massless scalar one-loop three-point function.
Thick internal lines denote the insertion of the infinitesimal mass $\mu$,
which generates $\mu$-massive propagators.}
\end{figure}
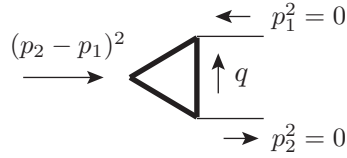

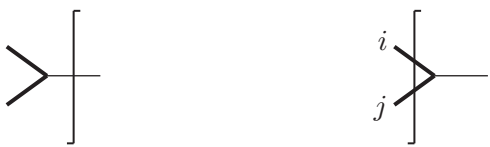
\begin{figure}[t]
\begin{center}
\begin{picture}(450,50)(40,0)
\SetScale{0.50}
\SetWidth{1}

\SetOffset(215,8)
\Line(-60,11)(-20,11)
\SetWidth{3}
\Line(-60,11)(-90,33)
\Line(-60,11)(-90,-11)
\SetWidth{1.5}
\Line(-40,60)(-35,60)
\Line(-40,-40)(-40,60)
\Line(-45,-40)(-40,-40)
\SetWidth{1}

\SetOffset(360,8)
\Line(-60,11)(-20,11)
\SetWidth{3}
\Line(-60,11)(-90,33)
\Line(-60,11)(-90,-11)
\Text(-95,30)[br]{\normalsize$i$}
\Text(-95,-23)[br]{\normalsize$j$}

%
\SetWidth{1.5}
\Line(-75,60)(-70,60)
\Line(-75,-40)(-75,60)
\Line(-80,-40)(-75,-40)

\end{picture}
\end{center}
\caption{\label{fig2}
Splitting regularized by $\mu$-massive (thick) unobserved particles. The
one-particle cut contributes to the virtual part, the two-particle cut to
the real radiation.}
\end{figure}

The modification of the propagators induced by \Eqn{eq:shift} also regularizes
soft and collinear divergences in the virtual integrals~\cite{Pittau:2013qla}.
As an example, the massless one-loop three-point function corresponding
to the Feynman diagram shown in Fig.~\ref{fig1} is interpreted in \FDR\ as
\begin{subequations}
\label{eq:FDRirExample}
\begin{align}
\label{eq:triang}
I_{\text{\FDR}}
=\int [d^4q]\frac{1}{\qbar^2 \bar D_1 \bar D_2}
=\lim_{\mu \to 0} \int d^4q \frac{1}{\qbar^2 \bar D_1 \bar D_2}\,,
\end{align}
with $\qbar^2= q^2 -\mu^2$ and $\bar{D}_i= (q+ p_i)^2-\mu^2$. It is worth
noticing that this is the same definition as given in \Eqn{eq:FDRdef1}.
In fact, there is no $[J_{\rm INF}]$ term to subtract in this case since
the integrand is UV finite. It is easy to compute
\begin{align}
\label{eq:eqn1}
I_{\text{\FDR}}
= \frac{i \pi^2}{s}\Big[
    \frac{\ln^2(\mu_0)}{2}
    +i \pi \ln(\mu_0)
    -\frac{\pi^2}{2}
    +\mathcal{O}(\mu_0)
    \Big]\,,
\end{align}
\end{subequations}
with $s\!=\!(p_2\!-\!p_1)^2$ and $\mu_0\!=\!\mu^2/s$.
Thus, IR divergences take the form of logarithms of $\mu_0$.
In the case at hand, the squared logarithm is generated by an overlap
of soft and collinear divergences when $q\!\to\!0$ and $q$ is
collinear to~$p_i$.

This prescription certainly allows one to assign a precise meaning to
virtual integrals also in the presence of IR singularities. Nevertheless,
the correct final result is obtained only if the real part of the radiative
corrections is treated likewise. This is obtained by carefully analyzing
the Cutkowsky rules~\cite{Cutkosky:1960sp} relating real and virtual
contributions with different cuts of diagrams at a higher perturbative level,
where cutting a propagator means going on-shell,
$\frac{i}{q^2+ i 0} \to (2\pi)\, \delta(q^2)\,\theta(q_0)$.
This correspondence is linked to the identity%
\footnote{This relation is also one of the starting points of the \FDU\ scheme
  described in Sec.~\ref{sec:fdu}.}
\begin{align}
\label{eq:funIR}
\frac{i}{q^2+ i 0}
= (2\pi)\, \delta(q^2)\,\theta(q_0)
+ \frac{i}{q^2- i 0\, q_0}\,.
\end{align}
In fact, IR singularities on the l.\,h.\,s.\ of \Eqn{eq:funIR} manifest
themselves as pinches of the integration path by two (or more) singularities
in the $q_0$ complex plane, which occur in the virtual part of the radiative
corrections. On the other hand, the first term on the r.\,h.\,s.\ gives end-point
singularities, typical of the real radiation, and the last term generates IR
finite contributions. It is then clear that the \FDR\ modification
$\frac{i}{q^2+ i 0} \to \frac{i}{\bar q^2+ i 0}$
in the virtual contribution is matched by the $\mu$-massive version of
\Eqn{eq:funIR}, namely
\begin{align}
\label{eq:funIRmu}
\frac{i}{\bar q^2+ i 0}
=(2\pi)\, \delta(\bar q^2)\,\theta(q_0)
 + \frac{i}{\bar q^2- i 0\, q_0}\,,
\end{align}
which in turn is responsible for the correspondence
$\frac{i}{\bar q^2+ i 0} \to (2\pi)\, \delta(\bar q^2)\,\theta(q_0)$
depicted in Fig.~\ref{fig2}, see also Ref.~\cite{Pittau:2013qla}.
For example, \Eqn{eq:funIRmu} can be used to rewrite the real part of
\Eqn{eq:triang} as an integral over an eikonal factor
\begin{align}
\label{eq:cancc}
\frac{\pi^2}{4}\, \operatorname{Re}\left(\frac{1}{i \pi^2}\int [d^4q]
  \frac{1}{\qbar^2 \bar D_1 \bar D_2} \right)
= \lim_{\mu \to 0} \int\displaylimits_{\bar \Phi_3}
  \frac{1}{\bar s_{13} \bar s_{23}}\,, 
\end{align}
where $\bar s_{ij}\!=\!(\bar p_i\!+\!\bar p_j)^2$,
$\bar p^2_{i,j}\!=\!\mu^2\to 0$ and
$\bar \Phi_3$ denotes the $\mu$-massive $3$-particle phase space. 

\begin{figure}[t]
\begin{center}
\begin{picture}(400,50)(0,0)
\SetScale{0.70}
\SetWidth{1}

\Text(105,23.71)[r]{$\sigma_{\mbox{\tiny  NLO}}~~=$}
\SetOffset(130,10) 
\SetWidth{1.5}
\Line(25,60)(30,60)
\Line(25,-40)(25,60)
\Line(20,-40)(25,-40)
\Text(15.7,32.8)[l]{\tiny 1}
\Text(15.7,21.4)[l]{\large .}
\Text(15.7,14.3)[l]{\large .}
\Text(15.7,7.14)[l]{\large .}
\Text(5.71,-10)[l]{\tiny m-1}
\Text(10,-25.7)[l]{\tiny m}
\SetWidth{1}
\Line(-60,11)(-20,11)
\SetWidth{2.5}
\Arc(-19,41)(30,270,360)
\Arc(10,11)(30,90,180)
\SetWidth{1}
\Arc(10,11)(30,0,90)
\Arc(10,11)(30,180,360)
\Line(40,11)(80,11)
\Text(112.66,10.666)[l]{\normalsize +}

\Text(-8.57,50)[b]{\normalsize $q$}
\LongArrow(7,47)(-10,42)

\Arc(10,35)(40,215,325)
\BCirc(-20,11){11}
\BCirc(40,11){11}

\SetOffset(285,10)

\SetWidth{1.5}
\Line(0,60)(5,60)
\Line(0,-40)(0,60)
\Line(-5,-40)(0,-40)
\SetWidth{1}
\Line(-60,11)(-20,11)
\SetWidth{2.5}
\Arc(-19,41)(30,270,360)
\Arc(10,11)(30,90,180)
\SetWidth{1}
\SetColor{Black}
\Arc(10,11)(30,0,90)
\SetColor{Black}
\SetWidth{1}
\Arc(10,11)(30,180,360)
\SetWidth{1}
\Line(40,11)(80,11)

\Text(-8.57,50)[b]{\normalsize $q$}
\LongArrow(7,47)(-10,42)

\Arc(10,35)(40,215,325)
\BCirc(-20,11){11}
\BCirc(40,11){11}
\end{picture}
\end{center}
\caption{\label{fig3}
IR divergences drop out when summing the $m$-particle virtual piece
$\sigma^{(v)}_{}$ and its real $(m\!+\!1)$-particle counterpart
$\sigma^{(r)}_{}$. Adding the contributions gives the fully inclusive
NLO cross section.
}
\end{figure}
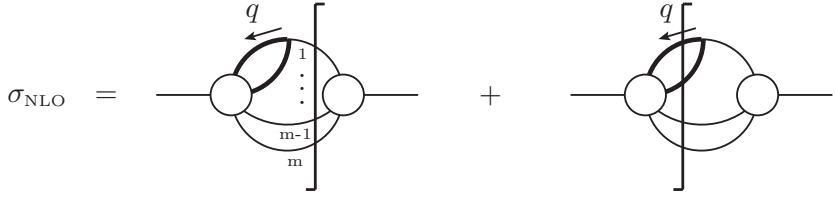

In summary, the IR divergent $1\!\to\!2 $ massless splitting gets regularized
by the introduction of an infinitesimal mass $\mu$ for all unobserved particles.
In the case of external particles, this is equivalent to trade a massless
phase space for a $\mu$-massive one. Furthermore, IR infinities cancel when
summing real and virtual contributions, for instance
\begin{align}
\label{eq:sigmatot}
\sigma_{}
=\int\displaylimits_{\Phi_{m}} d \sigma^{(v)}
  +\lim_{\mu \to 0} \int\displaylimits_{\bar \Phi_{m+1}}
    d \sigma^{(r)}(\{\bar s_{ij}\})
=\sigma^{(v)}_{}
+\sigma^{(r)}_{}\,,
\end{align}
as illustrated in Fig.~\ref{fig3}.
Finally, when $d \sigma^{(r)}_{}(\{s_{ij}\})$ is analytically known in
terms of massless invariants $s_{ij}=(p_i+p_j)^2$ with $p^2_{i,j}=0$,
\Eqn{eq:sigmatot} prescribes the replacement $s_{ij}\!\to\!\bar s_{ij}$.
If, instead, $d \sigma^{(r)}_{}$ is known only numerically, one can construct
a mapping from a massive to a massless phase space, $\bar \Phi_{m+1}
\to^{\hskip  -18.5pt \mbox{\tiny $\vphantom{A}^{\mbox {mapping}}$}}
\Phi_{m+1}$, use $\Phi_{m+1}$ to compute massless invariants and rewrite
the real contribution as
\begin{align}
\label{eq:sigmar}
\sigma^{(r)}_{}
= \lim_{\mu \to 0} \int\displaylimits_{\bar \Phi_{m+1}}
  d \sigma^{(r)}_{}(\{s_{ij}\})~
  \prod_{i < j} \frac{s_{ij}}{\bar s_{ij}}\,.
\end{align}
In this way, $d \sigma^{(r)}_{}(\{s_{ij}\})$ is gauge invariant
since it is computed with massless kinematics and the fudge factor
$\prod_{i < j} \frac{s_{ij}}{\bar s_{ij}}$ effectively replaces
$s_{ij}\!\to\!\bar s_{\ij}$ in all relevant IR singular configurations.
This is because $d \sigma^{(r)}_{}(\{s_{ij}\}) \sim {\frac{1}{s_{ij}}}$
when $s_{ij} \to 0$.

\subsection{Application example:
$e^{+}\,e^{-}\to\gamma^{*}\to\,q\bar q$ at NLO}
\label{sec:CSfdr}

\subsubsection*{Virtual contributions}
In this section we perform the computation of the total cross section of
the process $e^{+}\,e^{-}\!\to\!\gamma^{*}\!\to\!\,q\bar q$ in QCD to
illustrate a typical \FDR\ calculation. As for the virtual part of the
corrections, scaleless integrals vanish. More precisely, in \FDR\ they
are proportional to $\ln \mur/\mu$ (where $\mu$ is the IR regulator),
that gives zero when choosing $\mur=\mu$~\cite{Donati:2013voa}. 
Thus, only the vertex diagram where a virtual gluon connects the quark with
the anti-quark has to be considered. The only subtle point of the calculation
is the replacement
\begin{align}
\label{eq:gp}
\rlap/ q \gamma^\alpha \rlap/ q =  
- q^2\gamma^\alpha +2 \gamma_\beta q^\alpha q^\beta
\to 
-\qbar^2\gamma^\alpha +2 \gamma_\beta q^\alpha q^\beta
\end{align} 
in the fermion string, dictated by the global prescription.
Note that this is fully equivalent to the \IReg\ recipe of performing
simplifications \textit{before} introducing $\mu^2$ in the denominators.
In fact, the replacement in \Eqn{eq:gp} produces a contribution proportional to
\begin{align}
\label{eq:exgp}
\int [d^4q] \frac{-\qbar^2\gamma^\alpha +2 \gamma_\beta q^\alpha q^\beta}{
  \qbar^2 \bar D_1 \bar D_2}
=-\gamma^\alpha\int [d^4q] \frac{1}{\bar D_1 \bar D_2}
+2 \gamma_\beta\int [d^4q] \frac{q^\alpha q^\beta}{\qbar^2 \bar D_1 \bar D_2}\,,
\end{align}
which is the same result one would obtain by simplifying before introducing
$\mu$-massive propagators. In both cases, the gauge-preserving simplification
between the numerator and the denominator of the first integral on the
r.\,h.\,s.\ of \Eqn{eq:exgp} is achieved. Differences between \FDR\ and \IReg\
start when evaluating the second integral.  A customary Passarino-Veltman
tensor decomposition is possible in \FDR\ {\em before} using the definition
of the \FDR\ integral given in \Eqn{eq:FDRdef}%
\footnote{In \IReg, $C^{\alpha \beta}$ is directly computed by subtracting
  its UV divergent part.}:
\begin{align}
\label{eq:pasve}
C^{\alpha \beta}
\equiv  \int [d^4q] \frac{q^\alpha q^\beta}{\qbar^2 \bar D_1 \bar D_2}
= C_{00}\, (g^{\alpha\beta})
+ C_{11}\, (p_1^\alpha p_1^\beta)
+ C_{22}\, (p_2^\alpha p_2^\beta)
+ C_{12}\, (p_1^\alpha p_2^\beta+p_2^\alpha p_1^\beta)\,.
\end{align}
To obtain the coefficients $C_{ij}$, one needs to contract $C^{\alpha \beta}$
with $g_{\alpha\beta}$, resulting in
\begin{align}
C^{\alpha}_{\phantom{\alpha}\alpha}
= \int [d^4q] \frac{q^2}{\qbar^2 \bar D_1 \bar D_2}\,.
\end{align}
Since $q^2$ in the numerator is \textit{not} generated by Feynman rules, now it
would be incorrect to simplify it with the $\qbar^2$ denominator, in the sense
that one would not obtain the correct value of $C^{\alpha \beta}$. Here is the
place where the \FDR\ 'extra integrals' play an active role. In fact, by adding
and subtracting $\mu^2$, one rewrites
\begin{align}
\label{eq:alal}
C^{\alpha}_{\phantom{\alpha}\alpha}= 
\int [d^4q] \frac{\qbar^2 + \mu^2}{\qbar^2 \bar D_1 \bar D_2}=
 \int [d^4q] \frac{1}{\bar D_1 \bar D_2}
+\int [d^4q] \frac{\mu^2}{\qbar^2 \bar D_1 \bar D_2}\,,
\end{align}
which produces the correct answer in terms of a minimum set of scalar MIs.
In other words, thanks to the introduction of extra integrals, \Eqn{eq:FDRdef}
can be considered as a convenient way to define a loop integration for divergent
integrals that survives algebraic four-dimensional manipulations. This is a
peculiar property of \FDR. 

In the computation at hand, only $C_{00}$ and $C_{12}$ are needed.
The reduction gives
\begin{align}
C_{00} = \frac{I_{2,\text{\FDR}}}{4} + \frac{EI}{2}\,,\qquad
C_{12} = \frac{EI}{s}, 
\end{align}
\vspace*{-.3cm}
with%
\footnote{
  The value of the 'extra integral' $EI$ is the same as the one of
  $(2\pi)^4\,I_{3}^{\dim}[\mu^2]$ obtained in \FDF, see Eq.~\eqref{eq:I2muRes}.}
\begin{subequations}
\label{eq:BandEI}
\begin{align}
\label{eq:B}
I_{2,\text{\FDR}}  &=  \int [d^4q] \frac{1}{\bar D_1 \bar D_2} 
= \int_0^1 dx \int [d^4q] \frac{1}{[\qbar^2+s\,x (1\!-\!x)+i 0]^2}
= -\pi^2
  \Big(\ln \frac{-s\!-\!i 0}{\mur^2}\!-\!2\Big)\,,
\\*
\label{eq:EI}
EI &= \int [d^4q] \frac{\mu^2}{\qbar^2 \bar D_1 \bar D_2}
= \frac{i \pi^2}{2}\,,
\end{align}
\end{subequations}
see Eqs.~\eqref{eq:fin} and~\eqref{eq:eqextra}.
Analogously, one reduces the rank-one tensor
\begin{align}
C^{\alpha} \equiv  \int [d^4q] \frac{q^\alpha}{\qbar^2 \bar D_1 \bar D_2}=
  C_{1}^{\phantom{\alpha}}\,  p_1^\alpha
+ C_{2}^{\phantom{\alpha}}\,  p_2^\alpha,
\end{align}
obtaining
\vspace*{-.4cm}
\begin{align}
C_{1} = C_{2}= \frac{I_{2,\text{\FDR}}}{s}.
\end{align}
In summary, the virtual amplitude can be expressed as a linear
combination of the scalar integrals in Eqs.~\eqref{eq:eqn1} and%
~\eqref{eq:BandEI}. Multiplying with the Born amplitude and taking
the real part, one obtains%
\footnote{This result is identical to the one obtained in \IReg,
compare with Eq.~\eqref{crsv}.}
\begin{align}
\sigma^{(v)}_{\text{\FDR}}
=\sigma^{(0)}\,\Big(\frac{\alpha_s}{\pi}\Big)\,\CF\,\Big[
  -\frac{\ln^2(\mu_0)}{2}
  -\frac{3}{2} \ln(\mu_0)
  -\frac{7-\pi^2}{2}+\mathcal{O}(\mu_0)
  \Big]\,,
\label{eq:virt}
\end{align} 
where $\sigma^{(0)}$ is the Born total cross section given in
Eq.~\eqref{sig_yqq} and $\ln\mu_0$ is the IR logarithm. The process
at hand is UV finite, so that the dependence on the logarithms has to
drop in the final result. As a consequence, the effect of all\
scaleless integrals (nullified by our particular choice $\mur=\mu$)
is nothing but $\ln s/\mur^2 \to \ln s/\mu^2$ in \Eqn{eq:B}, as can be
easily checked with an explicit calculation.

\subsubsection*{Real contributions}
\label{sec:realFDR}

As for the bremsstrahlung contribution
$e^+e^-\!\to\!\gamma^{*}\!\to\!q(p_1)\,\bar q(p_2)\, g(p_3)$, a tensor
decomposition of the three-particle phase-space integrals produces the
matrix element squared\footnote{This corresponds to the usual matrix
  element squared for massless particles computed in four dimensions,
  as given in \Eqn{mrealfour}.}
\begin{align}
\label{eq:realme}
M_{\text{\FDR}}^{(0)}(s_{12},s_{13},s_{23})
= \frac{16 \pi\, \alpha_s}{s}\,\CF\,
  M^{(0)}_{\text{\FDR}}(s)
  \Big(
  \!-\!\frac{s}{ s_{13}}
  \!-\!\frac{s}{ s_{23}}
  \!+\!\frac{ s_{13}}{2\,s_{23}}
  \!+\!\frac{ s_{23}}{2\,s_{13}}
  \!+\!\frac{s^2}{s_{13}\,s_{23}}
  \Big)\,,
\end{align}
where $M^{(0)}_{\text{\FDR}}(s)$ is the fully inclusive Born matrix element
squared of $e^+ e^-\!\to\!\gamma^*\!\to\! q(k_1)\bar q(k_2)$,
\begin{align}
M^{(0)}_{\text{\FDR}}(s)
= \frac{2}{\pi} \int\displaylimits_{\Phi_2} M^{(0)}_{\text{\FDR}}(k_1,k_2)\,.
\end{align}
In accordance with \Eqn{eq:sigmatot}, we now replace all the invariants
by their massive counterparts, $s_{ij} \to \bar s_{ij}$, and integrate over a  
$\mu$-massive three-body phase-space,
\begin{align}
\int\displaylimits_{\bar \Phi_3}\!\!
  M_{\text{\FDR}}^{(0)}(\bar s_{12},\bar s_{13},\bar s_{23})
\!=\! \frac{4 \pi^3\, \alpha_s}{s^2}\, \CF\,
  M^{(0)}_{\text{\FDR}}(s)\!
  \int\displaylimits_{\bar R_3}\!\! d\bar s_{13} d\bar s_{23}
  \Big(
    \!-\!\frac{s}{ \bar s_{13}}
    \!-\!\frac{s}{ \bar s_{23}}
    \!+\!\frac{ \bar s_{13}}{2\, \bar s_{23}}
    \!+\!\frac{ \bar s_{23}}{2\,\bar s_{13}}
    \!+\!\frac{s^2}{\bar s_{13}\,\bar s_{23}}
    \Big)\,.
\end{align}
The quantity $\bar R_3$ represents the physical region of the Dalitz plot
for the $\mu$-massive three-particle phase-space parametrized in terms of
$\bar s_{13}$ and $\bar s_{23}$. The limit $\mu \to 0$ is understood
from now on.
The needed integrals can be expressed in terms of the scaled invariants
\begin{align}
\bar x= \frac{\bar s_{13}}{s}-\mu_0\,,\quad
\bar y= \frac{\bar s_{23}}{s}-\mu_0\,,\quad
\bar z= \frac{\bar s_{12}}{s}-\mu_0\,,\qquad{\rm with}\qquad
\mu_0=\frac{\mu^2}{s}\,,
\label{eq:barPar}
\end{align}
and are listed in Ref.~\cite{Pittau:2013qla}. We report them here for
completeness%
\footnote{Similar integrals have to be evaluated when using the
\IReg\ framework to determine the real contributions,
see Eqs.~\eqref{eq:RealIntegralsIReg}. Their counterparts in \DR\ are
given in Eqs.~\eqref{psintegrals}.}
\begin{subequations}
\label{eq:ints}
\begin{align}
\int\displaylimits_{\bar R_3}\!\! d\bar s_{13} d\bar s_{23}
\frac{1}{\bar s_{13}}
=\int_{\bar R_3}\!\! d\bar s_{13} d\bar s_{23}
  \frac{1}{\bar s_{23}}
=s\int\displaylimits_{\bar R_3} d \bar x d \bar y\, \frac{1}{\bar y+\mu_0}
&=s \, \Big[
  \!-\!\ln(\mu_0)
  -3
  +\mathcal{O}(\mu_0) \Big]\,,
  \phantom{\Big|}
\label{eq:inta}
\\
\int\displaylimits_{\bar R_3}\!\! d\bar s_{13} d\bar s_{23}
\frac{\bar s_{13}}{\bar s_{23}}
=\int\displaylimits_{\bar R_3}\!\! d\bar s_{13} d\bar s_{23}
  \frac{\bar s_{23}}{\bar s_{13}}
=s^2 \int\displaylimits_{\bar R_3} d \bar x d \bar y\, \frac{\bar y}{\bar x+\mu_0}
&=s^2\Big[
  \!-\!\frac{\ln(\mu_0)}{2}
  -\frac{7}{4}
  +\mathcal{O}(\mu_0)\Big]\,,
\label{eq:intb}
\\
\int\displaylimits_{\bar R_3}\!\! d\bar s_{13} d\bar s_{23}
\frac{1}{\bar s_{13}\,  \bar s_{23}}
=\int\displaylimits_{\bar R_3} d \bar x d \bar y\,
  \frac{1}{(\bar x+\mu_0)(\bar y+\mu_0)}
&=\frac{\ln^2(\mu_0)}{2}-\frac{\pi^2}{2}+\mathcal{O}(\mu_0)\,.
\phantom{\Big|}
\label{eq:intc}
\end{align}
\end{subequations}
The final result of the bremsstrahlung contribution reads%
\footnote{This result is identical to the one obtained in \IReg,
compare with Eq.~\eqref{ireg:real}.}
\begin{align}
\sigma_{\text{\FDR}}^{(r)}
= \sigma^{(0)}\left(\frac{\alpha_{s}}{\pi}\right)
  \CF\Big[\,
    \frac{\ln^{2}(\mu_{0})}{2}
    +\frac{3}{2}\ln(\mu_0)
    +\frac{17}{4}
    -\frac{\pi^{2}}{2} 
    + \mathcal{O}(\mu_0)
    \Big]\,.
\phantom{\Bigg|}
\label{eq:realpart}
\end{align} 
Adding the virtual contribution given in \Eqn{eq:virt}
produces the total NLO correction
\begin{align}
\sigma^{(1)}
= \sigma^{(0)} + \sigma^{(v)}_{\text{\FDR}} +
\sigma^{(r)}_{\text{\FDR}}\Big|_{\mu_0\to 0}
=  \frac{\eqSq\,N_c}{3\, s}\Big(\frac{e^4}{4\pi}\Big)
\Big[\,1+\Big(\frac{\alpha_s}{4\pi}\Big) \, 3\, \CF \Big] \, .
\phantom{\Bigg|}
\label{eq:totnlo}
\end{align}

Finally, we remark that it is possible to set up the entire calculation
in a fully local fashion. To achieve this, one has to rewrite the double
and single logarithms in \Eqn{eq:virt} as local counterterms to be added
to the real integrand. For instance, \Eqn{eq:intc} gives
\begin{align}
\ln^2(\mu_0)-\pi^2
= 2
  \int\displaylimits_{\bar R_3} d\bar s_{13} d\bar s_{23}
  \frac{1}{\bar s_{13} \bar s_{23}}\,.
\end{align}
The full counterterm needed for the case at hand can be inferred
uniquely from the factorization properties of the matrix element squared,
\begin{align}
\label{eq:ct}
M^{\text{\CT}}_{\text{\FDR}}(p_1,p_2,p_3)
= \frac{16 \pi\, \alpha_s}{s}\,\CF\,
  M^{(0)}_{\text{\FDR}}(\hat{p}_1,\hat{p}_2)
  \Big(
    \!-\!\frac{s}{\bar s_{13}}
    \!-\!\frac{s}{\bar s_{23}}
    \!+\!\frac{\bar s_{13}}{2\bar s_{23}}
    \!+\!\frac{\bar s_{23}}{2\bar s_{13}}
    \!+\!\frac{s^2}{\bar s_{13} \bar s_{23}}
    \!-\!\frac{17}{2}
    \Big)\,.
\end{align}
This equation is in agreement with \Eqn{eq:realme} when integrating over 
$\hat p_1$ and $\hat p_2$.  The constant $\frac{17}{2}$ is chosen in such
a way that only the logarithms and the $\pi^2$ term in \Eqn{eq:realpart}
are reproduced upon integration over $ \bar R_3$.
The quantity $M^{(0)}_{\text{\FDR}}(\hat{p}_1,\hat{p}_2)$ is computed
with mapped quark and anti-quark momenta defined as
\begin{align}
\hat{p}_1^{\alpha}
&= \kappa\, \Lambda^{\alpha}_{\phantom{\alpha}\beta}\,p_1^\beta\,
  \Big(1\!+\!\frac{s_{23}}{s_{12}}\Big),\quad
\hat{p}_2^{\alpha}
= \kappa\, \Lambda^{\alpha}_{\phantom{\alpha}\beta}\, p_2^\beta\,
  \Big(1\!+\!\frac{s_{13}}{s_{12}}\Big),\quad
\kappa
= \sqrt{\frac{s\,s_{12}}{(s_{12}\!+\!s_{13})(s_{12}\!+\!s_{23})}}\,,
\end{align}
where $\Lambda^{\alpha}_{\phantom{\alpha}\beta}$ is the boost that
brings the sum of the momenta back to the original center of mass frame,
$\hat{p}_1+\hat{p}_2= (\sqrt{s},0,0,0)$.
After subtracting $M^{\text{\CT}}_{\text{\FDR}}(p_1,p_2,p_3)$ from the 
exact matrix element squared, $\mu$ can be set to zero before integration.
In this case, an analytic knowledge of
$M_{\text{\FDR}}^{(0)}(s_{12},s_{13},s_{23})$ is not necessary.
A simple flat Monte Carlo with $10^5$ phase-space points reproduces the
result in \Eqn{eq:totnlo} at the 1 per mil level in a quarter of second.

\subsection{Established properties and future developments of FDR}
\label{sec:propertiesFDR}

\subsubsection*{Correspondence between integrals in FDR and DS}

At one loop, a one-to-one correspondence exists between integrals
regularized in \FDR\ and \DR. More precisely, according to the
definition of \FDR, any result of a loop integration is UV finite,
whereas IR divergences are expressed in powers of (logarithms of)
$\mu_0=\mu^2/s$. In \DR, on the other hand, results of an
integration in $\dim$ dimensions can be expanded in powers
of~$\epsilon$; UV and IR divergences are then parametrized as poles
$1/\epsilon^n$.

To provide an example for the relation between IR divergences
of integrals in \FDR\ and \DR, we consider the integral in
Eqs.~\eqref{eq:FDRirExample}. Using $\dim$-dimensional integration,
its result reads
\begin{align}
 I_{\text{\DR}}
  \!=\!c_{\Gamma}(\epsilon)\,\frac{i\pi^2}{s}\,\bigg[
    \frac{1}{\epsilon^2}
    +\frac{i\pi}{\epsilon}
    -\frac{\pi^2}{2}+\mathcal{O}(\epsilon)
    \bigg]\,.
\label{eq:FDRDSex}
\end{align}
The factor $c_{\Gamma}(\epsilon)$ is directly related to
integration in $\dim$ dimensions. It is given in Eq.~\eqref{cgammadef2}.
Comparing the result in Eq.~\eqref{eq:FDRDSex} with Eq.~\eqref{eq:eqn1},
the relation between the (regularized) IR divergences is given by
\begin{align}
 \frac{1}{\epsilon^2}\leftrightarrow\frac{1}{2}\,\text{ln}^2(\mu_0)\,,
 \qquad\quad
 \frac{1}{\epsilon}\leftrightarrow\text{ln}(\mu_0)\,.
\end{align}
Extending this to the 'finite' terms, the following generalized
relation for a (potentially UV and IR divergent) integral over a
generic integrand $F$ holds,
\begin{align}
\bigg[\frac{1}{(2\pi)^{4}}\int [d^4q]\,F(\qbar^2,q)\bigg]_{\mu^0}
=\bigg[c_{\Gamma}(\epsilon)^{-1}\,
  \mu^{4-\dim}\!\int\! \frac{d^d q}{(2\pi)^{\dim}}\,F(q^2,q)
  \bigg]_{\epsilon^0}\,.
\end{align}
Analogously, for the real contribution one finds
\begin{align}
\Bigg[\,
  \int\displaylimits_{\bar R_3}\!
    d\bar x\, d\bar y\, d\bar z\ F(\bar x,\bar y,\bar z)\
    \delta(1\!-\!\bar x\!-\!\bar y\!-\!\bar z)\Bigg]_{\mu^0}
=\Bigg[\,
  \Big(\frac{\mu^2}{s}\Big)^\epsilon\!
  \int\displaylimits_{R_3}\! dx\, dy\, dz\ F(x,y,z)\
    \frac{\delta(1\!-\!x\!-\!y\!-\!z)}{(x\,y\,z)^{\epsilon}}
    \Bigg]_{\epsilon^0}\,,
\end{align}
where $R_3,\,x,\,y$, and $z$ are the massless counterparts of
$\bar R_3,\,\bar x,\,\bar y$, and $\bar z$, respectively,
see also Eq.~\eqref{eq:barPar}.

Finally, there exists a connection between between the \FDR\
'extra integrals' and \FDF\ integrals containing powers of
the $(\!-2\epsilon)$-dimensional part of the loop momentum,
$q_{[-2\epsilon]}\equiv\tilde q$,
\begin{align}
\int [d^4q]\, F(\qbar^2,q,-\mu^2)
=\mu^{4-\dim}\!\int\! d^dq\,F(q^2,q,\,\tilde q^2)\,.
\end{align}
For more comments on the interplay $\!-\mu^2\leftrightarrow\tilde{q}^2$,
see also the discussion around Eq.~\eqref{eq:kSqFDF} and
Ref.~\cite{Pittau:2012zd}.

\subsubsection*{Gauge invariance, unitarity, and extra integrals}

Global prescriptions, such as the one described at one loop in \Eqn{eq:gp},
can be defined at any loop order. Their role is maintaining the needed gauge
cancellations. However, this is not enough to guarantee that results are
compatible with unitarity. In fact, in a unitary QFT, all perturbative orders
are linked by unitarity relations, and any renormalization procedure compatible
with unitarity has to fulfill the following two requirements:
\begin{itemize}
\item[a)] The UV divergences generated at any perturbative level should
  have no influence on the next perturbative orders.
\item[b)] The subintegration consistency in \Eqn{eq:subcon} should hold true.
\end{itemize}
Schemes based on \DR\ automatically respect subintegration consistency when
all objects (including $\gamma$~matrices) are treated in $d$ dimensions, while
requirement a) is fulfilled only if $1/\epsilon$ poles are subtracted
order-by-order by introducing counterterms in ${\cal L}$. This forbids one to
define \DR\ loop integrals beyond one loop by simply dropping $1/\epsilon$
poles. See the discussion is Section~2.5 of Ref.~\cite{Donati:2013voa} for
more details.

On the other hand, \FDR\ automatically respects requirement a) since
the UV subtraction is embedded in the definition of the \FDR\ integral,
so that there is no room for any UV divergence to have any influence at
higher perturbative levels. For instance, products of two one-loop \FDR\
integrals give the same result at any perturbative order, which is not the
case in~\DR.  

On the contrary, subintegration consistency is not automatically obeyed in 
\FDR. The reason for this can be traced back to the fact that the global
prescription needed at the level of divergent subdiagrams (sub-prescription)
clashes with the global prescription required at the level of the full diagram 
(full-prescription), so that one has to correct for this mismatch. 
However, this can be done directly at a diagrammatic level. This is possible
thanks to the \FDR\ extra integrals. They can be used to parametrize, in an
algebraic way, the difference between the result one gets when cancellations
do or do not take place between numerators and denominators, as illustrated,
for example, in \Eqn{eq:alal}. In practice, one looks at all possible UV
divergent subdiagrams, adds the piece needed to restore the 
sub-prescription and subtracts the {\em wrong} behaviour induced in the
subdiagram by the full-prescription. The net result of this process is the
addition of \FDR\ {\em extra-extra integrals} to the amplitude that enforce
requirement b) without the need of an order-by-order renormalization~%
\cite{Page:2015zca}. For example, a two-loop extra-extra integral can be
defined as the insertion of a one-loop extra integral into a two-loop \FDR\
integral. Thus, an \FDR\ calculation directly produces renormalized
quantities, which is a unique property of the \FDR\ formalism.

Work is in progress to find the connection between \FDR\ extra-extra
integrals and evanescent \FDH\ couplings. Preliminary results indicate that
the introduction of \FDR\ extra-extra integrals is equivalent to a restoration
of the correct behaviour under renormalization in an \FDH\ calculation in
which one sets equal gauge and evanescent couplings from the beginning. 

\section{FDU: Four-dimensional unsubtraction} 
\label{sec:fdu}

The four-dimensional unsubtraction (\FDU) \cite{Hernandez-Pinto:2015ysa,
Sborlini:2016gbr,Sborlini:2016hat,Rodrigo:2016hqc,Driencourt-Mangin:2016dpf}
approach constitutes an alternative to the traditional subtraction
method. It is based on the loop-tree duality (LTD) theorem \cite{Catani:2008xa,
Rodrigo:2008fp,Bierenbaum:2010cy,Bierenbaum:2012th}, which establishes a
connection among loop and dual integrals, the latter being similar to standard
phase-space integrals. In this way, the method provides a natural way to
implement an integrand-level combination of real and virtual contributions,
thus leading to a fully local cancellation of IR singularities. Moreover,
the addition of local UV counterterms allows to reproduce the proper results
in standard renormalization schemes.

In the following, we describe briefly the general facts about
the method, using the computation of the NLO QCD corrections to
$\gamma^{*}\!\to\! q \bar q (g)$ as a practical guideline.

\subsection{Introduction to LTD}
\label{ssec:FDUIntroduction}
The LTD theorem is based on Cauchy's residue theorem. Let us consider a
generic one-loop scalar integral for an $N$-particle process, where the
external momenta are labelled as $p_i$ with $i\in \{1,2,\ldots N\}$,
whilst the loop momentum is denoted by $\ell$. With these conventions,
the internal virtual momenta become $q_{i}\! =\! \ell\! + k_i$ where
$k_{i}\! =\! p_{1}\! +\! \ldots\! + p_{i}$ and $k_N\!=\!0$ because of
momentum conservation. If the mass of the internal particles is $m_i$,
a scalar integral can be expressed as
\beq
L^{(1)}(p_1, \dots, p_N) = \int_{\ell} \, \prod_{i=1}^{N} \,G_F(q_i)\,,
\label{eq:OneLoopScalar}
\eeq 
with the Feynman propagators $G_F(q_i) = (q_i^2-m_i^2+i 0)^{-1}$.
As usual, $q_i$ represents a four momentum which can be decomposed as
$q_{i,\mu} = (q_{i,0},\mathbf{q}_i)$, independently of the specific
space-time dimension\footnote{In other words, we could be working in
any of the \DR\ schemes mentioned in this article, with the only
requirement that the associated manifold is Lorentzian, i.\,e.\ that
it only contains \emph{time} component and an arbitrary number of
\emph{spatial} ones.}.
The energy component is $q_{i,0}$, whilst $\qb_{i}$ denotes the
spatial components.

At one-loop level, the dual representation of the loop integral is
obtained by cutting one by one all the available internal lines and
applying the residue theorem accordingly. The cut condition is
implemented by restricting the integration measure through the
introduction of
\beq
\td{q_i}
\equiv 2 \pi \, i \, \theta(q_{i,0}) \, \delta(q_i^2-m_i^2) \, ,
\label{eq:tddefinition}
\eeq
which transforms the loop integration domain into the positive energy
section (i.\,e.\ $q_{i,0}>0$) of the corresponding on-shell hyperboloid
(i.\,e.\ $q_i^2=m_i^2$). When the scattering amplitude under consideration
is composed by single powers of the propagators, the computation of the
residue simplifies to removing the cut propagator and replacing the
uncut ones with their \emph{duals}, i.\,e.\
\beq
G_D(q_i;q_j)
= \frac{1}{q_j^2 -m_j^2 - i 0 \, \eta \cdot k_{ji}} \, ,
\label{eq:DualPropagator}
\eeq
where $i,j \in \{1,2,\ldots N\}$, $k_{ji}\!=\! q_j\! - q_i$ and $\eta$ is
an arbitrary future-like or light-like vector, $\eta^2 \geq 0$, with positive
definite energy $\eta_0 > 0$. It is worth noticing that the dual prescription
takes care of the multiple-cut correlations introduced in the traditional
Feynman-tree theorem (FTT)~\cite{Feynman:1963ax,Feynman:1972mt},
thus allowing to prove their formal equivalence. 

In this way, the \emph{dual integrand} looks like a tree-level amplitude
whose building blocks are the same as in the standard theory with a
modified $i 0$ prescription. Thus, the one-loop scalar integral in
\Eq{eq:OneLoopScalar} reads
\beq
L^{(1)}(p_1, \dots, p_N)
= - \sum_{i=1}^N \, \int_{\ell} \, \td{q_i} \, \prod_{j\neq i} \,G_D(q_i;q_j)\,.
\label{eq:oneloopduality}
\eeq 
The existence of a dual representation for loop integrals straightforwardly
leads to a dual representation for loop scattering amplitudes. As explained
in Ref.~\cite{Catani:2008xa}, any loop contribution to scattering amplitudes
in any relativistic, local, and unitary quantum field theory can be computed
through the  decomposition into \emph{dual contributions}. Of course, this
idea generalizes to multi-loop amplitudes, where dual contributions
involve iterated single-cuts~\cite{Catani:2008xa,Bierenbaum:2010cy}.

For amplitudes containing higher powers of the propagators, the previous
result can be extended, as studied in Ref. \cite{Bierenbaum:2012th}.
It is worth appreciating that higher powers of the propagators explicitly
manifest when dealing with self-energy corrections at one loop, self-energy
insertions at higher orders, and when computing the local version of the
UV counterterms~\cite{Sborlini:2016gbr,Sborlini:2016hat}.

\subsection{Momentum mapping and IR singularities}
\label{ssec:FDUMapping}
The application of the LTD theorem to a virtual amplitude leads to a set
of dual contributions. From them, we can extract useful information about
the location of the singularities in the corresponding integration domain,
as well as the components (or cuts) that originate them. As explained in
Refs.~\cite{Buchta:2014dfa,Buchta:2015jea,Buchta:2015wna}, the intersection
of forward and backward hyperboloids defined by the on-shell conditions
allows to identify the IR (and threshold) singularities. Moreover, this study
is crucial to prove the compactness of the region developing IR divergences
\cite{Hernandez-Pinto:2015ysa,Sborlini:2016gbr,Sborlini:2016hat}, which
constitutes a very important result by itself. This is because the
real-radiation contributions are computed on a physical phase-space, which
is also compact\footnote{This assumption is true whenever the incoming
particles have fixed momentum, thus leading to a global constraint on the
energy available for generating final-state radiation.}. In consequence,
since the Kinoshita-Lee-Nauenberg (KLN) theorem states that there is a
cross-cancellation of IR singularities between real and virtual terms,
the compactness of the IR region inside the dual integration domain allows
to implement a local real-virtual cancellation of singularities by applying
a suitable momentum mapping. In this way, the singularities in the real
phase space (PS) are mapped to the dual integration domain where the
corresponding virtual singularities are generated; then, an integrand-level
cancellation takes place and there is no need of introducing any external
regulator to render the combination integrable.

\begin{figure}[t]
\begin{center}
\includegraphics[width=0.74\textwidth]{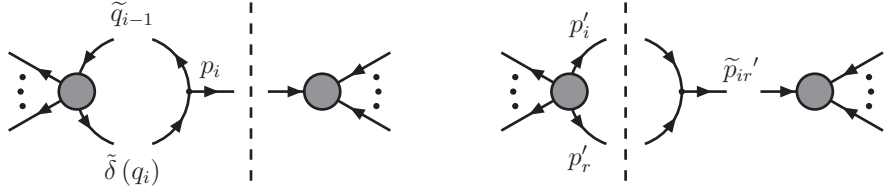}
\caption{\label{fig:factoriza}
Diagramatic contributions in the collinear limit, for both the dual
one-loop (left) and real-emission tree-level squared amplitudes (right).
The lines that are crossed by a dashed line correspond to on-shell states.
When particles are collinear, the parent becomes on-shell and the diagrams
factorize.}
\end{center}
\end{figure}

In order to connect the Born kinematics ($m$-particle PS) with the real-%
emission one (\mbox{$(m\!+\!1$)}-particle PS), we rely on techniques similar
to those applied for the dipole method~\cite{Catani:1996jh,Catani:1996vz}.
To be more concrete, let us start thinking about the virtual contribution.
After obtaining the dual amplitudes, we have a set of $m$ external momenta
and a free on-shell loop momentum. In this way, the dual amplitudes
introduce an \emph{extra} on-shell momentum. Since there are $(m\!+\!1)$
on-shell momenta available, the kinematics of the dual components exactly
matches the kinematics of the real contribution.

Then, it is necessary to isolate the real-emission IR singularities by
properly splitting the complete real PS. If $p_i'^{\mu}$ are the momenta
of the real-emission partons, we start by defining the partition
\begin{align}
{\cal R}_i = \{{y'}_{i r} < {\rm min}\, {y'}_{jk} \} \, , \quad \quad
\sum_{i=1}^{m} {\cal R}_i =1 \, ,
\label{eq:regionpartition} 
\end{align}
where ${y'}_{ij}=2\, p_i' \cdot p_j'/Q^2$, $r$ is the radiated parton
from parton $i$, and $Q$ is the typical hard scale of the scattering
process. It is important to notice that, inside ${\cal R}_i$, the only
allowed collinear/soft configurations are $i\parallel r$ or $p_r'^{\mu}\to 0$.
Thus, collinear singularities manifest in non-overlapping regions of the
real-emission PS which allows to introduce an optimized transformation
to describe the collinear configuration.

On the other hand, there are $m$ dual contributions, each one associated
with a single cut of an internal line. So, we can establish an identification
among partitions and dual amplitudes, based on the picture shown in
Fig.~\ref{fig:factoriza}. Concretely, the cut-line in the dual amplitude
must be interpreted as the extra-radiated particle in the real contribution;
i.\,e.\ $q_i \leftrightarrow p_r'$. Then, we settle in one of the partitions,
for instance ${\cal R}_i$. Because the only collinear singularity allowed
is originated by $i \parallel r$, we distinguish particle $i$ and call it
the {\it emitter}. After that, we single out all the squared amplitude-level
diagrams in the real contribution that become singular when $i \parallel r$
and cut the line $i$. These have to be topologically compared with the
dual-Born interference diagrams whose internal momenta $q_i$ are on-shell
(i.\,e.\ the line $i$ is cut), as suggested in Fig. \ref{fig:factoriza}.
In conclusion, the dual contribution $i$ is to be combined with the
real contribution coming from region ${\cal R}_i$.

The required momentum mapping is motivated by general factorization
properties in QCD \cite{Buchta:2014dfa,Catani:2011st} and the topological
identification in Fig.~\ref{fig:factoriza}. Explicitly, let us take the
$(m\!+\!1)$-particle real-emission kinematics, with $i$ as the emitter
and $r$ as the radiated particle, and we introduce a reference momentum,
associated to the spectator $j$. For the massless case, the generic
multi-leg momentum mapping with $q_i$ on-shell is given by 
\beqa
&& p_i'^\mu
= p_i^\mu - q_i^\mu + \alpha_i \, p_j^\mu\,, \quad
\quad p_j'^\mu = (1-\alpha_i) \, p_j^\mu\,, \quad \quad
p_k'^\mu = p_k^\mu \ \ k \ne i,j \ \, ,
\nn \\*
&& p_r'^\mu
= q_i^\mu~, \qquad \qquad \quad \quad \quad \ \,
\alpha_i = \frac{(q_i-p_i)^2}{2 p_j\cdot(q_i-p_i)}\,,
\label{eq:momentummapping}
\eeqa
with the primed momenta associated to the particles involved in the
real-emission process. In this case, note that $p_i'^2=p_j'^2=p_r'^2=0$ because
we restrict ourselves to massless particles. On the other hand, the initial-state
momenta ($p_a$ and $p_b$) are not altered by the transformation, neither is
$p'_k$ with $k\!\ne\!i,j$. Besides that, since
\begin{align}
p_i+ p_j +\sum_{k\ne i,j} p_k = p_i'+p_r'+p_j'+\sum_{k\ne i,j,r} p_k'\ \, ,
\label{eq:momentumconservationEXT}
\end{align}
the transformation preserves momentum conservation.
It is worth appreciating that this momentum mapping can be extended
to the massive case, even if the involved particles have different masses,
as we explained in Ref. \cite{Sborlini:2016hat}.

\subsection{Integrand-level renormalization and self-energies}
\label{ssec:FDURenormalization}
Besides dealing with IR singularities, any attempt to provide a
complete framework for higher-order computations must be able to treat
UV divergences. In this case, a suitable local version of the UV
counterterms is required. This topic is deeply discussed in
Ref.~\cite{Sborlini:2016gbr} for the massless case, whilst the massive
one is studied in Ref.~\cite{Sborlini:2016hat}. In the last case, the
self-energy and vertex corrections become non-trivial and some technical
subtleties arise: there are noticeable changes in the IR singular structure
compared to the massless case. On one hand, the mass acts as an IR
regulator, preventing collinear singularities to emerge. But,
on the other hand, soft singularities arising from gluon emissions
become non-vanishing because they are proportional to the mass of the
emitting leg. Since we are looking for a complete local cancellation
of singularities and a smooth massless transition, it is necessary that
the expressions for the massive case reduce to those already available
for massless processes, even at the \emph{integrand level}.

Let us start with the well-known expression for the wave-function
renormalization. Working in Feynman gauge with on-shell renormalization
conditions, its integrated form is given by%
\footnote{The result of the field renormalization coincides
  with the one of \CDR\ and not with the one of \FDH. In the latter scheme,
  the constant '$\!-\!4$' would be replaced by '$\!-\!5$', see e.\,g.\
  Eq.~(2.24) of Ref.~\cite{Gnendiger:2016cpg}.}
\beq
\Delta Z_2 = \Big(\aas\Big) \, \CF \left[
  -\frac{1}{\ep_{\text{\UV}}}-\frac{2}{\ep_{\text{\IR}}}
  + 3\,  \ln\left(\frac{M^2}{\mu^2}\right)
  -4\right]\,,
\label{eq:DeltaZ2expressionINTEGRADA}
\eeq
where we kept track of the IR and UV origin of the $\ep$ poles within \DR.
The unintegrated expression \cite{Sborlini:2016hat} is given by
\beqa
\nn \Delta Z_2(p_1) &=& -\g^2 \, \CF \,
\int_{\ell} G_F(q_1) \, G_F(q_3) \,
\left[(d-2)\frac{q_1 \cdot p_2}{p_1 \cdot p_2} \right.
+4\, M^2 \left. \left(1- \frac{q_1 \cdot p_2}{p_1 \cdot p_2}\right)
  G_F(q_3)\right]\,,
  \\ 
\label{eq:DeltaZ2expression}
\eeqa
which includes higher-order powers of the propagators, and where we define
$q_1\!=\!\ell\!+\!p_1$, $q_2\!=\!\ell\!+\!p_1\!+\!p_2$, and $q_3\!=\!\ell$.
It is worth appreciating that there are many equivalent integrand-level
expressions to describe $\Delta Z_2(p_1)$, but the one presented in
\Eq{eq:DeltaZ2expression} develops the proper IR behaviour to cancel
singularities coming from real-emission processes. Besides this, notice
that the corresponding formula for the massless case \cite{Sborlini:2016gbr}
is simply recovered by considering $M\!\to\! 0$ at the integrand level.
The term proportional to $M^2$ is responsible for soft divergences that
appear when $q_1$ is set on shell, and it vanishes as $M \to 0$ since soft
singularities are absent in the massless self-energy computation. On the
contrary, the collinear singularities that appear in $\Delta Z_2(M\!=\!0)$
manifest themselves as quasi-collinear divergences, i.\,e.\ terms that behave
like $\ln(M^2/\mu^2)$, as shown in \Eq{eq:DeltaZ2expressionINTEGRADA}.
Once we combine the self-energy contributions with the virtual matrix elements,
there are still UV singularities present. These have to be removed by performing 
an expansion around the UV propagator
$G_F(q_\text{\UV})
\!=\!(q_\text{\UV}^2\!-\!\mu_\text{\UV}^2\!+\!i 0)^{-1}$,
\begin{align}
G_F(q_i) = G_F(q_{\text{\UV}}) \, \left(1-
\frac{2q_\text{\UV} \cdot k_{i,\text{\UV}}+k_{i,\text{\UV}}^2+\mu_\text{\UV}^2-m_i^2}
  {q_\text{\UV}^2-\mu_\text{\UV}^2+i 0}+ \ldots\right)\, ,
\label{eq:ExpansionUV}
\end{align}
with the renormalization scale $\mu_\text{\UV}$ and
$k_{i,\text{\UV}}\!=\!q_i\!-\!q_\text{\UV}$.
A similar expansion is carried out in the numerator, which leads to the
UV counterterm for the wave-function renormalization,
\begin{align}
\nn\Delta Z_2^{\text{\UV}}(p_1) &= (2-d)\, \g^2 \, \CF \,
\int_{\ell} \big[G_F(q_\text{\UV})\big]^2 \,
  \left(1+\frac{q_\text{\UV} \cdot p_2}{p_1 \cdot p_2}\right) 
\Big[1\!-\!G_F(q_\text{\UV})(2\, q_\text{\UV} \cdot p_1 + \mu^2_\text{\UV})\Big]
\nn \\ 
&= - 
(4\pi)^{\epsilon}\,\Gamma(1+\epsilon)
\, \aas \, \CF \,
\left( \frac{\mu_\text{\UV}^2}{\mu^2}\right)^{-\ep}\,\frac{1-\ep^2}{\ep}
\,.
\label{eq:ParteUV}
\end{align}
The integrated form exactly reproduces the UV pole present in
\Eq{eq:DeltaZ2expressionINTEGRADA}. The subleading terms proportional
to $\mu_\text{\UV}^2$ are chosen to subtract only the pole part from
\Eq{eq:DeltaZ2expressionINTEGRADA} and, in this way, settle in the
$\MS$ scheme. Finally, we define the UV-free wave-function
renormalization 
\begin{align}
\Delta Z_2^{\text{\IR}} = \Delta Z_2 - \Delta Z_2^{\text{\UV}}\,,
\end{align}
that only contains IR singularities. To conclude this discussion, it is
important to emphasize that this construction is completely general and
that the subleading terms can be adjusted to reproduce the desired
scheme-dependent contributions.

Besides the wave-function renormalization, it is also necessary to
remove the UV singularities associated to the vertex corrections.
The corresponding renormalization counterterm in its unintegrated
form is given by
\begin{align}
{\Gamma}^{(1)}_{A, \text{\UV}} = \, \gs^2 \, \CF\, 
\int_\ell \big[ G_F(q_\text{\UV})\big]^3 \, 
\left(
  \gamma^{\nu} \, \bq_{\text{\UV}} \, {\Gamma}^{(0)}_A \, \bq_{\text{\UV}} \,
  \gamma_{\nu}
  - d_{A, \text{\UV}}^{\phantom{2}} \, \mu_{\text{\UV}}^2 \,
    {\Gamma}^{(0)}_{A}\right)\,,
\label{eq:UVGenericvertex}
\end{align} 
where ${\Gamma}^{(0)}_{A}$ represents the tree-level vertex. Again,
the term proportional to $\mu_{\text{\UV}}^2$ in the numerator is subleading
in the UV limit and its coefficient, $d_{A,\text{\UV}}$, must be adjusted in
order to implement the desired renormalization scheme \cite{Sborlini:2016hat}.

\subsection{Application example:
$e^{+}\,e^{-}\to\gamma^{*}\to\,q\bar q$ at NLO}
\label{ssec:FDUExample}
In order to compute the NLO QCD corrections to
$e^{+} e^{-}\!\to\!\gamma^{*}\!\to\!q\bar q$, we start from the
complete set of ${\cal O}(\as^2)$ real and virtual diagrams, including the
self-energy ones. The total \emph{unrenormalized} virtual cross section is
\begin{align}
\sigma^{(v)}_{\text{\FDU}}
= \frac{1}{2\,s_{12}} \, \int d\Phi_{1\to 2} \,
  \left\{
    2\, \operatorname{Re}
      \la {\cal A}^{(0)}_{\text{\FDU}}| {\cal A}^{(1)}_{\text{\FDU}} \ra
    + \big[\Delta Z_2(p_1)+\Delta Z_2(p_2)\big]\, M^{(0)}_{\text{\FDU}}\right\}\,,
\label{eq:virtualunrenormalized}
\end{align}
where we distinguish contributions originated in the triangle diagram
from those related to self-energies. After that, we must introduce the
local UV counterterms which implements the desired renormalization
scheme and replace the self-energy contributions by the wave-function
renormalization constants, $\Delta Z_2^\text{\IR}$. In this case, we
apply LTD to \Eq{eq:virtualunrenormalized} and obtain a set of three
dual contributions, $\widetilde \sigma^{(v)}_{i,\text{\FDU}}$.

Once the dual contributions are computed, we turn the attention to the
real-emission terms. As explained in Sec. \ref{ssec:FDUMapping}, we
need to isolate the different collinear singularities by introducing a
partition of the real phase space. This leads to
\begin{align}
\widetilde \sigma^{(r)}_{i,\text{\FDU}}
= \frac{1}{2\,s_{12}} \, \int d\Phi_{1\to 3} \,  \,
M_{\text{\FDU}}^{(0)}(\qqg) \, \theta(y_{jr}' -y_{ir}' )~
\qquad i,j \in \{ 1,2\} \, , \quad i \neq j \,,
\label{eq:realalqqg}
\end{align}
which fulfills $\widetilde \sigma^{(r)}_{1,\text{\FDU}}+
\widetilde \sigma^{(r)}_{2,\text{\FDU}} = \sigma^{(r)}_{\text{\FDU}}$.
After that, we apply the real-virtual mapping in each
partition. This converts the real terms into fully local IR counterterms
for the dual contributions; this guarantees a complete cancellation of IR
singularities at the integrand level, thus rendering the full expression
integrable in four dimensions. This is a really important fact, because
it allows to put aside \DR\ safely by directly considering the limit
$\ep\! \to\! 0$ at the integrand level \cite{Hernandez-Pinto:2015ysa}.
Finally, the master formula for computing the \emph{finite}
cross-section correction is
\begin{align}
\sigma^{(1)}
= {\cal T}\left(\sum_{i=1}^3 \, \widetilde \sigma^{(v)}_{i,\text{\FDU}}
  +\, \sum_{j=1}^2 \, \widetilde \sigma^{(r)}_{j,\text{\FDU}}\right)
- \widetilde \sigma^{\text{\UV}} \, ,
\label{eq:MasterFORMULA}
\end{align}
where $\widetilde \sigma^{\text{\UV}}$ is the dual representation
of the local UV counterterms and ${\cal T}$ is an operator that
implements the unification of dual coordinates at the integrand level
(with the corresponding Jacobians). If we add all the contributions at the
integrand level and deal with a single master integration, the expression
in \Eq{eq:MasterFORMULA} is directly implementable in four space-time
dimensions and leads to the correct result after numerical computation.
It is worth mentioning that, in order to improve the numerical
stability, it helps to compactify the integration domain, applying a
transformation as suggested in Ref.~\cite{Sborlini:2016hat}.

\subsection{Further considerations and comparison with other schemes}
\label{ssec:FDUOutlook}
As we depicted in the previous paragraphs, the \FDU\ approach is based on a
fully local cancellation of IR and UV singularities in strictly four dimensions.
In this way, we avoid many of the practical/conceptual problems related to
the extension of physical properties to $d$ space-time dimensions. In particular,
the $\gamma^5$ issue is naturally absent here. Moreover, the idea of using the
mapped real contributions as local IR counterterms for the dual part simplifies
the treatment of IR divergences, as well as it provides a better understanding
of their origin.

On the other hand, the application of the traditional renormalization procedure
within this framework implies to recompute the renormalization constants in an
unintegrated form (i.\,e.\ for the integrand-level implementation). In any case,
by fixing subleading terms in the UV expansion it is possible to specify the
finite part of the counterterms, thus reproducing the results in any scheme
(for instance, in $\overline{\rm MS}$). Moreover, this algorithm is completely
process-independent and, in consequence, fully compatible with higher-order
computations. In this sense, the treatment of UV divergences is similar to
the procedure proposed within \FDR. The main difference is that we transform
the local counterterms to the dual-space, in order to combine it with
virtual amplitudes.

Besides this, it is worth mentioning that LTD can handle loop amplitudes,
as any other method described in this report, but \FDU\ is designed to work
directly with physical observables. For instance in Ref.~%
\cite{Driencourt-Mangin:2017gop}, we applied our framework to deal with the
Higgs boson decay to massless gauge bosons, which although known to be finite
still requires a proper regularization due to the fact that the amplitudes are
UV singular locally.

Finally, we would like to emphasize that \FDU\ is compatible with the
desired requirements mentioned in the introduction. In fact, since it is a
four-dimensional approach which relies on proper physically motivated changes
of variables, \FDU\ does not alter the four-dimensional properties of the
underlying theory (i.\,e.\ unitarity, causality, and associated symmetries).
Moreover, it fulfills the crucial requirement of mathematical consistency
because singularities are completely removed by a local mapping. In this way,
all the singularities are cancelled before they manifest themselves in the
integration.

\section{Summary and outlook \label{sec:summary}}

The vast majority of higher-order calculations are done using
\CDR. While there is no doubt that this made possible impressive
progress in perturbative calculations, there is a certain danger that
this success stifles the progress of other methods. Whether such
alternative methods will ever result in a viable way to perform actual
computations can only be established by actually using them. In order
to facilitate this, this article provides an overview of recent (and
not so recent) developments of regularization schemes other than
\CDR. Some are very close to \CDR, for others the differences are much
larger. Using simple examples, we have illustrated the differences and
similarities of these methods and their relation to \CDR. Let us
summarize the key points by means of the following list.

\begin{description}
\item{\bf FDH and DRED} are perfectly consistent regularizations
  schemes, at least up to NNLO. However, they require the introduction
  of additional (evanescent) couplings with (in general) different
  counterterms. In non-supersymmetric theories, for \DRED\ this is already
  mandatory at NLO, for \FDH\ this is unavoidable only at NNLO and beyond.
  Supersymmetry might protect the equivalence of the couplings even beyond
  these approximations.
  Statements in the literature that \FDH\ is inconsistent always refer to
  `naive \FDH', i.\,e.\ \FDH\ without distinguishing the couplings.
\item{\bf Conversions} between results in \CDR, \HV, \FDH, and
  \DRED\ can be made for individual parts contributing to a cross
  section. For the virtual contributions this is known to NNLO and can
  be elegantly described solely through the scheme dependence of
  $\beta$~functions and anomalous dimensions. For real corrections and
  initial-state factorization terms the explicit scheme dependence is
  only known to NLO. These results have been used to explicitly
  demonstrate the scheme independence of a cross section at NLO.
\item{\bf FDF} is an adaption of the (naive) \FDH~scheme that can be
  used in strictly four dimensions. This enables the use of unitarity
  methods, writing loop integrands as products of tree-level
  amplitudes and performing numerical calculations with the components
  of spinors and momenta.  At NLO, \FDF\ gives results that are
  equivalent to \FDH.  How to extend this beyond NLO is currently under
  investigation. The scalars of \FDF\ are not identical to the
  $\epsilon$-scalars of \FDH.
\item{\bf GoSam} makes use of \FDF\ and other four-dimensional
  techniques. The one-loop virtual amplitudes that are called `\DRED'
  and '\CDR' in GoSam correspond to what we call `naive \FDH' and
  '\HV', respectively, in this article. Virtual one-loop amplitudes
  in other schemes are obtained indirectly through conversion formulas.
\item{\bf SDF} is based on the same idea as \FDF. However, having
  two-loop amplitudes in mind, the integer dimension is set to
  $\dsix\!=\!6$. Hence, the spinor formalism has to be extended to 6
  dimensions.
\item{\bf UV singularities in IREG and FDR:} The basic idea of
  \IReg\ and \FDR\ is similar and based on the observation that UV
  singularities are independent of the kinematics. This is used to
  isolate the UV singular part of loop integrals. In \IReg, the UV
  singular part is expressed in terms of (implicit) integrals
  $I_{\text{log}}$ and boundary terms (that have to be set to zero to
  respect gauge invariance), whereas in \FDR\ they are set to zero.
  The resulting UV finite integrals are evaluated in (strictly) four
  dimensions.  
\item{\bf IR singularities in IREG and FDR} are also treated in
  strictly four dimensions. The matrix elements squared are computed for
  massless particles (in four dimensions) and the phase space integration
  is also carried out in four dimensions. IR singularities are
  regularized by modifying the phase-space boundaries through a shift
  $q\!\to\!q\!+\!\mu$ and result in logarithms $\ln(\mu_0)\!=\!\ln(\mu^2/s)$.
  In this sense the method is similar to the introduction of a photon or
  gluon mass. However, the procedures used by \IReg\ and \FDR\ are
  superior as they preserve gauge invariance.
\item{\bf Differences between IREG and FDR:} In \IReg, gauge invariance
  is achieved by performing first the Dirac algebra in the numerator
  and then cancel terms in the numerator and denominator before the
  shift $q\!\to\!q\!+\!\mu$. In \FDR, the shift is done universally in the
  numerator and denominator. Then additional terms with $\mu^2$ in the
  numerator  (called `extra integrals') are included. \IReg\ produces
  expressions where the UV singularities are still present in the form
  of implicit integrals $I_{\text{log}}$. They have to be removed by a
  suitable renormalization procedure, as in \DR. Applying \FDR, on the
  other hand, results directly in UV renormalized quantities.
\item{\bf Relation between IREG/FDR and dimensional schemes:} In
  \IReg\ and \FDR, 'singularities' related to real contributions are
  encoded in powers of $\ln(\mu_0)$. At NLO, there is a direct mapping
  between these terms and the $1/\epsilon^n$ singularities in the \FDH~scheme,
  namely $1/\epsilon^2 \leftrightarrow 1/2\,\ln^2(\mu_0)$ and
  $1/\epsilon \leftrightarrow \ln(\mu_0)$. The extension to NNLO of
  such a correspondence between the four-dimensional schemes and the
  traditional dimensional schemes is under active investigation. This
  also includes on how to compensate for the absence of evanescent
  couplings in \IReg\ and \FDR.
\item{\bf FDU} is an even more radical method in that it does not
  split a cross section into (potentially IR divergent) virtual and
  real parts. Rather, the combination of the two parts (and thus the
  cancellation of IR singularities) is done at the integrand
  level. Local counterterms are used to perform $\MS$
  renormalization.  The extension to 
initial-state singularities is also possible; the application of a
slightly modified momentum mapping allows to cancel the soft
singularities. The remaining initial-state collinear singularities
can be canceled by adding unintegrated initial-state counterterms.
This is currently under investigation.
\item{\bf Evanescent couplings} are a fact of life! Even though they
  can be avoided at NLO in some four-dimensional formulations (like
  \FDH) or do not show up in some particular processes even at NNLO
  (like $g g\to g g$ in \FDH), they are present in all (partly)
  four-dimensional regularizations of QED and QCD. In particular, they
  have an effect at NNLO in \FDH\ (like e.\,g.\ for $g g\to
  q\bar{q}$). The connection of these effects to the  'extra-extra
  integrals' in \FDR\ is under investigation.
\end{description}
The list above illustrates that there are promising alternatives
available that at least at NLO are well understood. They can and have
been used for NLO calculations and in some cases have proved to be
more efficient.

Currently, a huge effort in perturbative calculations is dedicated to
going beyond NLO towards automated computations at NNLO. Many of the
schemes above have been revisited in the hope they provide a smoother
road towards this goal. We are convinced that this deserves to be
investigated more thoroughly. In any case, for an alternative scheme
to be consistent, there must -- at least in principle~-- exist a
well-defined relation to \CDR. At NNLO, these relations are fairly well
established for other traditional dimensional schemes like \HV, \FDH,
and \DRED. Regarding new formulations of dimensional schemes like
\FDF\ or non-dimensional schemes such as \IReg\ and \FDR, first steps
towards establishing such relations have been made. \FDU\ has the
advantage that a separate regularization of the final-state IR 
singularities is not required, but only the UV singularities have to be
treated in a well-defined way, such as $\MS$.

Comparing to the impressive list of NNLO calculations for physical
cross sections that have been made using \CDR, it is fair to say, that
none of the other methods has had a similar impact so far. Since
\CDR\ is the best established scheme, it is tempting to keep using
it. However, it is not clear at all, if \CDR\ is really the most
efficient scheme. Hence, the investigation of other regularization
schemes is an important aspect of making further progress in
perturbative computations. Are there more efficient dimensional
schemes? Or is it ultimately advantageous to work completely in
four dimension? 

That is the question.


\vskip 7mm
\centerline{\bf Acknowledgments}
\vskip 2mm

\noindent This article is a result of a Workstop/Thinkstart that took
place on 13.--16.~September~2016 at the Physik-Institut of the
University of Z\"urich (UZH). We gratefully acknowledge support of UZH
in general and its Physics Department in particular. A special thank
you to M.~R\"ollin and C.~Genovese for their help in organizing the
workshop.
We thank S.~C.~Borowka for carefully reading the manuscript and helpful
comments.

\noindent G.~M.~Pruna, Y.~Ulrich and A.~Visconti are supported by the
Swiss National Science Foundation (SNF) under contracts 200021\_160156
and 200021\_163466, respectively.

\noindent W.~J.~Torres Bobadilla is supported by Fondazione 
Cassa di Risparmio di Padova e Rovigo (CARIPARO).

\noindent A.~Cherchiglia is supported by CAPES (Coordena\c{c}{\~a}o
de Aperfei\c{c}oamento de Pessoal de N{\'i}vel Superior) - Brazil.
B.~Hiller acknowledges partial support from the FCT (Portugal) project
UID/FIS/04564/2016.
M.~Sampaio acknowledges financial support from the Brazilian institutions
CNPq (Conselho Nacional de Desenvolvimento Científico e Tecnol{\'o}gico) and
FAPEMIG (Funda\c{c}{\~a}o de Amparo {\`a} Pesquisa do Estado de Minas Gerais).

\noindent R.~Pittau was supported by the Research Executive Agency
(REA) of the European Union under the Grant Agreements
PITN-GA2012-316704 (HiggsTools) and ERC-2011-AdG No 291377
(LHCtheory),\,and\,by\,the\,MECD\,Proyects\,FPA2013-47836-C3-1-P\,and\,%
FPA2016-78220-C3-3-P.

\noindent F.~Driencourt-Mangin, G.~Rodrigo, G.~Sborlini, and 
W.~J.~Torres Bobadilla have been supported by CONICET Argentina,
by the Spanish Government, by ERDF funds from the European Commission
(Grants No. FPA2014-53631-C2-1-P and SEV-2014-0398) and by Generalitat
Valenciana (Grants No. PROMETEOII/2013/007 and GRISOLIA/2015/035).
G.~Sborlini was supported in part by Fondazione Cariplo under the grant
number 2015-0761.

\bibliography{schemes}
\bibliographystyle{JHEP}

\end{document}